\documentclass[useAMS,usegraphicx]{mn2e}

\title[On the highly reddened members in young clusters]{On the highly reddened members in 6 young galactic star 
    clusters - a multiwavelength study \thanks{Figures 5 and 8 are available electronically and can be obtained from 
    authors} }

\author[Brijesh Kumar et al.]{Brijesh Kumar$^{1}$\thanks{E-mail: brij@upso.ernet.in}, Ram Sagar$^{1}$, B. B. Sanwal$^{1}$ 
and M. S. Bessell$^{2}$\\ 
$^{1}$State Observatory, Manora Peak, Nainital, 263 129, Uttaranchal, India\\
$^{2}$ Research School of Astronomy and Astrophysics, Mount Stromlo Observatory, Cotter Road, Weston, ACT 2611, Australia}
\begin{document}

\date{December 2003}

\pagerange{\pageref{firstpage}--\pageref{lastpage}} \pubyear{2004}

\maketitle

\label{firstpage}

\begin{abstract} 
The spectral and reddening properties of 211 highly reddened proper motion members 
with $V < 15$ mag in 6 young galactic  star clusters are investigated using low resolution spectroscopic, broad-band 
$UBVRIJHK$ and mid-IR data. We report emission features in CaII HK and HI lines for a sample of 29 stars including 
11 stars reported for the first time and also provide either a new or more reliable spectral class for a sample of 24 stars. 
CaII triplet width measurements are used to indicate the presence of an accretion disk for a dozen stars and to hint luminosity for 
a couple of stars. On the basis of spectral features, near-IR excesses, dereddened color-color diagrams and mid-IR spectral 
indices we identify a group of 28 pre-main sequence cluster members including 5 highly probable Herbig 
Ae/Be and 6 classical T Tauri star. A total of 25 non-emission MS stars, amounting to $\sim$ 10 \% early type MS members, 
appears to show Vega-like characteristics or are precursors to such a phenomenon. The various membership indicators suggest 
that $\sim$ 16\% of the PM members are non-members. A significant fraction ($>$70\%) of program stars in 
NGC 1976, NGC 2244, NGC 6530 and NGC 6611 show anomalous reddening with $R_{V}$ = $5.11\pm0.11$, $3.60\pm0.05$, $3.87\pm0.05$ 
and $3.56\pm 0.02$, respectively, indicating the presence of grain size dust larger than that typical to the diffuse medium. A 
small number of stars in NGC 1976, NGC 2244 and NGC 6611 also show normal behavior while the cluster NGC 6823 appears to have 
a normal reddening. Three highly luminous late type giants, one in NGC 2244 and two in NGC 6530, appears to be member and are 
in post-hydrogen-core-burning stages suggesting a prolonged duration ($\sim$ 25 Myrs) of star formation. 
\end{abstract}


\section{Introduction}

Study of differential extinction and the distribution of dust and gas 
in young clusters (age $<$ 10 Myr) have played an important role in understanding the star formation processes in them 
(Elmegreen \& Lada 1977; Krelowski \& Strobel 1979; Margulis \& Lada 1984). 
Study of the variation of reddening, $E(B-V)$, across the cluster face, with the spectral type and luminosity indicates that the
observed variation of reddening in young open clusters may not be explained by a "simple" or even "relatively" simple physical
scenario (Sagar 1987). What factors determine the non-uniform extinction in young star clusters and relatively
large value of $E(B-V)$ observed for some members is not understood on the basis of existing observations. Is it the patchiness 
in distribution of gas and dust or the circumstellar shells around individual stars or the dust shell around the cluster or is it
due to some emission features in  the individual members or a combination of any of these ? 

The early type stars, being intrinsically luminous and associated with dusty regions, provide 
invaluable help in probing properties of extinction in the interstellar (He et al. 1995) as well as in the intracluster 
medium (Sagar \& Qian 1993). In order to determine accurate physical properties of young star clusters, extinction contributions 
from interstellar, intracluster and circumstellar material should be known reasonably well (Pandey et al. 2003). The anomalous 
extinction behavior of the 
interstellar medium is usually characterised by the ratio of the total to selective extinction $R_{V} = A_{V}/E(B-V)$, with a 
normal value of 3.1 for the diffuse dust/clouds (Mathis 1990; Martin \& Whittet 1990; He et al 1995). However, $R_{V}$ is found 
to deviate significantly in many directions particularly towards young clusters embedded in dust and gas clouds 
(Krelowski \& Strobel 1987; Pandey et al. 2003 and references therein). In addition to intracluster dust and
gas, the anomalous extinction is also caused by the nature of circumstellar material around young stars for example Herbig Ae/Be,
T Tauri and Vega-like stars (Meeus et al. 2001; Hillenbrand 2002; Manoj et al. 2003 and references therein). Moreover the 
extinction law is found to be uniform in all directions for $\lambda \geq 0.9 \mu m$.  

To gain insight into the above questions, 
systematic spectrophotometric data of highly reddened cluster members are required. Analysis of such data will not only help 
in studying the role of emission features, if present in these members but also help in characterising the star's intrinsic 
properties. In general, the previous extinction studies were mainly based on $UBV$ data taken primarily with either 
photographic plates or single channel photometers in combination with the spectroscopic information derived photometrically. 
On the other hand, current optical photometry and astrometry, as well as mid-IR surveys allows us to derive multi-wavelength 
information on cluster members. In this paper we present a low dispersion spectroscopic data of a sample of 211 highly reddened 
proper motion cluster members mostly in their main-sequence (MS) phase.  The brighter members ($V<15$ mag) with mostly early 
spectral type are selected from six young galactic clusters. We collected their available $UBVRI$ broad-band data from the 
WEBDA\footnote{http://obswww.unige.ch/webda} database, JHK data from 
2MASS\footnote{http://www.ipac.caltech.edu/2mass} (2 Micron All Sky Survey) and mid-IR data from 
ISOGAL\footnote{http://www-isogal.iap.fr} (Infrared  Space Observatory GALactic survey), 
MSX\footnote{http://www.ipac.caltech.edu/ipac/msx} (Midcourse Space Experiment) and 
IRAS\footnote{http://vizier.u-strasbg.fr} (InfraRed Astronomical Satellite) databases. Sect. 2 describes the selection of
objects and their observational data for the study, while the results derived from the present 
analysis and the discussions are presented in the remaining part of the paper.

\section{Observations}

\subsection{Sample selection}

Our sample consists of 211 stars in the direction of six young (age $<$ 13 Myr) galactic clusters 
namely NGC 1976, NGC 2244, NGC 2264, NGC 6530,
NGC 6611 and NGC 6823 (see Table 1). These clusters lie in a highly obscured star forming regions, viz M42 - OB1 Ori
- in Orion, OB2 Mon - Rosette nebula - in Monoceros, OB1 Mon - in Monoceros, M8 - Lagoon nebula - in Sagittarius, 
M16 - Eagle nebula - in Serpens and OB1 Vul - in Vulpecula, respectively. Their galactocentric
distances are in the range of 0.4 kpc to 2.1 kpc. A histogram of the magnitudes of the selected stars is shown in Fig. 1(a). 
The sample consists of about 72\% of early type stars. There are 22, 109, 20, 20, 24 and 16 cluster members 
of O, B, A, F, G and K spectral types respectively. More than 95\% of the stars have $E(B-V)$ larger than the mean $E(B-V)$ 
for their respective cluster. A histogram of the $E(B-V)$ is plotted in Fig. 2 for each cluster. The proper motion (PM) 
membership of most of the stars are more than 90\%. Around a dozen of the sample have either low PM membership 
probability or belong to the field star population, chosen deliberately to represent interstellar properties in the cluster 
directions. Thus a large fraction of objects under study are highly reddened PM cluster members. 
Further details of their spectroscopic and photometric data are described in the following subsections.

\subsection{Spectroscopy}

The long slit spectroscopic data were obtained from the Siding Spring Observatory, Australia. The clusters NGC 6530 and NGC 6611 
were observed in August 1989 on the Australian National University (ANU) 1-m telescope using a CCD and spectrograph with a 
dispersion of $\sim$ 5.7 \AA/pixel and covering a range of $\lambda\lambda$ 3150 to 6500 \AA.  The remaining clusters were
observed in October 1995 on the ANU 2.3-m telescope using a CCD and the double beam spectrograph with a dispersion
of $\sim$ 4.0 \AA/pixel, covering the wavelength ranges from 3150 to 6500 \AA~ in the blue and from 5700 to 9800 \AA~ in the red. 
A sample of 77 stars mostly from NGC 6611 and NGC 6530 have therefore spectra only in the blue region.
A consolidated log of observations, spectrographs and grating are given in Tables 2 \& 3. Sufficient numbers of bias, flat and 
arc frames were obtained for calibration purposes. Apart from the cluster stars, several bright and faint spectrophotometric 
standards were also observed. During the 1995 observations, stars with near 
blackbody and smooth spectra from Tables 3 \& 4 of Bessell (1999) were also observed in each night. These are used to remove 
atmospheric absorption features and any large variations along the spectrum due to detector sensitivity, grating efficiency 
and the spectrograph vignetting. 

Data reduction is done using the spectroscopic software packages of IRAF\footnote{ftpsite - iraf.noao.edu; IRAF is 
distributed by the National Optical Astronomy Observatories, which are operated by the Association of Universities for 
Research in Astronomy, Inc., under cooperative agreement with the National Science Foundation.} and 
FIGARO\footnote{ftpsite - ftp.aao.gov.au}. Arc frames were used for wavelength calibration. 
A typical RMS uncertainty in wavelength and flux calibrations are 1 \AA~ and 0.05 mag respectively for a typical blue region 
(3500 - 6500 \AA) spectrum with S/N ratio of 20 (see Fig. 1(b)). Whenever the arc spectra were not taken (see Table 3), 
the wavelength calibrations were done using Balmer lines of bright A type standards, which may introduce
a further uncertainty in $\lambda$ calibration of a few Angstroms. A set of six flux and wavelength calibrated spectra are 
given in Fig. 3 while the continuum normalised spectra for all the stars are presented in Fig. 5. The flux is given in AB 
magnitudes with the zero-point flux of 3.63 $\times$ 10$^{-9}$ ergs/cm$^{2}$/sec/\AA. 
In stars 6, 17, 18, 20, 24, 28, 31, 91, 115 and 179 - the nebular background was very 
strong or saturated making for imprecise background subtraction. Seven of these stars belong to the nearest and most fertile
star forming region (i.e. Orion nebula - NGC 1976). The effects of cosmic rays can also be seen in a few spectra (see Fig. 5). 

The 1995 observations were reduced following the strategies described by Bessell (1999), which
has indeed helped us to examine very weak spectral lines of equivalent width (EW)  $\leq$ 1 \AA~ such as NaI D, CaII T and OI T,
where D and T stands for doublet and triplet respectively (see Section 3 for details). The zig-zag continuum can be 
seen in the blue part of the 1989 spectra while it is absent in 1995 spectra, this effect is more apparent in the normalised
spectra (see Fig. 5). The telluric features have not been removed completely for a few stars as they
are much dependent on night conditions. In order to check the flux accuracy and correctness of reduction procedure, the synthetic 
photometric indices were determined.  We found that for 95\% of the samples the differences of observed $V$, and $B-V$ with the 
synthetic ones were lying in the range $\pm 0.1$ mag and $\pm 0.05$ mag respectively. The residuals higher than this range may 
be due to either variability or a low S/N ($<$10) or a close by star or emission features or a combination of these. 
However, we may conclude that the flux accuracy and reduction procedure are correct within observational uncertainties.     

\subsection{Photometry}

\subsubsection{Optical}

The broad-band Johnson $UBV$ and Cousins $RI$ data are collected from earlier $UBV$ photometry done by 
us (Sagar \& Joshi 1978, 1979, 1981 and 1983 for clusters NGC 6530, NGC 6611, NGC 6823 and NGC 2264 respectively) and from
the WEBDA database (Mermilliod \& Paunzen 2003 and references therein). For 45 stars, we had only Johnson $RI$ data which 
were converted to the Cousins system using transformations given by Bessell (1979). We could collect simultaneous $UBV$ data for 
all the samples except 9 stars which have only
$BV$ data. There are 83 stars for which we do not have data in either $R$ or $I$ or in both bands (see Table 4).    
The maximum uncertainly in the magnitude determination is considered to be $\sim$ 0.05 mag. Our sample also contains many known 
and suspected optical variables (see Table 4). The star 138 is a double star and we consider the brighter component (Walker 1957)
in our analysis. 

\subsubsection{Near-infrared (NIR)}

The NIR $JHK$ data are taken from the ground based 2MASS Survey (Skrutskie et al. 1997; Cutri 1998), which provides 
magnitudes for about 471 million point sources observed at the bands $J$ (1.25 $\mu$m), $H$ (1.65 $\mu$m)
and $K_{s}$ (2.17 $\mu$m) with the limiting magnitudes of 15.8, 15.1 and 14.3 respectively. We had JHK data available 
for about 40\% of the program stars in the literature for example by 
Qian \& Sagar (1994) for NGC 1976, Hillenbrand et al. (1993) for NGC 6611 and a few more stars from 
scattered sources. We collected $JHK$ data from 2MASS as it contained homogeneous measurement for most of the sample stars. 
The $K_{s}$ magnitude were transformed to the $K$ magnitude using relations given by Carpenter (2001) or the 
subsequent updates posted on the 2MASS website. There are 12 stars for which the quality of the 2MASS data are either poor or 
the data were not recorded. For six of these stars we took data from the literature i.e. for stars 17, 18 and 22 
from Qian \& Sagar (1994); for star 26 from Hillenbrand et al. (1998); for star 56 from P\'{e}rez et al. (1987) and for 
star 115 from van den Ancker et al. (1997). For the stars 87, 131, 133, 136, 137 and 141 we had either poor S/N ($<$ 10) data or
had only upper magnitude estimate in $J$ band. As we had bright NIR sources ($\langle K \rangle$ $\sim$ 9 mag) in our samples, 
almost all stars have good S/N ($>$ 10) ratio. A Gaussian fit to the frequency distribution of uncertainties in the $JHK$ bands 
give a mean of 0.02, 0.03 and 0.03 mag with a $\sigma$ of 0.018, 0.022 and 0.022 mag respectively.

As a further check on the consistency of the NIR data, we compared 2MASS $JHK$ magnitudes with that available in 
the literature for a total of 90, 89 and 91 stars in $J$, $H$ and $K$ band respectively. The difference in $J$, $H$ and $K$ 
magnitude has a mean $\pm$ $\sigma$ (s.d.) of $-0.01\pm0.08$, $-0.04\pm0.11$ and $-0.05\pm0.15$ respectively. We observed
more than 3$\sigma$ deviation for a total of 16 stars altogether and their numbers are 11, 6 and 7 in $J$, $H$ and $K$ bands 
respectively out of which only 5 have the deviation in at least 2 of the $JHK$ bands. Most of these stars have either emission 
features or show NIR variability (Qian \& Sagar 1994; Carpenter 2001). Thus we could secure
almost a complete broad-band $UBVRIJHK$ data for entire sample of 211 program stars (see Table 4).

\subsubsection{Mid-infrared (MIR)}
The MIR (7 to 100 $\mu$m) data are taken from ISOGAL (Schuller et al. 2003; Omont et al. 2003), MSX (Price et al. 2001) 
and IRAS (Beichman et al. 1988) Surveys. A total of 65 stars were found to have associations at least at one band (see Table 5).

ISOGAL provides data mainly at two bands centered around 7 $\mu$m, with filters LW2, 6.7 (3.5) $\mu$m; LW5, 6.8 (0.5) $\mu$m; 
and LW6, 7.7 (1.5) $\mu$m,  and around 15 $\mu$m, with filters LW3, 14.3 (6.0) $\mu$m and LW9, 14.9 (2.0) $\mu$m with the 
detector sensitivity down to 0.01 Jy. We found 28 sources within a search radius of 10$^{\prime\prime}$. Twelve of these 
stars (169, 175, 190, 191, 192, 195, 197, 198, 200, 204, 203, 107) have a separation of around 5$^{\prime\prime}$ while the rest 
have a separation of around 1$^{\prime\prime}$. We consider these as reliable identifications as ISOGAL's positional accuracy for 
DENIS associated sources are $\sim$ 0.5$^{\prime\prime}$ and $\sim$ 8$^{\prime\prime}$ for non-DENIS associated sources. 
The stars 190, 191, 192, 195, 196, 197, 198, 200, 203, 204 and 207 were observed in broad-band (LW2, LW3) filters and most 
of them have fluxes near detector sensitivity (see Table 5). Moreover, most of them are 
identified within a search radius of $\sim$ 5$^{\prime\prime}$. Another 17 stars were observed with narrow-band (LW6 and LW9) 
filters and were identified within a search radius of $\sim$ 1$^{\prime\prime}$. A typical mean uncertainty of these stars are 
about 0.01 Jy with a quality flag of 4 for most of the associated sources (Schuller et al. 2003). 

MSX provides data in four bands namely 
A (8.28 $\mu$m), C (12.13 $\mu$m), D (14.3 $\mu$m) and E (21.34 $\mu$m) with the highest sensitivity at band A of 0.1 Jy. 
A total of 23 sources (Table 5) are found to have an MSX counterpart and twelve of these (87, 90, 95, 98, 126, 132, 160, 
172, 173, 175, 176, 178) have separations within 2$^{\prime\prime}$ while others have a mean separation of 6$^{\prime\prime}$. 
A typical mean 1$\sigma$ flux uncertainty for the selected sources are $\sim$ 7\%, 10\%, 11\% and 12\% respectively at A, C, D 
and E bands. 

IRAS provides fluxes at 12, 25, 60 and 100 $\mu$m with a typical detector sensitivity of 0.4, 0.5, 0.6 and 1.0 Jy respectively. 
The typical positional uncertainties are $\sim$ 25$^{\prime\prime}$. We identified 27 IRAS sources to be associated 
with our program stars out of which thirteen (1, 5, 13, 30, 41, 57, 65, 71, 87, 111, 126, 199) have separations 
within 30$^{\prime\prime}$ while the rest have a mean separations of 45$^{\prime\prime}$. Most of these objects
are found to have identifiers in the SIMBAD\footnote{http://simbad.u-strasbg.fr} database. However a few of these may be a 
spurious associations considering the positional uncertainties. 

Table 5 lists all the stars identified with these catalogues. Nine of these stars (87, 97, 111, 126, 160, 173, 175, 176, 178) have 
been observed with two surveys. Their fluxes are comparable within observational uncertainty except for stars 111 and 126 where 
the differences are considerable. This may be either due to wrong association or due to source variability. In many cases, while 
studying spectral energy distribution, we averaged and considered the mean fluxes representing at 7, 12, 15 and 25 $\mu$m. 
For example ISOGAl's bands around 7 $\mu$m and MSX's A band fluxes were averaged to represent the flux at 7 $\mu$m.

\section{Spectral properties}

\subsection{Spectral classification}

A technique based on the cross-correlation method was used to classify the spectra. A set of 161 template spectra having different
spectral type and luminosity class were taken from the optical library of Jacoby et al. (1984) but rebinned to the same
resolution in the wavelength range (3800 to 5000 \AA) as the present data. The continuum of template as well as 
programme spectra were
approximated by filtering the high frequency component in Fourier space. The continuum divided (LaSala \& Kurtz 1985) 
spectra were then used for cross-correlation. Each programme star was cross-correlated with the template stars and R-values 
defined as the ratio of peak height to rms noise (Tonry \& Devis 1979) were examined. The spectral type corresponding to 
the highest R-value was selected. 

A comparison of our spectral types with those in the literature is plotted in Fig. 4. Most of the stars lie within 2 sub-spectral 
types while a few of them have larger deviations. This is due to the presence of Balmer line emission in early type spectra and 
CaII HK line emission in late type spectra. Moreover for late type stars it may also be due to wrong classification in the 
literature. In order to confirm the spectral type determined by cross-correlation, we stacked the Fourier filtered spectra, by 
eliminating high frequency (noise) and low frequency (continuum) components, in the sequence of their spectral type (see Fig. 5).
This helped us to examine and reclassify the individual spectrum visually using the literature classification as many of the stars 
have accurately determined spectral types. A large range in spectral types exist in the literature for some such as 
star 24 (B0 Vp by Johnson (1965), B8 by Greenstein \& Struve (1946), B2 by Levato \& Abt (1976)), we adopt the ones
matching with our type. The final adopted spectral types was chosen from the literature, if it was based on higher dispersion, 
otherwise it is adopted from the present classification. The luminosity class is mostly taken from literature, if available, 
otherwise it is considered to be a MS star as per their location in the color-magnitude and color-color diagrams of the
respective clusters. However, for a few stars new luminosity classification is also being provided (see Section 3.2). Table 6 
lists the finally adopted spectral class including either a new or a more reliable classification for a sample of 24 stars 
indicated with an asterisk.     

\subsection{Spectral features - CaII HK, HI, CaII T, OI T and NaI D lines}

Emission features in the Balmer and CaII HK lines were reported for a total number of 43 stars on the basis of optical spectra 
taken mainly in the classical MK region (3800 to 5000 \AA) and/or H$_{\alpha}$ region (see Table 4 and references therein). We 
could not confirm the reported emission features for 25 of them. Some of these emission features are in the H$_{\alpha}$ region 
where we do not have data for 77 stars and for some stars the reported features are not seen at our resolution or the features may 
be absent. We confirm the spectral peculiarities for 18 stars reported earlier and find spectral peculiarities for another 11 stars 
for the first time. The prominent spectral peculiarities for all these 29 stars are listed in Table 7 and the emission
features are shown in Figs. 6 and 7. In several stars, namely 6, 18, 20, 24, 26, 28, 31 115 and 179, it could not be ascertained 
whether certain emission features were of nebular or stellar origin.

The spectral peculiarities associated with relatively weaker lines such as Paschen, CaII T (8498, 8542 and 8662 \AA), 
OI T (7772, 7774 and 7775.4 \AA) and the interstellar NaI D (5896 and 5890 \AA) have
also been investigated. We have determined the integrated equivalent width for CaII T lines only for F, K and G type 
as these lines  merge with Paschen lines of Hydrogen in early type stars. The equivalent width of the NaI D lines are determined 
for all the stars.  A typical uncertainty of 5 to 10\% is assigned to these determinations as is seen from repeating the width 
measurements.  The values obtained in this way are tabulated in Table 6. The EW of the OI T lines are too 
weak ($\leq$ 0.3 \AA) to estimate with low resolution spectra and in a few cases due to low S/N ratio of the spectra. However, 
individual cases of enhanced OI T absorptions are remarked in the following subsections along with the Paschen line peculiarities. 

The CaII T lines have been used to explore the disk frequency and disk accretion rate in late type pre-main sequence (PMS) 
candidates (see Hillenbrand et al. 1998). It can be inferred using transformations by Gullbring et al. (1998) that an 
EW (CaII T) $\approx -3$ \AA~ corresponds to $\dot{M}_{acc} \leq 10^{-8} M_{\odot} yr^{-1}$ and at lower resolution as in 
present case a $\dot{M}_{acc} < 10^{-8} M_{\odot} yr^{-1}$ should correspond to a filled-in CaII T lines.
The CaII T widths were determined for 54 stars. The mean value, best represented by 32 non-emission MS stars, 
is $2.9\pm0.3$ (s.d.) \AA, which is comparable to the value quoted for CaII T widths ($\sim$ 3 \AA) for late type MS stars by 
Hillenbrand et al. (1998). There are stars having larger than 3$\sigma$ discrepancy (see Fig. 8(a)). The 
stars 87, 114, 149 and 211 have widths greater than 3.9 \AA, two of these are classified supergiant in the literature, 
therefore we assign the supergiant (I) luminosity to the other two stars (114, 149). There are 11 stars with widths less 
than 2 \AA~ indicating the presence of a circumstellar disk, moreover all of these except star 48 also show weak to strong emission
features. For stars 5, 27 and 100, the width is near zero or negative indicating the presence of an accretion 
disk ($\sim$ 10$^{-8}$ M$_{\odot}$ yr$^{-1}$), while stars 6, 23, 32, 33, 34, 48, 92 and 99 have values between 1 to 2 \AA, 
qualifying them for having circumstellar disks with low accretion rate $\leq$ 10$^{-8}$ M$_{\odot}$ yr$^{-1}$. 

The dependence of NaI D line widths on spectral types indicates higher values for late 
type stars (F, G, K) as the measure is affected by the stellar components (Fig. 8(b)). A large 
scatter (0.3 to 1.0 \AA) for B type stars is measured, which indicates surplus circumstellar and intra-cluster gas components 
in addition to the normal extinction effects. The largest scatter in NaI D width is seen for the cluster NGC 1976 and NGC 2264. 
The average values and standard deviation of NaI D width determined from early 
type PM members for clusters NGC 1976, NGC 2264, NGC 2244, NGC 6530, NGC 6823 and NGC 6611 are $0.18\pm0.04$, $0.22\pm0.02$, 
$0.45\pm0.07$, $0.60\pm0.15$, $0.71\pm0.20$ and $0.70\pm0.18$ \AA~ respectively. These values appear to show a weak correlation
with the corresponding galacto-centric distances to the clusters (see Fig. 8(c)), however it is difficult 
to discuss the extinction effects on the NaI D widths due to circumstellar and intracluster dust and gas.

\subsubsection{Early types - O, B and A stars}

There are 22 O type stars. Eleven of these stars (107, 142, 154, 155, 156, 157, 161, 164, 169, 170, 171) have been
observed only in the blue. H$_{\alpha}$ emission is reported for star 22 by Conti (1974) and for star 58 by Lesh (1968).
The H$_{\alpha}$ appears to be partially filled in star 22 (indicating a weak core emission) while comparing its strength 
with the normal spectra of the stars 67 and 202. No indication of any emission was seen for star 58. The spectral features 
characteristic of Of stars are seen clearly in 
our spectra (see Fig 5) for example the stars 81, 110, 67, 155 and 156 show NIII 4630-34 \AA~ line in emission, moreover, 
HeII 4686 \AA~ is also seen in emission for star 161 (a brightgiant, IIb) confirming the reported spectral features from 
dispersions much higher than ours. 

In a sample of 109 B type stars (86 early type - up to B4), 29 stars are reported to have weak to strong emission features. 
Seventeen of these are reported by Hiltner et al. (1965) on MK region observations for NGC 6530 with a caution for 
effects of nebulosity. We have H$_{\alpha}$ region spectra for only four stars namely 78, 97, 188 and 205. Altogether
we could confirm Balmer emissions only in 7 of them viz 97, 122, 126, 160, 173, 174 and 175. Of these, stars 97, 122, 160
and 175 are reported to have Herbig Ae/Be (HAB) characteristics. A strong OI T absorption in 
star 97 (EW $\sim$ 0.83 \AA) indicates that it is not a dwarf. The width of OI T features comes out to be around
0.3 \AA~ for dwarfs. Star 97 is the only one which shows Paschen lines in weak emission. We find H$_{\beta}$ in strong 
emission for star 109 and in filled emission for star 115 (see Fig. 6). About 8\% of our sample show characteristics of Be stars
which is well in accordance with the general percentage of Be stars i.e. 11\% (Jaschek and Jaschek 1987). The luminosity 
effects can be seen as sharper balmer lines for the stars 176 (B2.5 I), 195 (B0.5Ib), 52 (B2.5 II-III) and 198 (B1 III), 
however it is less evident for stars 1, 79, 183, 184 and 188 reported as subgiants in the literature.

A total of 20 A type stars were observed. Out of these, stars 35 and 98 are reported to have emission features by Herbig (1960) and 
by Young (1978) respectively. Our spectra shows H$_{\alpha}$ in strong single peak emission for star 35 and H$_{\alpha}$ totally
filled up to continuum for star 98. A strong OI T in absorption in the star 35 (EW $\sim$ 0.85 \AA) according to the 
relation obtained by 
Danks and Dennefield (1994) indicates that these are candidate HAB star. The star 8 is reported to be of Ap type by Levato and 
Abt (1976). Our spectrum shows sharp Balmer lines with enhanced CaII T absorption, very weak H$_{\alpha}$ absorption and large 
NaI D absorption, indicating a PMS star with weak core emission at H$_{\alpha}$ having a circumstellar shell and making
it another candidate HAB star. 

\subsubsection{Late types - F, G and K stars}

Our sample contains 20 stars of F spectral type. The star 5 was reported to have CaII HK in emission by Walker (1983). Our spectrum
also shows H$_{\alpha}$ in strong emission in addition to CaII HK. The star 99 has totally filled H$_{\alpha}$ and partially 
filled CaII HK. The star 92 also appears to have a very weak H$_{\alpha}$ emission. These stars (5, 92, 99) have CaII T width 
of $-0.2$, 1.93 and 1.95 \AA, respectively, indicating them to have a circumstellar disk, characteristics of a PMS star. The CaII T
width indicates that star 114 is a supergiant. Of the 24 G type stars, five are reported to have H$_{\alpha}$ emission viz 
star 31 (Penston et al. 1975), 100 (P\'{e}rez 1988), 101, 106 (Young 1978) and 27 (Herbig and Bell 1988). We confirm emission 
features for stars 27, 31, 100 and 101, however star 106 does not have H$_{\alpha}$ in emission during present observations as 
found by P\'{e}rez (1988). The star 100 shows a strong H$_{\alpha}$ 
emission, this star is reported to show weak and partially filled H$_{\alpha}$ emission by P\'{e}rez (1988). In addition to
these we also find totally filled H$_{\alpha}$ for stars 23 and 29. The star 6 shows a clear CaII HK core emission while star
91 shows partially filled CaII HK lines. A very weak EW ($\sim$ 0.2 \AA~) of NaI D line for star 91 indicates that it may be a 
field star. 

There are only 16 K type stars in our sample, of which four stars (32, 33, 34 by Walker (1983) and 145 by van den 
Ancker et al. (1997)) are reported to have either Balmer or CaII line emissions. The star 33 is also reported to have
more Li (6704 \AA) absorption by Walker (1983). We confirm CaII emissions for stars 32, 33 and 34, in addition we report 
H$_{\alpha}$ emission also for stars 32 and 34. No emission features were seen for star 145. Moreover we report 2 more stars to
show CaII emissions viz 15 and 87. The stars 149 and 151 are of special interest as they are reported to have
post-AGB characteristics. The TiO bands around 6651 \AA~ and 7054 \AA~ can be seen very clearly in both of these stars along with a 
weak appearance in star 199. The MK spectra of star 149 matches best with K7 III (BD+590128 - a MK standard from the 
Jacoby library). The CaII T width for this star indicates that it is a supergiant. Therefore, we classify these stars (149, 151) as 
giants of type K7, however the star 149 may be a candidate supergiant.
  
Table 8 summarises the distribution of all the program stars according to their spectral type, luminosity class 
and emission line peculiarities.

\section{Extinction Properties}

\subsection{Intrinsic colors and color excesses}

To study the behavior of reddening and extinction properties, we 
derive $E(U-V)$, $E(B-V)$, $E(V-R)$, $E(V-I)$, $E(V-J)$, $E(V-H)$ and $E(V-K)$ color excesses of the cluster members using the MKK 
spectral-type-luminosity-class-colors relation given by FitzGerald (1970) for $(U-V)$ and $(B-V)$; by Johnson (1966) for $(V-R)$ 
and $(V-I)$, transformed to the Cousins one using Bessell's transformation (1979); and by Koornneef (1983) 
for $(V-J)$, $(V-H)$, and $(V-K)$. Maximum uncertainties in the color excess determination in $UBVRI$ is considered to 
be about $\sim$ 0.08 mag and in $JHK$ to be $\sim$ 0.15 mag. The color excesses in some bands are found to be negative for 
stars 48, 59, 96, 100, 102 and 106 but the values are within the uncertainty limit. Most probably, this is due to 
non-simultaneous observations and/or due to variability of the star. 

We have determined the value of $R_{V}$ using $A_{V}$ which is approximated as 1.1 $E(V-K)$. 
The ratio of $A_{V}$ to $E(V-K)$ does not change appreciably with $R_{V}$ value (Whittet \& van Breda 1980). He et al. (1995) 
estimated the $A_{V}$ by different methods and concluded that they agree to better than 3\% for normal stars. However, the value 
of $R_{V}$ is very sensitive to the excess radiation at the band K as well as to very small values of $E(B-V)$. Therefore
the value of $A_{V}$ for a sample of 41 stars, found to have anomalous radiation at NIR (see Section 5), are determined
using $E(V-J)$ assuming a normal reddening. Table 6 contains values of $R_{V}$ estimated in this way for all the program stars 
except for those 23 stars which have very low values ($<$ 0.2 mag) of $E(B-V)$.

In order to understand the properties of interstellar matter in the direction of clusters, we plot the ratios of color 
excesses $E(U-V), E(B-V), E(V-R), R(V-I), E(V-H)$ and $E(V-K)$ against $E(V-J)$ in Fig. 9. These shall be referred as 
color excess diagrams (CEDs) hereafter. As our sample belong to the star 
forming regions and is likely to have young stellar objects, the color excess ratios are plotted relative to $E(V-J)$ primarily 
with the aim to minimise contributions from the blueing effect and ultraviolet excess on the band $V$ as well as to minimise 
contributions from circumstellar dust and gas shells at the $J$ band.  Moreover, $E(V-J)$ does not depend on properties such as the 
chemical composition, shape, structure and degree of alignment of the interstellar 
matter (Qian \& Sagar 1994; Yadav \& Sagar 2002). In Fig. 9, the dotted lines are drawn to represent the 
normal interstellar extinction law (Mathis 1990). Errors expected from the observational uncertainties are shown with crosses.
Most of the sample stars are lying on the normal reddening line for all the clusters
in the NIR CEDs, while a scatter of varying amount can be seen in diagrams with $E(U-V)$ and $E(B-V)$.
It is maximum for NGC 1976 and NGC 6611 and minimum for NGC 2244 and NGC 6530. 

We divided the stars into three groups depending upon their deviation in the CEDs with respect to the normal line. Group 1 contains 
all the stars following the normal law within observational uncertainty, while Group 2 stars follow the normal law at NIR and 
deviate towards a lower ratio in the ultraviolet (UV). In addition to these, the sample also contains peculiar stars in the sense 
that they do not follow a particular trend and deviate at either one or two bands in the NIR and UV. These usually show emission 
features, chemically peculiarities and variability at visible or NIR wavelengths. All of these stars along with the ones with very 
low membership were
kept in Group 3 and were not considered in estimating mean reddening for the cluster. The last column of Table 6 indicate the 
categorisation of Groups 1, 2 and 3 stars, while a number distribution for Group 1 and 2 stars are given in Table 9. 
A cluster-wise description of the reddening properties is given below.

\subsection{Extinction law towards clusters}

For NGC 1976, a wide variations in the reddening values e.g., $R_{V}$ = 4.8 (Mendez 1967), 5.0 (Johnson 1968),
3.1 (Walker 1961) have been reported. It has been argued that a part of the large value of $R_{V}$ may also be contributed by the 
far-infrared brightness of stars (Johnson 1967). A recent study by Qian and Sagar (1994) concludes that the cluster presents both 
normal as well as anomalous behavior. The UV extinction study also suggests an anomalous galactic curve towards higher values 
of $R_{V}$ (Bohlin \& Savage 1981). Among most of our sample having high values of $E(B-V)$, there exists 6 stars in Group 1 
which follow a normal extinction law and have a mean value of $R_{V} = 3.18\pm 02$ with a range from 3.07 to 3.30 
(Tables 6 and 9). Group 2 stars, 28 in total, have a mean value of $R_{V}$ of $5.11\pm0.11$ with a range from 3.77 to 9.94. 
The extinction properties do indicate a shift in the grain size distribution towards larger than normal sized
particles in these regions.

The cluster NGC 2244 has a mean reddening of $E(B-V)=0.47$ mag with a normal reddening law and a small amount of differential 
reddening (Park \& Sung 2002), though individual values towards the cluster field ranges from 0.08 to 0.98 mag 
(Massey et al. 1995; Bergh\"{o}fer \& Christian 2002). 
At the time of observations we selected highly probable members based on the study by Marshall et al. (1982),
though a recent refined PM study on brighter members (Sabogal-Mart\'{i}nez et al. 2001) dubs many of the high PM 
member as non members. In our sample we have stars 48, 54, 59 and 69 with $E(B-V)$ $\sim$ 0.05 mag. We adopt a synthetic $E(B-V)$ 
of 0.05 and 0.07 mag for stars 48 and 59 as the V magnitude estimate in the literature were highly uncertain and the color 
excesses could not be determined at all bands for these stars. A re-estimated PM membership assigns a value less than 0.5 for 
stars 54, 59 and 69 (Sabogal-Mart\'{i}nez et al. 2001). Similarly the stars 51, 55, 63, 85 and 86 have $E(B-V)$ $\sim$ 0.15 mag and 
have low membership according to the new study. Therefore all of these stars were not considered for determining the extinction law 
for the cluster. On the basis of 10 PM members in Group 1, we determine $R_{V}$ as 3.16 and a value of $R_{V} = 3.60$ is obtained 
on the basis of 19 members in Group 2. Therefore a significant number of the reddened stars do show anomalous extinction 
contrary to the earlier finding that the cluster has a normal reddening law i.e. $R_{V}$ $\sim$ 3.2 (Ogura \& Ishida 1981; 
Park \& Sung 2002).   

NGC 2264 has the lowest mean reddening i.e. $E(B-V) = 0.07\pm0.03$  mag in our sample (Sung et al. 1997). For stars 88, 89, 
91, 96 and 102, $E(B-V)$ is less than 0.02 mag. The accurate value of $R_{V}$ for these stars therefore could not be determined.
The CEDs have a large 
scatter with most of them showing peculiar extinction.  The remaining stars 92, 93, 99 and 103, all of late spectral type, 
show abnormal behavior with a mean $R_{V} = 3.91\pm0.17$. The star 104 show a normal behavior with $R_{V}$ $\sim$ 3.03. 
Moreover our sample contains a very small number of early type stars. The abnormal extinction for the late type stars may also be 
due to circumstellar material. 

In NGC 6530, $E(B-V)$ has a mean of 0.35 mag with abnormal reddening reported for a number of embedded 
members (Sung et al. 2000; van den Ancker et al. 1997). The only two stars which follow normal law in our sample are 107 and 135. 
A total of 32 stars fall in Group 2 and show abnormal extinction behavior with $R_{V}$ = $3.87\pm0.05$.

NGC 6611 has a mean $R_{V}$ of 3.75 (3.5 to 4.8) and $E(B-V)$ of 0.86 mag with variation from 0.4 to 1.8 mag 
(Hillenbrand et al. 1993). We have also included 8 BD stars on the basis of PM data from Kamp (1974), however five of 
these stars (181, 182, 183, 184, 186) have $E(B-V) \le 0.6$ mag and hence have low membership probabilities. The stars 164, 169 
and 171 show the normal law with a mean of $R_{V} = 3.03$ with the remaining 23 showing abnormal reddening with a mean 
of $R_{V} = 3.56$. 

NGC 6823 has a mean reddening of $E(B-V) = 0.85$ mag with a range of 0.64 to 1.16 mag and follows a normal reddening law 
(Sagar \& Joshi 1981; Guetter 1992). The stars 206 and 208 have reddenings of $E(B-V) < 0.22$ mag making them unlikely to 
be cluster members. The stars 194, 195 and 198 show variability while stars 199, 209 and 211 occupy anomalous position
in the color-magnitude diagram (Sagar and Joshi 1981). The remaining stars, 
16 in total show a normal reddening with a mean of $R_{V}$ = 3.0, although a closer look at the $UBV$ CEDs may give the impression 
that the cluster region contains material representing a shallower as well as a steeper $R_{V}$ than the ISM.

Using data from highly reddened stars with no emission features and high membership probabilities (i.e. Group 1 + Group 2 stars),
the clusters present diverse extinction properties. The clusters NGC 6530 and NGC 6823 
are represented by single extinction laws with the value of $R_{V}$ $3.87\pm0.02$ and $3.00\pm0.02$ respectively. A small fraction 
of the stars (18\%, 37\% and 14\% respectively) in the clusters NGC 1976, 2244 and 6611 show normal extinction with $R_{V}$ 
equal to $3.18\pm0.02$, $3.16\pm0.02$ and $3.03\pm0.03$ respectively, while the major fraction of stars show anomalous extinction 
with the corresponding $R_{V}$ of $5.11\pm0.11$, $3.60\pm0.05$ and $3.56\pm0.02$ respectively. We could not derive a representative 
value of $R_{V}$ for NGC 2264 due to small statistics. The clusters NGC 2264 and NGC 1976 have the largest scatter in the value of 
$R_{V}$ i.e., from 2.8 to 6.7 and from 3.07 to 9.9 respectively. The above observations based on reddened members led us to 
conclude that the young clusters, 4 out of 6, appears to have anomalous extinction in general, indicating the prevalence of larger 
than normal grain size dust distributions in the star forming regions. This supports the earlier finding of steeper $R_{V}$ than for
the ISM in star forming regions (Terranegra et al. 1994). The mean normal value of $R_{V}$ for the clusters NGC 1976 and 
NGC 2244 (with distances, D, $<$ 2 kpc) is $3.17\pm0.03$ while for the clusters NGC 6611 and NGC 6823 (D $>$ 2 kpc) 
$R_{V} = 3.01\pm0.03$. These figures are in excellent agreement with the distance dependence of $R_{V}$ for the diffuse dust 
reported in literature (see Table 5 of He et al. 1995). This indicates that the early type stars with high $E(B-V)$ does 
provide a reliable estimate of $R_{V}$.

\section{Infrared properties}

\subsection{NIR excess}

The interstellar as well as intracluster material show normal reddening for wavelengths $\lambda \geq \lambda_{J}$ irrespective
of the grain properties of the matter. Therefore the nature of extinction due to circumstellar material can be indicated by an 
excess or a deficit of radiation at wavelength longward of 1 $\mu$m. In order to probe this aspect, we derive $(V-H)$ and $(V-K)$ 
color excesses individually for all program stars from the color excess $E(V-J)$ assuming a normal extinction law. The differences 
between these and the observed color excesses derived on the basis of the spectra are plotted in Fig. 10 and show a random scatter 
around the mean indicating the independence of the extinction at $J$. The scatter is best represented by the average values of 
$0.02\pm0.09$ (s.d.) mag for $\Delta (V-K)$ and $0.01\pm0.07$ (s.d.) mag for $\Delta(V-H)$. These estimates were determined 
by ignoring highly deviant points and are based on 195 and 204 data points in $H$ and $K$, respectively. We expect the star to have
a NIR excess if it has differences greater than 2$\sigma$ at either $H$ or $K$ or at both. There are 10, 11 and 20 stars which have 
differences $> 5\sigma$, 3-5$\sigma$ and 2-3$\sigma$ respectively at either $H$ or $K$ or both. Table 10 provides the list of all 
these stars with their differences. Several of these stars are reported to have large or moderate NIR excess e.g. for stars 5, 8, 
14, 24, 27 and 33 by Qian \& Sagar (1994); for stars 160 and 175 by Hillenbrand et al. (1993); for stars 115, 122, 136, 149 and 
151 by van den Ancker et al. (1997). Stars 17, 18, 119, 194, 198 and 133 also show a NIR deficiency at $H$ or $K$ or both. 

\subsection{Dereddened color-color diagram}

In order to understand the nature of a NIR excess we generate the dereddened color-color (($J-H$) versus ($H-K$)) diagram. For this 
we chose $E(V-J)$ to estimate $A_{V}$ ($\sim 1.1 E(V-K) \sim 1.364 E(V-J)$ - assuming normal reddening) for stars showing a NIR 
excess (see subection 4.1 for detailed discussion). The color excesses $E(J-H)$ and $E(H-K)$ were derived as $0.11A_{V}$ and 
$0.065A_{V}$ respectively (Reike \& Lebofsky 1985). A typical uncertainty in these colors is $\sim$ 0.1 mag with a maximum up 
to 0.15. The dereddened NIR color-color diagram is shown in Fig. 11 and is used to identify non-MS stars and their 
nature in combination with the spectral properties.

Most of the 14 stars showing only Balmer emissions, are of early type (F0) except star 92 which is of F8 type, and occupy 
positions towards right of the reddened MS and possess moderate to strong NIR excesses. These are 
probable HAB or classical Be stars. Four of these stars (8, 35, 97, 98) have strong NIR excesses 
and lie to the extreme right in the color-color diagram. These are late B or early A type stars and appears to show Group I HAB 
characteristics (Hillenbrand et al. 1992, 1993). While another group of 5 stars (109, 122, 126, 160, 175) have moderate NIR 
excesses and lie isolated towards the right and lower down in the diagram. These early type stars ($\leq$ B5) are possible 
Group II or III HAB candidate. Two of these stars (160, 175) have rigorously been argued to have HAB characteristics and were kept 
in Group III by Hillenbrand et al. (1993). A further group of 4 stars (22, 115, 173, 174) shows weak emission features and lie 
within the reddened MS region are probably the so called classical Be stars. The list of these 13 probable PMS or Be stars are 
given in Table 11. 

A group of 18 late type stars ($\geq$ F8) are probable T Tauri stars (see Table 12). 
Six of these stars (5, 27, 32, 33, 100, 101) have high to moderate NIR excesses and H$_{\alpha}$ emission above the continuum 
with a mean EW of $\sim$ 5\AA. In the ($J-H$) vs ($H-K$) diagrams they occupy the position just below the zero-age MS (ZAMS) 
track of classical T Tauri (CTT) stars. These stars have both CaII HK and Balmer lines in emission. Moreover, star 27 also has 
Paschen line emission; stars 27 and 33 have emission like spectrum; the stars 33 have strong Li absorption while stars 27 
and 32 are reported to vary in the NIR. Therefore these stars appear 
to have characteristics of CTT stars. Another group of 12 stars (6, 15, 23, 29, 31, 34, 53, 90, 91, 92, 99, 178) occupy the
location of weak-line T Tauri (WTT) stars. Six of these show weak excesses and the rest do not show any excess. Two of 
them (92 and 99 - both F8), probably belong to the post-T Tauri phase and are about to reach the ZAMS.

The stars 87, 149 and 151 are either giants or supergiants of K spectral type. Star 87 also shows CaII HK emissions and 
moderate IR excess while the star 151 is a variable (Sagar \& Joshi 1978) and shows strong IR excess. Their distinct location a 
little to the left and above the reddened MS suggests that these may be supergiants. In fact two of these stars (149, 151) have 
been reported in the literature as probable post-asymptotic giant branch (AGB) candidates (van den Ancker et al. 1997). 

\subsection{Spectral energy distributions (SEDs)}

The presence of circumstellar material, their geometry and the nature of radiation is best studied by SEDs covering the MIR region.
The SEDs are derived using reddening corrected broad-band fluxes. The reddened broad-band optical and NIR fluxes were taken from
Table 4 while MIR fluxes are taken from Table 5. The dereddened fluxes were derived using  $R_{V}$ dependent analytical expression
given by Cardelli et al. (1989). The values of T$_{eff}$ (effective temperature) and log($g$) (gravity) corresponding to the 
adopted spectral type are taken from Schmidt-Kaler (1982). The reddened, dereddened and synthetic KURUCZ (1993) spectra based on 
T$_{eff}$ and log($g$) were plotted and found to match in all the cases authenticating the determination of temperature, gravity 
and reddening. Fig. 12 provides these plots for stars having fluxes at MIR bands. The synthetic spectra are normalised at the $V$ 
band. The MIR region contains fluxes at 6 bands (7, 12, 15, 25, 60, 100 $\mu$m). Of 65 stars with MIR fluxes, 32 have fluxes at 
more than 2 bands while another 13 have fluxes at 2 bands only. 
We determined the spectral index $s$ defined as ($\lambda F_{\lambda}$ $\sim$ $\lambda^{s}$), in the 
region (2.2 to 25 $\mu$m) as most of the known IRAS fluxes at 60 and 100 $\mu$m are only upperlimits. The value of $s$ is 
listed in Table 5. It is seen that 17 stars have value of $s$ close to -3.0 
representing a black-body (see Fig. 12). Another group of 17 stars have fluxes only at one, two  or three bands but their $s$ 
values deviate significantly, however these determination may be considered uncertain. The remaining 31 stars have fluxes at 
more that 3 MIR bands and may be considered a reliable determination of spectral index (Table 5). The values of $s$, if available, 
deviate significantly from blackbodies for the probable PMS stars with circumstellar material (Table 12). In addition, there is
a group of MS and post-MS stars which also appears to deviate from blackbodies and are considered to be probable candidate 
with circumstellar material. The list of these stars, 37 in total, are provided in Table 12. 

\section{Evolutionary status and nature of individual stars}

The luminosity was derived using log(L/L$_{\odot}$) = 1.9 - 0.4M$_{bol}$, where M$_{bol}$ is written as M$_{J}$ + $(V-J)_{0}$ 
+ BC$_{V}$. The BC$_{V}$ is taken from Massey et al. (1989), Code et al. (1976)  and Bessell \& Brett (1988). M$_{J}$ is derived 
using the cluster distances as given in Table 1 and
the A$_{J}$ is given as 0.28 A$_{V}$. The uncertainty in luminosity is mainly contributed by uncertainty in J magnitude apart from
the uncertainties in bolometric corrections, distances and extinction. All added together amounts to a maximum
uncertainty of $\sim$ 0.06 in log(L). The uncertainties in log(T$_{eff}$) are generally below 0.02. The Hertzsprung-Russell 
diagram (HRD) of all the program stars is shown in Fig. 13. The selected clusters have 
PMS (turn-on) age in the range 1 to 3 Myr and age spreads $\sim$ 8 Myr (see Table 1), where PMS age denotes the median value of
the age distribution of individual stars while the value of age spread contains 95\% of the stars in the cumulative age 
distribution (Park et al. 2000). The turn-on ages usually agree with the 
turn-off ages and provide clues on the duration of star formation. The derived PMS age and in particular, the age spread, depends on
the treatment of convection and opacity in stellar atmospheres and is found to vary widely from model to model (see Hartigan 
et al. 1994; Park et al. 2002; Wolff et al. 2003). We have chosen the frequently used PMS tracks by D'Antona \& Mazittelli (1994) 
as none of our stars have masses below 1 M$_{\odot}$. 
In the HRD of sample clusters, the stars with masses $\ge$ 3 M$_{\odot}$ are on the MS whereas the low mass stars 
($<$ 2-3 M$_{\odot}$, i.e. $\sim$ 25\% of the sample) are still in their PMS stage. In the present sample, most of the identified 
PMS stars lie around the 2 Myr evolutionary track and are below the birthline (Fig. 13), however, some of the stars are non-members
and lie above the birthline. The spread seen in the MS is probably due to the effect of binarity, variability and 
uncertainty in the flux conversions than the real evolutionary effects on the MS. The HRDs of the clusters 
have been used to infer the nature and evolutionary status of individual stars as described below.

\subsection{Clusters}

NGC 1976 is a an open cluster with a high proportion of low-mass stars, a median age $\sim$ 1 Myr and an age spread $\sim$ 8.5 Myr 
as derived from
PMS stars (Park et al. 2002). Fifty five to 99\% of the low-mass stars in NGC 1976 are found to have circumstellar disks from NIR
data by Hillenbrand et al. (1998). Our sample contains 19 stars with masses ranging from 1 to 3 M$_{\odot}$ and ages from 1 to
3 Myr except stars 4, 10, 38 and 39 which have low luminosity and show ages $\sim$ 10 Myr. The latter are PM members with $R_{V}$
typical to NGC 1976 and have masses below 1.5 M$_{\odot}$ with no PMS characteristics. Therefore, these are probably in the post-T 
Tauri phase with a nearly edge-on disk (Park et al. 2002). Eleven of these stars show PMS characteristics. The stars 5, 27, 
32 and 33 show CTT characterisctics as is seen from strong NIR excesses, their location in dereddened NIR color-color 
diagram (as all of them lie 
along the CTT loci), H$_{\alpha}$ emission (EW $\sim$ 6 \AA) and a  circumstellar disk as seen from the CaII T line widths. These 
are identified as PMS objects in the literature.  The stars 6, 12, 15, 23, 29, 31 and 34 have weak NIR excess, low or absent 
emission activities and some show light variability in the NIR suggesting that these are WTT stars, though a further investigation 
is needed to know their true nature. The CaII T measurements for stars 6 and 23 indicate them to have a weak accretion disk 
supporting the recent finding by Littlefair et al. (2003) that WTT stars 
also possess accretion disks in contrary to the earlier conception that they are non-accreting diskless objects. A recent PM study 
by Tian et al. (1996) makes star 6 a non-member (zero probability of membership) whereas the star has weak NIR excess, shows 
Ca II emission 
and is variable in the NIR light (Carpenter 2001) suggesting that it is a PMS object. The stars 6, 15, 23 and 29 are reported 
to show emission features for the first time. Another group of three intermediate mass stars (8, 22, 35) show PMS characteristics 
and 
probably belong to the HAB group. The star 35 is a well known HAB star (Herbig 1994; Hillenbrand et al. 1992; Maheswar et al. 2002; 
Leinert et al. 1997). The stars 8 and 22 have H$_{\alpha}$ filled-in with emission and show weak NIR excesses. As star 8 is
located very close to star 35 in the HRD and is of early A type, it is most likely a HAB star. Further study is
required to probe the true nature of star 8 and 22.
The remaining group of 22 stars lie on the MS. Of these, stars 1, 12, 13, 14, 24, 30, 36 and 41 are found to have 
circumstellar material as is seen either from a weak NIR excess or above zero value of the MIR spectral index, $s$. The mean value 
of $s$ for these stars is above zero, suggesting that they contain cool circumstellar dust. These are probable Vega-like stars or 
are precursors to such a phenomenon. The Vega-like stars are characterised by substantial far infrared excesses due to cool dust, 
relatively low NIR excesses, low polarization and a lack of emission lines in their spectra. In fact, one of these stars (13) has 
recently been studied in detail and was identified to have Vega-like characteristics by Manoj et al. (2002).

NGC 2244 is a young cluster and the recent studies suggest on-going star formation in the region 
(P\'{e}rez 1991, Berh\"{o}fer \& Christian 2002). It has a median PMS age $\sim$ 1.9 Myr and an age spread of $\sim$ 6 Myr 
(Park \& Sung 2002). Though Massey et al. (1995)
suggests the existence of intermediate mass PMS stars, but a recent study by Park \& Sung (2002) found a gap near 2M$_{\odot}$, 
and the most massive PMS stars reported are of spectral type G2. Our sample contains 16 stars of spectral type later than A0, many 
of these are non-members as judged by more than one membership indicator supporting the scarcity of PMS objects near A0. The stars 
48, 51, 54, 55, 59, 63, 69 and 86 are 
non-members (see Sect. 4.2). The stars 47, 53, 60 and 85 are PM members but their locations in the HRD suggests them to 
be non-members. Moreover, the star 60 has a value of $s$ $\sim$ 0.45 indicating that it is surrounded by
cold circumstellar material, therefore it may also be a candidate PMS star. The stars 45 and 46 are PM non-members while the
stars 62 and 87 appear to be members. The star 62 is probably an early A type Vega-like MS star with cool circumstellar
dust as indicated by index $s$ while star 87 is probably
a red-supergiant with mass $\sim$ 12 M$_{\odot}$ and age $\sim$ 20 Myr. The remaining group of 28 stars are probably on
the MS with masses in the range 3 to 80 M$_{\odot}$. Of these, star 71 shows Vega-like characteristics.  

Being a nearby cluster, NGC 2264 harbors a number of peculiar stars and has frequently been used to test PMS evolutionary 
models, to study the IMF and to understand the role of stellar variability in stellar evolution (Walker 1956; Kippenhahn 1965;
Flaccomio et al. 1999; Park et al. 2000, 2002; Rebull et al. 2002). It is observed to have an age of $\sim$ 1.5 Myr and age spread 
$\sim$ 9 Myr. Our sample contains 19 stars and most of them are of spectral type A0 and later. The identified PMS stars 
(92, 99, 100, 101) do lie in the expected age range. The star 91 lies on the ZAMS and shows CaII HK in weak emission, therefore 
it is probably a foreground MS star or a BMS (below or near ZAMS) star in post T Tauri phase. The BMS stars are T Tauri stars with 
nearly edge-on disks which obscures the light from the central star and makes them low luminosity. Observations indicate that 
$\sim$ 3 to 5\% of the stars with disks are edge-on systems (Park et al. 2002). The star 90 shows light-variability and is a field 
star. The stars 93, 103, 105 and 106 lie above the birthline and have large value of $E(B-V)$, therefore they are probably 
background non-members. The star 104, of A type, also appears to be a foreground non-member located far below the ZAMS. 
Among the intermediate mass objects, stars 97 and 98 are reported to be PMS candidates (Sung et al. 1997; Park et al. 2002).
These are HAB stars. The stars 94 and 95 have weak NIR excesses and are located on the MS, therefore we consider them as candidates 
for Vega-like cluster members.  

Star formation and PMS stars in NGC 6530 have been studied in detail by van den Ancker et al. (1997) and by Sung et al. 
(2000). It has a PMS age of $\sim$ 1.5 Myr with an age spread of $\sim$ 5 Myr. In HRD, the stars 117, 143, 145, 149 and 151 
are located well above the birthline and are foreground non-members, though stars 149 and 151 are more luminous, show NIR excesses 
and are reported to be probable post-AGB candidates (van den Ancker et al. 1997). The star 151 also shows light variability.  
The post-MS track for a 12 M$_{\odot}$ appears to follow their location in the HRD. These as well as the star 87 in NGC 2244 
occupy red-giant branch (RGB) locations in the HRD, however, a 10 M$_{\odot}$ evolutionary track would place them in an AGB 
phase pushing the 
age a little older to $\sim$ 30 Myr. So, the indicators do support these being cluster members in either RGB, AGB or post-AGB 
phases. If this scenario for cluster NGC 2244 and 6530 is true then these are the stars formed $\sim$ 20 to 30 Myrs 
ago extending the duration 
of star formation to a few tens of Myrs, although a further spectroscopic investigation is needed to confirm the nature of these 
stars. The star 114 is a foreground giant with a very low $E(B-V)$. Among the remaining objects, stars 109, 115, 122 and 126 have 
weak to moderate excesses and show H$_{\beta}$ in emission. Moreover, the stars 109, 122 and 126 also occupy separate positions in 
the dereddened NIR color-color diagram. They are probable Group II HAB stars as the values of $s$ for star 109 and 126 are above 
or near zero. Star 122 is suggested to be a HAB star by Boesono et al. (1987). Star 115 lies on the reddening line track in the 
NIR color-color diagram, hence this is a candidate Be star. Another group of 8 stars (110, 111, 119, 121, 129, 134, 136, 146) 
show Vega-like characteristics most showing substantial MIR excesses.

NGC 6611 is also well studied and is reported to contain hundreds of low-mass PMS stars with a median age of $\sim$ 2 Myr and 
an age spread $\sim$ 7 Myr (Hillenbrand et al. 1993; Belikov et al. 2000). In the HRD, the stars 177, 178 and 185 lie well above the
birthline and have $E(B-V) < 0.6$ mag, lower than the cluster mean $\sim$ 0.86 mag. A recent PM study by Belikov et al. (1999) 
assigns P$_{\mu}$ = 0.08 for star 177, so these stars
may be non-members. Another group of 5 BD stars (181, 182, 183, 184, 186) have $E(B-V) < 0.4$ mag and are located
well above the birthline, therefore these stars also could be non-members. The stars 174 and 176 occupy positions away from the MS.
The former is just evolving off the MS and is probably a member while the latter may be affected by binarity. Star 174 with
H$_{\beta}$ filled-in with emission and a candidate for a classical Be star is therefore more likely to be a cluster member. Stars 
180 and 187, both of B type, are quite interesting and are member as indicated from their PM, $E(B-V)$, $R_{V}$ and HRD 
location, however these are located diagonally $\sim$ 20 arcmin away
from the cluster center. The projected corona radius for the cluster is $\sim$ 15 arcmin (Belikov et al. 1999). Therefore if these
are members then they probably formed in the outer region. Their location also 
indicates that the region of cluster formation would have been quite extended. Further kinematical data is required to authenticate 
these possibilities. Among remaining objects, stars 160, 173 and 175 have HAB characteristics. Star 160 has value of $s \sim -2.3$, 
near blackbody and shows no intrinsic polarisation (Orsatti et al. 2000), therefore it is probably a group III HAB or a classical 
Be star. Star 175 has $s \sim -0.47$ and shows strong intrinsic polarisation suggesting it to be a group I HAB star. 
The stars 154, 155, 166, 171, 172 and 179 show MIR excesses, stars 155 and 171 are reported to be spectroscopic 
binary (Bosch et al. 1999) while stars 155 and 166 show intrinsic polarisation (Orsatti et al. 2000). Therefore all these stars 
appears to have circumstellar material and the stars 154, 166 and 172 are probable Vega-like.

NGC 6823 has an age $\sim$ 3 Myr and an age spread $\sim$ 9 Myr (Guetter et al. 1992; 
Pigulski et al. 2000). A recent study by Pigulski et al. (2000) indicates that the stars with spectral classes later than A0 are 
in their PMS phase. Stars 194, 199, 206, 208, 209 and 211 are non-members, stars 206 and 208 have PM membership 
$<$ 0.5 while others lie above the birth line. The spectrum obtained by us indicates that star 209 is a late A type star, though 
it is reported to be of G8 type by Shi \& Hu (1999).  The remaining stars 18 in total are cluster members. Star 
195 occupies a position slightly away from the MS and is reported to show light variability. It is either a binary or a single star
about to leave the MS. Of these, stars 188, 200, 203 and 204 have circumstellar material as indicated from MIR spectral indices. The
value of $s$ for star 210 is above zero and it is a probable candidate Vega-like star. Star 188 has recently 
been argued to be a hot post-AGB candidate (Gauba et al. 2003) based on its IR excess, but, being a PM member this 
is unlikely to be in a post-AGB phase.  

\subsection{Stars}

Membership indicators suggest that the sample contains 34 non-members and 10 probable members while the 
remaining are members. These are indicated in the last column of Table 6. Around 16\% of the PM members were found to be 
non-members in the present study and this is supported in a few cases by recent PM studies. The converse is also true in a few 
cases, for example star 6 is a PM non-member, but other indicators suggest it to be a bona-fide member.  
Out of the total 28 identified emission line stars, only 13 are of early type.
Their characteristics suggests that they belong to the classical Be, Herbig Ae/Be or T Tauri populations.
A group of 36 stars are reported to have weak to strong NIR excesses (29 members, 5 probable members and 2 non-members). Most of 
these are observed to be PMS objects. Another group of 6 stars (all members except one), show IR deficits. 
A group of 37 non-emission stars were identified to have circumstellar material as seen either from weak NIR excesses 
or MIR spectral indices. Of these, 25 (17 have above zero values of $s$ and 8 have weak NIR excesses) are Vega-like or precursors 
to such stars (see Table 12) - one of these stars (13) has recently been shown to have Vega-like characteristics by 
Manoj et al. (2002). Three of these stars (87, 149, 151) are probably in a highly evolved stage with ages $\sim$ 20 
to 30 Myr and with masses $\sim$ 10 to 12 $M_{\odot}$.

\section{conclusions}

We present the spectral and reddening properties of 211 highly reddened proper motion members (mostly 
early type with $V < 15$ mag) in six young galactic clusters. The main conclusions of the study are given below.

(i) Emission features in CaII HK and HI lines are observed for a sample of 29 stars including 11 reported for the first time.
We also provide either a new or more reliable spectral class for a sample of 24 stars. CaII triplet width measurements were used to 
indicate the presence of disks for a dozen stars, a few of them were found to be candidate weak-line T Tauri stars.

(ii) A significant fraction ($>$70\%) of cluster members in NGC 1976, 
NGC 2244, NGC 6530 and NGC 6611 show anomalous reddening with $R_{V}$ = $5.11\pm0.11$, $3.60\pm0.05$, $3.87\pm0.05$ 
and $3.56\pm 0.02$, respectively, indicating the presence of larger grain size dust in these star forming regions. A small 
number of stars in NGC 1976, NGC 2244 and NGC 6611 also show normal behavior while the cluster NGC 6823 appears to have 
normal reddening. 

(iii) On the basis of spectral features, NIR excesses, dereddened color-color diagrams and MIR spectral indices we identify a 
highly probable group of 5 Herbig Ae/Be and 6 classical T Tauri stars. Moreover, a further probable group of 3 classical Be, 5 
Herbig Ae/Be and 9 weak line T Tauri stars are also identified. These PMS population amount to $\sim$ 15\% of the cluster members.

(iv) A total of 37 non-emission line stars, mostly of early type, were identified to have circumstellar material as seen from 
weak NIR excesses or MIR spectral indices. Of these, 25 or 10 \% of the early-type MS members, appear to
show Vega-like characteristics or are precursors to such a phenomenon.  

(v) Three highly luminous late type giants in two of the six clusters appear to be members and are in 
the post-hydrogen-core-buring stage suggesting a prolonged duration ($\sim$ 25 Myrs) of star formation in these clusters. 

\section*{acknowledgements} 

We are grateful to Mount Stromlo Observatory, Australia for generous allotment of observing time. One of us (RS) is thankful 
to the IAU and the Anglo-Australian Observatory, Epping, Australia, for the financial support during the observations. We are 
thankful to Dr. A. K. Pandey for useful discussions. 
The present research makes use of data from (i) the open cluster data base at the website {\bf http://obswww.unige.ch/webda/} 
maintained by Dr. J. C. Mermilliod (ii) Two Micron All Sky Survey, which is a joint project of the
university of Massachusetts and the Infrared Processing and Analysis center/California Institute of Technology, funded
by the National Aeronautics and Space Administration and the National Science Foundation.

\clearpage


\begin{table*}
\caption{General information on age, distance and mean reddening which are taken from recent 
studies by Hillenbrand (1997) for NGC 1976,  by Park \& Sung (2002) and Massey et al. (1995) for NGC 2244, by Sung et al. (1997) 
for NGC 2264, by Sung et al. (2000) for NGC 6530, by Hillenbrand et al. (1993) and Belikov et al. (1999) for NGC 6611 and 
by Sagar \& Joshi (1978), Guetter et al. (1992) and Pigulski et al. (2000) for NGC 6823. The likely age and reddening spread 
for cluster members are given in columns 2 \& 4. Columns 5 and 6 contain cluster members with $V<15$ mag and proper motion 
probability (p) $>$ 50\% along with the source of proper motion data. The last column provides the number of selected 
sample stars in the cluster.}

\begin{tabular}{lrccclc} \hline 
  Cluster&     Age& Distance&   $E(B-V)$&       Memb.& Source& Samples\\
         &   (Myr)&     (kpc)&  (mag)&   (p $>$ 50\%)&       &        \\ \hline  
 NGC 1976&     1.0(9)&           0.47&        0.06 (0.02-1.00)&        50& McNamara and Huels (1983)  &      44\\ 
 NGC 2244&     1.9(6)&           1.70&        0.47 (0.40-0.56)&       100& Marshall et al. (1982)     &      43\\ 
 NGC 2264&     1.5(9)&           0.76&        0.07 (0.06-1.20)&       140& Vasilevskis et al. (1965)  &      19\\
 NGC 6530&     1.5(5)&           1.80&        0.35 (0.25-0.50)&        88& van Altena and Jones (1972)&      45\\
 NGC 6611&     2.0(7)&           2.14&        0.86 (0.40-1.60)&        50&  Kamp (1974)               &      36\\
 NGC 6823&     3.0(9)&           2.10&        0.85 (0.60-1.16)&        41&  Erickson (1971)           &      24\\ \hline         
\end{tabular} 
\end{table*} 


\begin{table*}
\caption{Spectrographs, gratings, CCDs and central wavelengths ($\lambda_{c}$). The Boller Chivens Spectrograph (BCS)
was used at ANU 1-m Cassegrain focus and the Double Beam Spectrograph with blue and red arms (DBS-B, R) was used at 
ANU 2.3-m Nasmyth focus.} 

\begin{tabular}{lccccc} \hline 
 Spectrographs&  Grating& \multicolumn{2}{c}{Dispersion}&  $\lambda_{c}$&  CCDs\\ \cline{3-4} 
              &   (g/mm)& \AA/mm& \AA/pixel&  (\AA)&    \\  \hline
           BCS&      258&    260&       5.7&   4825&  GEC (576 $\times$ 380, 22$\mu$)\\
         DBS-B&      300&    140&       4.5&   4500&  SITe(1752 $\times$ 532, 15$\mu$)\\
         DBS-R&      316&    280&       4.1&   8400&  SITe(1752 $\times$ 532, 15$\mu$)\\
\hline
\end{tabular}           
\end{table*} 


\begin{table*}
\caption{Consolidated log of observations taken with the Siding Spring Observatory (SSO) telescopes.} 

\begin{tabular}{ccclccc} \hline 
           Date&  Telescope& Instrument&  Clusters                  & Objects& Standards& Arcs\\ \hline
 13/14 Aug 1989&        1-m&        BCS&  NGC 6611                  & 36& 10& 1\\
 13/14 Aug 1989&        1-m&        BCS&  NGC 6530                  & 42& 14& -\\
 08/09 Oct 1995&       2.3m&        DBS&  NGC 2244                  & 12& 15& 2\\ 
 09/10 Oct 1995&       2.3m&        DBS&  NGC 1976, 2244            & 31& 16& 1\\ 
 10/11 Oct 1995&       2.3m&        DBS&  NGC 1976, 2244, 6823      & 38&  7& -\\ 
 11/12 Oct 1995&       2.3m&        DBS&  NGC 6530, 6823            &  8&  1& -\\ 
 12/13 Oct 1995&       2.3m&        DBS&  NGC 1976, 2264, 6530, 6823& 38&  9& -\\ 
 13/14 Oct 1995&       2.3m&        DBS&  NGC 1976, 2264            & 25&  4& -\\  \hline
\end{tabular}           
\end{table*} 

\clearpage 


\begin{table*}
\caption{The first three columns provide identifications. A running number is adopted and is given in column 1. The ``N'' in the 
object column denote NGC numbers followed by the numbering 
from Parenago (1954) for NGC 1976 - "P", from Ogura \& Ishida (1981) for NGC 2244 - "OI", from Vasilevskis et al. (1965) 
for NGC 2264 - "VAS", from Walker (1957) or van Altena \& Jones (1972) for NGC 6530 - "W" or "VA", from Walker (1961) and 
Kamp (1974) for NGC 6611 - "W" or "K" and from Erickson (1971) for NGC 6823 - "E". The membership probability (p) is taken from
the sources given in Table 1. Spectral type (SpT) as available in literature with their references given in columns 5 and 6
respectively.  Optical and NIR colors are taken from the literature (see text). Further information about asterisked
stars and the references to SpT are given at the end of the table.}   

\scriptsize
\begin{tabular}{llcclcrrrrrrrr} \hline
  ID& Object & Others&p(\%)&SpT& Ref.&\it(U-V)&  \it(B-V)& \it V& \it(V-R)&  \it(V-I)& \it(V-J)&     \it(V-H)&    \it(V-K)\\ \hline 
  1*& N1976 P1044 &      HD 36629    & 96& B2.5 IV     & 28&    $-$0.64&  0.01&  7.69&  0.10&     -&  0.19&  0.13&  0.17\\
  2 & N1976 P1049 &                  & 61& K2 IV       & 22&    $ $3.14&  1.60& 11.87&     -&     -&  3.11&  3.90&  4.11\\
  3*& N1976 P1212 &      HD 294224   & 92& B8 V        & 26&    $ $1.15&  0.69& 11.39&  0.49&  0.98&  1.62&  1.87&  2.04\\
  4*& N1976 P1360 &                  & 87& G8 V        & 32&    $ $1.36&  0.94& 13.81&     -&     -&  1.60&  2.03&  2.16\\
  5*& N1976 P1409 &      EZ Ori      & 94& F8 Vn(e)    & 32&    $ $1.16&  0.86& 11.57&  0.48&  0.98&  1.90&  2.61&  3.15\\
  6*& N1976 P1469 &                  & 73& G9 IV-V     & 24&    $ $1.38&  1.19& 11.92&  0.58&  1.13&  2.29&  2.96&  3.23\\
  7 & N1976 P1539 &                  & 92& B8:         & 17&    $ $1.00&  0.71& 10.77&  0.54&  1.06&  1.69&  1.94&  2.04\\
  8*& N1976 P1623 &  BD -05$^{o}$1306& 93& A2 Vp       & 25&    $ $1.00&  0.57& 10.13&  0.44&  0.84&  1.35&  2.13&  2.88\\
  9 & N1976 P1683 &  BD -05$^{o}$1309& 92& A0          & 17&    $ $0.85&  0.46& 10.93&  0.27&  0.53&  0.89&  0.97&  1.08\\
 10*& N1976 P1699 &                  & 83& G0 V        & 32&    $ $0.87&  0.81& 13.04&  0.46&  0.96&  1.70&  2.12&  2.19\\
 11 & N1976 P1712 &  BD -05$^{o}$1310& 91& B9          & 17&    $ $0.83&  0.57& 10.47&  0.40&  0.68&  1.30&  1.54&  1.70\\
 12 & N1976 P1736 &                  & 95& G5          & 17&    $ $2.01&  1.27& 11.11&  0.90&  1.64&  2.94&  3.57&  3.96\\
 13*& N1976 P1772 &       HD 36982   & 96& B2 V        & 25&    $-$0.52&  0.09&  8.46&  0.18&  0.42&  0.72&  0.82&  1.01\\
 14 & N1976 P1798 &      HD 294264   & 94& B3 Vn       & 18&    $-$0.05&  0.36&  9.47&  0.29&  0.69&  1.38&  1.73&  1.91\\
 15*& N1976 P1799 &      LT Ori      & 96& K0? IV,V?   & 32&    $ $1.98&  1.36& 12.76&  0.74&  1.82&  3.22&  3.90&  4.17\\
 16*& N1976 P1854 &      HD 294263   & 95& A0          &  8&    $ $0.49&  0.33& 10.10&     -&     -&  0.80&  0.97&  1.09\\
 17*& N1976 P1863 &       HD 37021   & 95& B0 V        & 16&    $ $0.04&  0.23&  7.72&     -&     -&  2.09&  2.39&  2.37\\
 18*& N1976 P1865 &       HD 37020   & 96& B0.5 V      & 25&    $-$0.85&  0.03&  6.74&  0.22&  0.42&  1.94&  2.13&  2.01\\
 19 & N1976 P1881 &      HD 294262   & 85& A0          & 17&    $ $0.11&  0.21&  9.81&     -&     -&  1.05&  1.27&  1.44\\
 20*& N1976 P1885 &      MR Ori      & 93& A2: V       &  1&    $ $0.49&  0.34& 10.55&  0.28&  0.69&  1.31&  1.57&  1.71\\
 21*& N1976 P1889 &       HD 37023   & 96& B0.5 V      & 25&    $-$0.73&  0.08&  6.68&  0.18&  0.38&  0.62&  0.79&  0.95\\
 22*& N1976 P1891 &       HD 37022   & 96& O6:         & 25&    $-$0.97& -0.02&  5.14&  0.14&  0.32&  0.63&  0.81&  0.96\\
 23*& N1976 P1955 &                  & 90& G0-1 IV-III & 26&    $ $1.69&  1.09& 10.91&  0.61&  1.28&  2.26&  2.85&  3.06\\
 24*& N1976 P1956 &       V1230 Ori  & 90& B2: V       & 25&    $-$0.11&  0.32&  9.61&  0.13&  0.60&  1.54&  2.01&  2.26\\
 25*& N1976 P1993 &       HD 37041   & 93& O9.5 V      & 25&    $-$1.04& -0.08&  5.06&  0.03&     -&  0.15&  0.14&  0.24\\
 26*& N1976 P2001 &       V358 Ori   & 94& G8 V, K3    &  8&    $ $1.24&  0.90& 12.44&  0.55&  1.32&  2.21&  2.82&  2.94\\
 27*& N1976 P2029 &       AI Ori     & 93& G3: IV-V    & 24&    $ $2.64&  1.81& 12.26&  0.79&  1.92&  3.26&  4.37&  5.31\\
 28*& N1976 P2033 &       V1232 Ori  & 94& G3 IV-V     & 24&    $ $1.07&  0.91& 11.73&  0.61&  1.14&  2.03&  2.58&  2.79\\
 29*& N1976 P2069 &                  & 97& G8 V        &  8&    $ $2.43&  1.15& 12.20&  0.52&  1.36&  2.25&  2.90&  3.08\\
 30*& N1976 P2074 &       HD 37061   & 94& B0.5 V      & 25&    $-$0.43&  0.26&  6.83&  0.28&  0.57&  1.08&  1.24&  1.36\\
 31 & N1976 P2100 &                  & 96& G9: IV-Ve:  & 24&    $ $2.04&  1.20& 11.77&  0.56&  1.47&  2.46&  3.16&  3.40\\
 32*& N1976 P2115 &       TV Ori     & 77& K2 IV(e)    & 32&    $ $1.92&  1.17& 13.40&     -&  1.36&  3.09&  3.83&  4.31\\
 33*& N1976 P2181 &       V500 Ori   & 95& K2: V (Li)  & 32&    $ $2.07&  1.20& 12.93&  0.63&  1.67&  2.94&  3.64&  4.06\\
 34*& N1976 P2244 &                  & 97& K1-2 V(e)   & 32&    $ $2.05&  1.19& 12.62&  0.69&  1.38&  2.67&  3.34&  3.53\\
 35*& N1976 P2247 &       T Ori      & 93& B8-A3 Vp    & 14&    $ $0.82&  0.41& 10.00&  0.08&  0.40&  1.73&  2.76&  3.85\\
 36 & N1976 P2248 &                  & 95& B4 V        & 24&    $ $0.47&  0.63& 11.31&  0.59&  1.23&  2.44&  2.97&  3.27\\
 37 & N1976 P2302 &       HD 37130   & 87& B9          & 17&    $ $0.12&  0.19&  9.94&  0.15&  0.36&  0.70&  0.75&  0.88\\
 38*& N1976 P2317 &                  & 59& F6: IV      & 32&    $ $0.94&  0.72& 12.49&  0.40&  0.93&  1.58&  1.92&  2.04\\
 39 & N1976 P2367 &                  & 92&             &   &    $ $1.40&  0.83& 13.24&  0.47&  1.11&  1.48&  1.86&  1.97\\
 40*& N1976 P2424 &  BD -05$^{o}$1337& 88& A9-F0 IV    & 14&    $ $0.33&  0.26& 10.74&     -&     -&  0.58&  0.67&  0.74\\
 41 & N1976 P2425 &                  & 96& B5 V        & 34&    $ $0.67&  0.71& 10.67&  0.66&  1.23&  2.05&  2.33&  2.48\\
 42 & N1976 P2476 &                  & 95& G0-G5       & 17&    $ $2.60&  1.60& 12.83&  0.75&     -&  3.43&  4.15&  4.41\\
 43 & N1976 P2500 &                  & 96& A3          & 36&    $ $0.58&  0.37& 11.34&  0.26&  0.45&  0.96&  1.13&  1.18\\
 44 & N1976 P2519 &  BD -04$^{o}$1193& 95& A0          & 36&    $ $0.89&  0.69& 11.27&  0.40&  0.90&  1.52&  1.75&  1.82\\
 45*& N2244 OI14  &                  & 97& G0 III      & 37&    $ $1.47&  1.05& 12.80&     -&     -&  2.01&  2.44&  2.61\\
 46*& N2244 OI20  &                  & 98& F9 V        & 37&    $ $0.79&  0.75& 12.62&     -&     -&  1.34&  1.63&  1.70\\
 47 & N2244 OI29  &                  & 91&             &   &    $ $2.28&  1.25& 13.13&     -&     -&  2.86&  3.52&  3.78\\
 48 & N2244 OI44  &                  & 95& F6 V        & 37&    $ $0.43&  0.46& 13.02&     -&     -&  1.02&  1.29&  1.33\\
 49 & N2244 OI45  &       HDE 258859 & 92& A1 IV       & 37&    $ $0.47&  0.21& 10.55&     -&     -&  0.56&  0.56&  0.67\\
 50 & N2244 OI46  &                  & 62& B9 V        & 37&    $ $0.73&  0.51& 12.37&     -&     -&  1.28&  1.38&  1.51\\
 51 & N2244 OI48  &                  & 89& F1 V        & 37&    $ $0.65&  0.48& 12.53&     -&     -&  1.07&  1.22&  1.29\\
 52 & N2244 OI62  &                  & 53& B2.5 II-III & 37&    $ $0.57&  0.71& 12.93&     -&  1.17&  1.95&  2.22&  2.41\\
 53 & N2244 OI63  &                  & 93& F4          & 36&    $ $3.41&  1.66& 14.11&     -&  1.89&  3.46&  4.29&  4.61\\
 54*& N2244 OI72  &                  & 00&             &   &    $ $0.53&  0.47& 12.53&     -&  0.66&  1.08&  1.34&  1.45\\
 55 & N2244 OI78  &                  & 50& A7 V        & 37&    $ $0.58&  0.33& 12.27&     -&     -&  0.69&  0.78&  0.82\\
 56*& N2244 OI79  &                  & 94& B2 V        & 10&    $-$0.39&  0.16& 10.70&  0.21&  0.35&  0.43&  0.48&  0.49\\
 57*& N2244 OI80  &       HDE 259012 & 97& B0.5 V      & 10&    $-$0.50&  0.14&  9.39&  0.28&  0.46&  0.49&  0.55&  0.61\\
 58*& N2244 OI84  &        HD 46056  & 92& O8 V(e)     & 21&    $-$0.58&  0.15&  8.22&  0.17&  0.33&  0.38&  0.39&  0.41\\
 59*& N2244 OI91  &       HDE 258986 & 62& F4 V        & 37&    $ $0.53&  0.49& 10.15&     -&     -&  0.66&  0.85&  0.92\\
 60 & N2244 OI102 &                  & 74&             &   &    $ $1.23&  0.84& 13.49&     -&     -&  2.00&  2.41&  2.54\\
 61 & N2244 OI108 &                  & 83& B8 V        & 37&    $ $0.06&  0.22& 11.42&     -&  0.33&  0.60&  0.66&  0.67\\
 62 & N2244 OI109 &                  & 82& A0:         & 36&    $ $0.77&  0.49& 13.80&     -&  0.69&  1.22&  1.36&  1.50\\
 63 & N2244 OI110 &                  & 58& G7 III      & 37&    $ $1.85&  1.09& 10.78&     -&  1.14&  1.97&  2.46&  2.67\\
 64*& N2244 OI114 &        HD 46149  & 95& O8.5 V      & 21&    $-$0.43&  0.18&  7.72&  0.28&  0.41&  0.48&  0.47&  0.48\\
 65*& N2244 OI115 &        HD 46106  & 93& B0 V        & 15&    $-$0.67&  0.08&  8.03&  0.22&  0.33&  0.42&  0.44&  0.42\\
 66*& N2244 OI116 &                  & 90& B8 V        & 37&    $ $0.20&  0.32& 12.79&     -&  0.49&  0.88&  0.99&  1.09\\
 67*& N2244 OI122 &        HD 46150  & 86& O5 V((f))   & 19&    $-$0.70&  0.16&  6.74&  0.13&  0.24&  0.29&  0.27&  0.32\\
 68*& N2244 OI125 &                  & 87& B4 V        & 37&    $-$0.07&  0.27& 12.01&  0.45&  1.17&  0.75&  0.85&  0.97\\
 69*& N2244 OI127 &        HD 46107  & 61& A2 V        &  9&    $ $0.17&  0.08&  8.76&  0.20&  0.24&  0.24&  0.22&  0.30\\
 70 & N2244 OI128 &       HDE 259105 & 97& B1.5 V      & 15&    $-$0.54&  0.14&  9.39&  0.15&  0.29&  0.39&  0.40&  0.45\\
\hline                                                               
\end{tabular}                                                        
\end{table*}                                                          
                                                                    
\begin{table*}                                                       
{{\bf Table 4.}  Continued.}                                       
                                         
\scriptsize                                                                   
\begin{tabular}{llcclcrrrrrrrr} \hline
  ID& Object & Others&p(\%)&SpT& Ref.&\it(U-V)&  \it(B-V)& \it V& \it(V-R)&  \it(V-I)& \it(V-J)&     \it(V-H)&    \it(V-K)\\ \hline 
 71 & N2244 OI130 &                  & 94& B2.5 V      & 15&    $-$0.17&  0.26& 11.65&     -&  0.39&  0.65&  0.71&  0.79\\
 72*& N2244 OI133 &                  & 94& B9 V        & 37&    $ $0.48&  0.40& 11.73&  0.85&  1.50&  1.06&  1.20&  1.35\\
 73 & N2244 OI167 &       HDE 259172 & 97& B3 V        & 65&    $-$0.40&  0.19& 10.70&  0.14&  0.29&  0.52&  0.61&  0.63\\
 74 & N2244 OI172 &                  & 95& B2.5 V      & 15&    $-$0.17&  0.27& 11.27&  0.26&  0.49&  0.72&  0.78&  0.86\\
 75*& N2244 OI180 &        HD 46202  & 93& O9 V        & 21&    $-$0.53&  0.14&  8.21&  0.17&  0.32&  0.43&  0.43&  0.50\\
 76 & N2244 OI190 &                  & 98& B2.5 Vn     & 15&    $-$0.20&  0.24& 11.25&  0.19&  0.37&  0.60&  0.67&  0.72\\
 77 & N2244 OI192 &                  & 87& B7 V        & 37&    $ $0.31&  0.45& 12.55&     -&  0.49&  1.15&  1.32&  1.49\\
 78*& N2244 OI194 &                  & 95& B6 Vne      & 15&    $ $0.03&  0.31& 12.02&  0.30&  0.56&  0.87&  1.02&  1.08\\
 79*& N2244 OI197 &                  & 87& B8 IV       & 37&    $ $0.17&  0.33& 12.64&     -&  0.48&  0.93&  1.12&  1.21\\
 80*& N2244 OI200 &       HDE 259135 & 98& B0.5 V      & 15&    $-$0.56&  0.14&  8.54&  0.12&  0.27&  0.39&  0.41&  0.43\\
 81*& N2244 OI203 &        HD 46223  & 96& O4 V((f))   & 19&    $-$0.59&  0.21&  7.24&  0.16&  0.35&  0.50&  0.54&  0.58\\
 82*& N2244 OI253 &       HDE 259300 & 96& B3 Vp       & 15&    $ $0.01&  0.30& 10.78&  0.19&  0.46&  0.91&  1.13&  1.47\\
 83 & N2244 OI256 &                  & 92& B6 V        & 37&    $ $0.23&  0.41& 12.82&     -&  0.60&  1.07&  1.19&  1.27\\
 84 & N2244 OI376 &       HDE 258691 & 81& O9.5 V      & 37&    $ $0.07&  0.54&  9.71&  0.36&  0.78&  1.49&  1.69&  1.80\\
 85 & N2244 OI377 &                  & 94& F3 V        & 37&    $ $0.99&  0.62& 12.28&     -&     -&  1.25&  1.45&  1.57\\
 86*& N2244 OI393 &       HDE 259635 & 99& G1 V        & 37&    $ $1.02&  0.79&  9.74&     -&     -&  1.46&  1.81&  1.93\\
 87*& N2244 OI397 &       HDE 259696 & 97& K3 Ib       & 37&    $ $4.58&  2.19&  8.85&     -&     -&  3.74&  5.02&  5.35\\
 88*& N2264 VAS1  &                  & 95& F4 V        & 30&    $ $0.85&  0.62& 12.58&  0.35&  0.71&  1.19&  1.47&  1.58\\
 89 & N2264 VAS2  &                  & 98& F7 V        & 30&    $ $0.88&  0.74& 13.33&  0.49&  0.91&  1.52&  1.93&  2.04\\
 90*& N2264 VAS10 &                  & 98& K1 V        & 30&    $ $5.27&  2.52& 11.67&  1.20&  2.80&  4.72&  5.76&  6.20\\
 91 & N2264 VAS11 &                  & 84&             &   &    $ $0.68&  0.63& 14.22&  0.47&  0.70&  1.18&  1.47&  1.58\\
 92 & N2264 VAS22 &                  & 96& F5 V        & 30&    $ $1.07&  0.78& 13.22&  0.44&  0.94&  1.51&  1.97&  2.24\\
 93*& N2264 VAS32 &                  & 94&             &   &    $ $2.77&  1.59& 13.94&  0.93&     -&  3.44&  4.14&  4.42\\
 94 & N2264 VAS46 &          V780 Mon& 96& A0 V        & 30&    $ $0.83&  0.71& 12.34&  0.58&  1.43&  2.69&  3.34&  3.79\\
 95*& N2264 VAS47 &                  & 93& B2 V        &  5&    $ $0.19&  0.62& 10.83&  0.45&  1.20&  2.42&  2.99&  3.37\\
 96 & N2264 VAS48 &                  & 93& G0 IV-V     &  5&    $ $0.77&  0.62& 11.71&  0.30&  0.74&  1.18&  1.49&  1.59\\
 97*& N2264 VAS62 &          HRC 219 & 92& B4 V        & 30&    $ $0.22&  0.14& 12.76&  0.35&  0.84&  1.32&  2.31&  3.50\\
 98*& N2264 VAS72 &         HD 261841& 94& A2 IV       &  5&    $ $0.23&  0.14& 10.04&  0.10&  0.28&  0.52&  0.91&  1.60\\
 99 & N2264 VAS86 &                  & 92& F8 V        & 30&    $ $1.05&  0.86& 12.47&  0.42&  0.90&  1.69&  2.16&  2.29\\
100*& N2264 VAS92 &                  & 96& G5 V        & 30&    $ $1.35&  0.94& 12.30&  0.35&  0.86&  1.65&  2.24&  2.64\\
101*& N2264 VAS122&          V360 Mon& 96& G8 V        & 30&    $ $1.33&  0.95& 13.39&  0.49&  0.96&  1.75&  2.38&  2.88\\
102*& N2264 VAS188&                  & 65& F5 V        & 30&    $ $0.83&  0.66& 11.53&  0.27&  0.61&  1.08&  1.42&  1.49\\
103*& N2264 VAS192&                  & 86&             &   &    $ $3.53&  1.95& 14.43&  1.14&  2.30&  4.03&  4.82&  5.20\\
104 & N2264 VAS227&                  & 86&             &   &    $ $1.09&  0.61& 15.04&  0.18&     -&  1.36&  1.56&  1.69\\
105*& N2264 VAS228&                  & 67&             &   &    $ $2.92&  1.67& 13.53&  0.70&     -&  3.11&  3.83&  4.09\\
106*& N2264 VAS238&                  & 93& K5 V        & 30&    $ $2.12&  1.01& 11.21&  0.37&  0.82&  1.56&  2.02&  2.16\\
107 & N6530 W2    &      HD 164536   & 00& O7          & 23&    $-$0.95& -0.03&  7.11&     -&     -& -0.01& -0.04& -0.01\\
108 & N6530 W3    & CD -24$^{o}$13785& 72& B0.5        & 40&    $-$0.76&  0.01&  8.68&  0.04&  0.09&  0.13&  0.04&  0.09\\
109 & N6530 W5    &      HD 314900   & 02& B5          &  2&    $-$0.37&  0.13& 10.42&     -&     -&  0.47&  0.63&  0.84\\
110 & N6530 W7    &      HD 164794   & 08& O4 V((f))   & 19&    $-$0.88&  0.03&  5.97&  0.13&  0.19&  0.22&  0.22&  0.25\\
111*& N6530 W9    &      HD 164816   &   & O9.5 IVn    & 28&    $-$0.89&  0.00&  7.07&     -&  0.08&  0.06&  0.02&  0.01\\
112 & N6530 W15   &                  & 57& B7          &  6&    $-$0.20&  0.17& 11.60&     -&  0.31&  0.57&  0.66&  0.68\\
113 & N6530 W19   &      HD 315025   & 00& B8          &  2&    $-$0.23&  0.25& 10.78&     -&  0.50&  0.94&  1.21&  1.35\\
114*& N6530 W27   &                  & 79& F0:         & 40&    $ $0.90&  0.66& 12.95&  0.58&  1.20&  1.54&  1.82&  1.94\\
115 & N6530 W29   &                  & 20& F5-G0       & 11&    $ $0.93&  0.81& 13.19&  0.53&  1.23&  2.58&  3.36&  3.77\\
116 & N6530 W32   &                  & 84& B3 Ve       & 12&    $-$0.48&  0.11& 10.51&  0.19&  0.34&  0.48&  0.56&  0.63\\
117 & N6530 W35   & CD -24$^{o}$13822& 76& K0 III      & 12&    $ $1.94&  1.14&  9.84&     -&  1.22&  2.17&  2.66&  2.87\\
118*& N6530 W42   &      HD 315032   & 84& B2 Vne      & 12&    $-$0.71&  0.04&  9.18&  0.08&  0.14&  0.18&  0.17&  0.22\\
119*& N6530 W43   &      HD 315026   &   & B2          & 40&    $-$0.64&  0.09&  9.02&     -& $-$0.05&  0.23&  0.09&  0.16\\
120*& N6530 W46   &                  & 60& B8e:        & 40&    $-$0.03&  0.19& 11.63&     -&  0.49&  0.95&  1.18&  1.33\\
121*& N6530 W56   & CD -24$^{o}$13829& 86& B1.5 Vne    & 12&    $-$0.61&  0.10&  9.03&  0.06&  0.20&  0.39&  0.42&  0.50\\
122*& N6530 W58   & CD -24$^{o}$13830& 81& B2 Ve       & 12&    $-$0.47&  0.18&  9.86&  0.14&  0.39&  0.77&  1.00&  1.32\\
123*& N6530 W59   &      HD 315033   & 75& B2 Vp       & 12&    $-$0.56&  0.09&  8.93&  0.05&  0.19&  0.51&  0.56&  0.67\\
124*& N6530 W60   &                  & 81& B1 Ve       & 12&    $-$0.59&  0.07&  9.66&  0.07&  0.20&  0.34&  0.41&  0.46\\
125 & N6530 W61   &                  & 25& B2 Ve       & 12&    $-$0.49&  0.12& 10.29&     -&  0.24&  0.42&  0.46&  0.48\\
126 & N6530 W65   &      HD 164906   & 01& B0 IVpne    & 12&    $-$0.60&  0.16&  7.42&  0.20&  0.46&  0.77&  0.94&  1.28\\
127*& N6530 W66   & CD -24$^{o}$13831& 79& B2 Vpe      & 12&    $-$0.54&  0.11& 10.14&  0.06&  0.21&  0.35&  0.41&  0.45\\
128 & N6530 W70   &                  & 83& B2 Ve       & 12&    $-$0.46&  0.15& 10.45&  0.04&  0.18&  0.48&  0.54&  0.62\\
129*& N6530 W73   &      HD 315031   & 48& B2 IVn      & 12&    $-$0.62&  0.08&  8.24&     -&     -&  0.31&  0.27&  0.35\\
130 & N6530 W74   &                  & 86& B2.5 Ve     & 12&    $-$0.46&  0.10& 10.69&     -&  0.27&  0.43&  0.53&  0.58\\
131 & N6530 W76   &      HD 315024   & 85& B2.5 Ve     & 12&    $-$0.72&  0.06&  9.56&  0.11&  0.18&  0.31&  0.32&  0.39\\
132*& N6530 W80   & CD -24$^{o}$13837&   & B1 Ve       & 12&    $-$0.71&  0.07&  9.40&     -&  0.23&  0.35&  0.38&  0.42\\
133*& N6530 W83   &                  & 75& B2.5 Vne    & 12&    $-$0.48&  0.13& 10.49&  0.12&  0.24&  0.44&  0.66&  0.64\\
134 & N6530 W85   &      HD 164933   & 00& B0.5 V      & 28&    $-$0.66&  0.13&  8.55&     -&     -&  0.38&  0.39&  0.43\\
135*& N6530 W86   & CD -24$^{o}$13840& 77& B2 Vne      & 12&    $-$0.42&  0.16&  9.75& -0.04&  0.12&  0.33&  0.42&  0.55\\
136*& N6530 W93   &      HD 315021   & 84& B2 IVn      & 12&    $-$0.77&  0.02&  8.63&  0.13&  0.21&  0.13&  0.32&  0.42\\
137 & N6530 W99   & CD -24$^{o}$13844& 50& B2.5 Vne    & 12&    $-$0.43&  0.09& 10.81&     -&  0.28&  0.44&  0.47&  0.55\\
138*& N6530 W100  &      HD 164947   &   & B2 IVe      & 12&    $ $   -&  0.09&  9.48&     -&     -&  0.19&  0.22&  0.26\\
139 & N6530 W105  &      HD 315022   & 56& B6          & 40&    $-$0.31&  0.17& 10.54&     -&  0.32&  0.51&  0.53&  0.58\\
140 & N6530 W110  &                  & 70& B3 V        & 33&    $-$0.33&  0.16& 11.22&  0.17&  0.31&  0.59&  0.64&  0.70\\
141*& N6530 W117  &      HD 315035   & 71& B2.5 Ve     & 12&    $-$0.20&  0.25& 10.81&  0.20&  0.43&  0.80&  0.99&  1.08\\
142*& N6530 W118  &      HD 165052   & 39& O6.5 V      & 19&    $-$0.76&  0.09&  6.87&  0.11&  0.21&  0.39&  0.40&  0.41\\
143 & N6530 VA92  &                  & 85& K5:         & 40&    $ $2.34&  1.28& 12.05&     -&  1.26&  2.28&  2.84&  3.01\\
144 & N6530 VA94  &                  & 80& B6          & 40&    $-$0.02&  0.24& 11.86&     -&  0.48&  0.97&  1.15&  1.28\\
145*& N6530 VA97  &                  & 45& K2e:        & 40&    $ $2.03&  1.24& 10.94&     -&  1.29&  2.28&  2.76&  2.96\\
\hline                                                                
\end{tabular}                                                         
\end{table*}                                                           
                                                                      
\begin{table*}                                                         
{{\bf Table 4.}  Continued.}

\scriptsize                                                                      
\begin{tabular}{llcclcrrrrrrrr} \hline
  ID& Object & Others&p(\%)&SpT& Ref.&\it(U-V)&  \it(B-V)& \it V& \it(V-R)&  \it(V-I)& \it(V-J)&     \it(V-H)&    \it(V-K)\\ \hline 
146*& N6530 VA107 &                  & 86& B2e         & 40&    $-$0.38&  0.16& 10.01&  0.15&  0.27&  0.77&  0.88&  0.99\\
147 & N6530 VA247 &                  & 81& B3          & 40&    $-$0.12&  0.28& 11.51&     -&     -&  0.74&  0.91&  1.01\\
148 & N6530 VA259 &CD -24$^{o}$13858 & 84& B8          & 40&    $ $0.17&  0.41& 11.61&     -&     -&  1.24&  1.46&  1.59\\
149*& N6530 VA304 &                  & 73& K5:         & 40&    $ $4.49&  2.32&  9.89&     -&     -&  4.87&  6.17&  6.67\\
150*& N6530 VA330 &                  & 45& B6+?        & 40&    $ $1.00&  0.86& 10.82&  0.67&  1.27&  2.30&  2.59&  2.79\\
151*& N6530 VA338 &                  & 69& K5:         & 40&    $ $4.89&  2.45& 10.27&     -&     -&  4.98&  6.26&  6.63\\
152*& N6611 W125  &  BD$-13^{o}~4920$& 94&         B1 V& 38&    $-$0.09&  0.47& 10.05&  0.33&  0.73&  1.23&  1.36&  1.47\\
153*& N6611 W150  &  BD$-13^{o}~4921$& 93&       B0.5 V& 38&    $-$0.03&  0.47&  9.89&  0.36&  0.76&  1.29&  1.47&  1.54\\
154 & N6611 W161  &                  & 94&       O8.5 V& 38&    $ $0.84&  1.02& 11.21&  0.74&  1.63&  3.05&  3.48&  3.79\\
155*& N6611 W175  &  BD$-13^{o}~4923$& 94&  05.5 V((f))& 41&    $ $0.41&  0.80& 10.02&  0.60&  1.34&  2.43&  2.80&  3.04\\
156*& N6611 W197  &  BD$-13^{o}~4925$& 94&  O6-7 V((f))& 41&    $-$0.24&  0.44&  8.77&  0.35&  0.74&  1.24&  1.34&  1.50\\
157*& N6611 W205  &  BD$-13^{o}~4926$& 94&   O4 V((f+))& 41&    $-$0.25&  0.45&  8.21&  0.39&  0.84&  1.32&  1.52&  1.66\\
158*& N6611 W223  &                  & 69&         B1 V& 38&    $ $0.22&  0.65& 11.19&  0.45&  1.02&  1.65&  1.88&  2.02\\
159*& N6611 W227  &                  & 92&      B1.5 Ve& 38&    $ $0.48&  0.69& 12.87&  0.42&  0.93&  1.82&  2.14&  2.30\\
160*& N6611 W235  &                  & 94&    Herbig Be& 38&    $ $0.36&  0.83& 11.00&  0.62&  1.38&  2.44&  2.88&  3.32\\
161*& N6611 W246  &  BD$-13^{o}~4927$& 94&    O7 IIb(f)& 41&    $ $0.52&  0.87&  9.55&  0.63&  1.31&  2.33&  2.69&  2.87\\
162*& N6611 W254  &                  & 94&         B1 V& 38&    $-$0.03&  0.44& 10.80&  0.29&  0.69&  1.24&  1.43&  1.54\\
163*& N6611 W259  &                  & 94&       B0.5 V& 38&    $ $0.41&  0.73& 11.61&  0.54&  1.17&  2.00&  2.30&  2.44\\
164*& N6611 W280  &  BD$-13^{o}~4928$& 94&       O9.5 V& 41&    $-$0.17&  0.49& 10.05&  0.29&  0.68&  1.09&  1.18&  1.31\\
165*& N6611 W301  &                  & 94&         B2 V& 38&    $ $0.28&  0.53& 12.27&  0.57&  1.05&  1.64&  1.87&  1.99\\
166*& N6611 W306  &                  & 89&      B1.5 Ve& 38&    $ $0.40&  0.63& 12.72&  0.48&  1.06&  1.86&  2.13&  2.20\\
167*& N6611 W314  &  BD$-13^{o}~4929$& 92&         B0 V& 38&    $ $0.08&  0.62&  9.92&  0.45&  0.96&  1.67&  1.84&  2.00\\
168*& N6611 W343  &                  & 93&         B1 V& 38&    $ $0.62&  0.82& 11.83&  0.70&  1.40&  2.35&  2.71&  2.96\\
169*& N6611 W367  &  BD$-13^{o}~4930$& 92&       O9.5 V& 41&    $-$0.52&  0.28&  9.43&  0.19&  0.38&  0.64&  0.66&  0.69\\
170*& N6611 W401  &  BD$-13^{o}~4932$&   &       O8.5 V& 41&    $-$0.34&  0.40&  8.96&  0.32&  0.66&  1.13&  1.28&  1.39\\
171*& N6611 W412  &  BD$-14^{o}~4991$& 93&       O9.5 I& 41&    $-$0.40&  0.34&  8.26&  0.25&  0.52&  0.73&  0.81&  0.89\\
172*& N6611 W468  &  BD$-13^{o}~4934$& 93&        B1 Vp& 38&    $-$0.42&  0.29&  9.44&  0.17&  0.42&  0.73&  0.79&  0.84\\
173*& N6611 W469  &  BD$-13^{o}~4933$& 94&      B0.5 Ve& 38&    $-$0.19&  0.42& 10.71&  0.30&  0.66&  1.13&  1.26&  1.42\\
174*& N6611 W483  &  BD$-13^{o}~4935$& 93&           Be& 27&    $ $0.20&  0.43& 10.96&  0.29&  0.62&  1.09&  1.20&  1.37\\
175*& N6611 W503  &  BD$-13^{o}~4936$& 94&    Herbig Be& 38&    $-$0.30&  0.32& 10.35&  0.96&  1.45&  1.62&  1.92&  2.27\\
176*& N6611 K556  &  BD$-13^{o}~4912$& 93&       B2.5 I& 38&    $ $1.27&  1.15&  9.91&  0.72&  1.58&  2.96&  3.38&  3.68\\
177*& N6611 K576  &                  & 72&           G5&  7&    $ $2.45&  1.31& 10.21&     -&     -&  2.63&  3.20&  3.43\\
178 & N6611 K599  &                  & 90&       Late G& 41&    $ $2.43&  1.28& 10.13&  0.70&  1.39&  2.52&  3.11&  3.38\\
179 & N6611 K601  &  BD$-13^{o}~4937$& 93&       B1.5 V& 38&    $-$0.14&  0.37& 10.66&  0.28&  0.62&  0.96&  1.14&  1.28\\
180*& N6611       &  BD$-13^{o}~4908$& 94&           B0&  7&    $ $   -&  0.38& 10.26&     -&     -&  1.15&  1.20&  1.35\\
181 & N6611       &  BD$-13^{o}~4909$& 94&         B9 V& 35&    $ $   -&  0.34&  9.61&     -&     -&  0.94&  1.03&  1.11\\
182 & N6611       &  BD$-13^{o}~4913$& 66&       A0 III& 35&    $ $   -&  0.38& 10.17&     -&     -&  1.18&  1.29&  1.45\\
183 & N6611       &  BD$-14^{o}~4967$& 86&  B9.5 III/IV& 35&    $ $   -&  0.16&  9.78&     -&     -&  0.64&  0.69&  0.77\\
184 & N6611       &  BD$-14^{o}~4972$& 69&    B9 III/IV& 35&    $ $   -&  0.10& 10.40&     -&     -&  0.78&  0.90&  1.01\\
185 & N6611       &  BD$-14^{o}~4974$& 61&           G3&  7&    $ $   -&  1.24& 10.06&     -&     -&  2.25&  2.77&  2.98\\
186 & N6611       &  BD$-14^{o}~4981$& 85&        A0 IV& 35&    $ $   -&  0.17& 10.13&     -&     -&  0.77&  0.88&  0.97\\
187 & N6611       &  BD$-14^{o}~4994$& 75&      B1/2(n)& 35&    $ $   -&  0.34&  9.41&     -&     -&  1.00&  1.13&  1.17\\
188*& N6823 E4    &                  & 91&    B0 IVe   & 31&    $ $0.55&  0.82& 10.28&     -&     -&  1.70&  1.88&  2.00\\
189 & N6823 E5    &                  & 97&    B2 V     & 42&    $ $0.47&  0.72& 12.67&     -&     -&  1.69&  1.87&  1.96\\
190*& N6823 E8    &       HD 344777  & 98&    O9.5 III & 31&    $ $0.33&  0.78&  9.48&     -&     -&  1.71&  1.88&  1.95\\
191 & N6823 E18   &                  & 98&    B2 V     & 42&    $ $0.55&  0.75& 11.99&     -&     -&  1.66&  1.84&  1.97\\
192*& N6823 E24   &                  & 94&    B1.5 V   & 39&    $ $0.28&  0.70& 11.85&     -&     -&  1.39&  1.58&  1.66\\
193 & N6823 E30   &                  & 85&    B1.5 V   & 42&    $ $0.51&  0.70& 12.61&     -&     -&  1.40&  1.60&  1.72\\
194*& N6823 E36   &                  & 97&    G8 III   & 42&    $ $3.08&  1.70& 12.50&     -&     -&  3.03&  3.63&  3.90\\
195*& N6823 E46   &       HD 344776  & 88&    B0.5 Ib  &  3&    $ $0.68&  0.86&  8.90&  0.56&  0.97&  1.68&  1.85&  1.92\\
196 & N6823 E50   &                  & 96&    B4 V     & 42&    $ $0.32&  0.59& 12.94&     -&     -&  1.54&  1.78&  1.90\\
197*& N6823 E68   &       HD 344783  & 95&    B0 IV    &  4&    $-$0.16&  0.44&  9.79&     -&     -&  1.01&  1.08&  1.15\\
198*& N6823 E77   &       HD 344775  & 97&    B1 III   &  4&    $ $0.64&  0.81& 10.53&     -&     -&  1.52&  1.64&  1.75\\
199*& N6823 E78   &                  & 93&    K8 V     & 31&    $ $3.95&  1.97& 10.45&     -&     -&  3.51&  4.32&  4.67\\
200 & N6823 E80   &                  & 96&    B2.5 V   & 42&    $ $0.36&  0.66& 12.38&     -&     -&  1.37&  1.57&  1.73\\
201*& N6823 E81   &                  & 98&    B1.5 V   & 31&    $-$0.02&  0.57& 11.07&     -&     -&  1.38&  1.48&  1.61\\
202*& N6823 E83   &       HD 344784  & 97&    O7 V((f))& 20&    $-$0.03&  0.56&  9.39&  0.40&  0.82&  1.36&  1.50&  1.56\\
203*& N6823 E84   &                  & 97&    B0.5 V   & 13&    $ $0.44&  0.76& 11.60&  0.51&  1.05&  1.80&  1.99&  2.11\\
204*& N6823 E86   &                  & 96&    O9 V     & 31&    $ $0.35&  0.74& 11.89&  0.50&  1.04&  1.67&  2.00&  2.16\\
205*& N6823 E88   &                  & 97&    B0 V:pe  & 31&    $ $0.40&  0.73& 11.83&     -&     -&  1.81&  2.02&  2.16\\
206 & N6823 E92   &                  & 51&    F8 V     & 42&    $ $0.80&  0.75& 12.65&     -&     -&  1.46&  1.77&  1.87\\
207*& N6823 E93   &                  & 93&    O9.5 V   & 31&    $ $0.54&  0.80& 12.72&     -&     -&  2.03&  2.27&  2.43\\
208 & N6823 E96   &                  & 55&    F2 IV    & 42&    $ $0.44&  0.48& 11.98&     -&     -&  1.00&  1.20&  1.27\\
209*& N6823 E103  &       HD 344789  & 95&    G8 III   & 42&    $ $2.21&  1.38& 12.95&  0.81&  1.52&  2.61&  3.21&  3.39\\
210*& N6823 E104  &                  & 98&    B0.5 V   & 31&    $ $0.65&  0.87& 11.74&     -&     -&  2.05&  2.29&  2.44\\
211*& N6823 E110  &       HD 344787  & 79&    F8 Ib    & 31&    $ $1.84&  1.19&  9.25&     -&     -&  2.23&  2.66&  2.84\\
\hline                                                              
\end{tabular}                                                       

\footnotesize
{\bf References to SpT}\\
1 - Greenstein \& Struve (1946), 2 - Cannon \&  Mayall (1949), 3 - Morgan et al. (1955), 4 - Hiltner (1956),
5 - Walker (1956), 6 - Walker (1957), 7 - Pronik (1958), 8 - Strand (1958), 9 - Hoag \& Smith (1959), 10 - Johnson (1962),
11 - Johnson \& Borgman (1963), 12 - Hiltner et al. (1965), 13 - Hoag \& Applenquist (1965), 14 - Johnson (1965),
15 - Morgan et al. (1965), 16 - Guetter (1968), 17 - Walker (1969), 18 - Smith (1972), 19 - Walborn (1972),
20 - Walborn (1973), 21 - Conti \& Leep (1974), 22 - Breger \& Rybski (1975), 23 - Houk et al. (1975),
24 - Penston et al. (1975), 25 - Levato \& Abt (1976) 26 - McNamara (1976), 27 - Schild \& Romanishin (1976),
28 - Abt \& Levato (1977), 29 - Garrison et al. (1977), 30 - Young (1978), 31 - Turner (1979), 32 - Walker (1983),
33 - Buscombe (1984), 34 - van Altena et al. (1988), 35 - Houck \& Smith-Moore (1988),
36 - Egret et al. 1991 (SIMBAD database), 37 - Verschueren (1991), 38 - Hillenbrand et al. (1993),
39 - Massey et al. (1995), 40 - van den Ancker et al. (1997), 41 - Bosch et al.(1999), 42 - Shi \& Hu (1999)

\end{table*}

\clearpage
\footnotesize
\centerline {\bf \large Notes to Table 4}

 {\bf   1.} Suspected variable (Walker 1969); B2 (Breger \& Rybski 1975). 
       
 {\bf   3.} V$_{r}$ = +25 km/s (McNamara 1976). 
       
 {\bf   4.} V$_{r}$ = +23 km/s based on metallic absorptions, H$_{\gamma}$+H$_{\delta}$ V$_{r}$ = $-$140 km/s, these may be 
            affected by nebular emission (Walker 1983).
       
 {\bf   5.} Variable (Kholopov et al. 1998); V$_{r}$ = $-$13 km/s,  CaII HK in emission only, large rotational broadening 
            V$_{sini}$ = 60 $\pm$ 20 km/s, suspected velocity variable (Walker 1983); V$_{r}$ = $-$51 km/s (Abt 1970); 
            NIR variable (Carpenter 2001).

 {\bf   6.} Non-member, $P_{\mu}$ = 0.0 (Tian et al. 1996); NIR variable (Carpenter 2001). 
       
 {\bf   8.} CaII K as in A0.5 (Levato \& Abt 1976). 
       
 {\bf  10.} V$_{r}$ = +68 km/s based on H$_{\gamma}$ plus metallic absorptions (Walker 1983).
       
 {\bf  13.} LP Ori; variable (Kholopov et al. 1998); B1.5Vp (Sharpless 1952); polarized source (Breger 1976); 
            youngest Vega-like star (Manoj et al. 2002).
       
 {\bf  15.} Variable (Kholopov et al. 1998); absorption spectrum veiled by a stellar blue continuum, line depths are 0.66 times 
            the normal value, strong nebular lines and continuous emission (Walker 1983); G5: (Breger \& Rybski 1975).
       
 {\bf  16.} LZ Ori; variable (Kholopov et al. 1998); forms a double with 1881 (Jeffers et al. 1963).
       
 {\bf  17.}  $\theta^{1}$(D) Ori = ADS 4186 B = MB Ori; variable (Kholopov et al. 1998).
       
 {\bf  18.} $\theta^{1}$(C) Ori = ADS 4186 A; broad HI lines probably due to high surface gravity (Levato \& Abt 1976).
       
 {\bf  20.} Variable (Kholopov et al. 1998); strong nebular spectrum, broad H wings and K line visible (Greenstein \& Struve 1946).
       
 {\bf  21.} $\theta^{1}$(B) Ori = ADS 4186 D; broad HI lines probably due to high surface 
            gravity (Levato \& Abt 1976); B0.5 Vp (Borgman 1960).
       
 {\bf  22.} $\theta^{1}$(A) Ori = ADS 4186 C; O6 (Johnson \& Hiltner 1956); O6pe (Lesh 1968).
       
 {\bf  23.} V$_{r}$ = +17 km/s (McNamara 1976).
       
 {\bf  24.} Variable (Kholopov et al. 1998); shell, the fairly narrow shell lines are due to FeI and MgII 4481 \AA~ but they may 
            represent a second star (Levato \& Abt 1976); B0 Vp (Johnson 1965); B8 (Walker 1969); B8 IV or V (Greenstein 
            \& Stuve 1946).
       
 {\bf  25.} $\theta^{2}$ Ori = ADS 4188 A.
       
 {\bf  26.} Variable (Kholopov et al. 1998); two spectral types from two different sources (Strand 1958).
       
 {\bf  27.} Variable, K0:e (Kholopov et al. 1998); PMS object, emission line star, HBC 141 (Herbig \& Bell 1988); 
            NIR variable (Carpenter 2001).

 {\bf  28.} Variable (Kholopov et al. 1998).

 {\bf  29.} NIR variable (Carpenter 2001).
       
 {\bf  30.} NU Ori; Variable (Kholopov et al. 1998); broad HI lines probably due to high surface gravity (Levato \& Abt 1976); 
            polarized source (Breger 1976).
       
 {\bf  32.} Variable (Kholopov et al. 1998); CaII HK in emission only, V$_{r}$ = +31 km/s based on metallic absorptions, 
            H$_{\gamma}$+H$_{\delta}$ V$_{r}$ = +218 $\pm$ 30 km/s,
            this measure may be affected by faint nebular emission visible at H$_{\gamma}$, H$_{\delta}$, H$_{\beta}$.
            CaII HK emission V$_{r}$ = +94 km/s, suspected velocity variable (Walker 1983); V$_{r}$ = $-$31 km/s (Wilson 1953);
            K6 (Cohen \& Kuhi 1979); K0e $\alpha$ (Herbig \& Rao 1972).
       
 {\bf  33.} Variable (Kholopov et al. 1998); V$_{r}$ = +82 km/s from metallic absorptions, HI and CaII HK in emission, strong 
            LiI 6707 \AA~ absorption feature, CaII K emission V$_{r}$ = +1 km/s, nebular continuum
            contamination, absorption lines veiled by stellar blue continuum, line depths are
            0.49 $\times$ normal, H$_{\alpha}$ emission V$_{r}$ = +17 km/s (Walker 1983).
       
 {\bf  34.} V$_{r}$ = 58 km/s based on metallic absorptions, H$_{\gamma}$ absorption V$_{r}$ = $-$154 km/s, measurement
            affected by nebular emission, CaII HK emission only, CaII HK emission V$_{r}$ = +36 km/s (Walker 1983).
       
 {\bf  35.} Variable (Kholopov et al. 1998); V$_{r}$ = +73 km/s (Johnson 1965); 9.9 to 12.2 mag. variability, the absorption 
            velocities from different elements differ significantly (Greenstein \& Struve 1946, Herbig 1960); possible
            spectroscopic binary, strong H lines with sharp cores, strong broad K line of type A3, CaI 4227 \AA~ as strong 
            as HeI 4472 \AA~ indicating a late B-type or a binary influence or disturbance of nebular material 
            (Greenstein \& Struve 1946); A3e (Herbig 1960); Herbig Ae/Be, SB (Leinert et al. 1997, Maheswar et al. 2002). 
       
 {\bf  38.} Suspected variable (Walker 1969); V$_{r}$ = $-$36 km/s, member, F4 to F8 (Walker 1983). 
       
 {\bf  40.} V$_{r}$ = +19 km/s, broad Balmer lines suspected (Johnson 1965).

 {\bf  45.} $P_{\mu} = 0.0$ (Sabogal-Mart\'{i}nez et al. 2001).

 {\bf  46.} $P_{\mu} = 0.11$ (Sabogal-Mart\'{i}nez et al. 2001).
       
 {\bf  54.} In observing log it is OI74 (an emission line B type star) but in our spectrum it is actually a
            F6 type star with no emission and hence we assumed that we observed OI72 instead of OI74; 
            $P_{\mu} = 0.0$ (Sabogal-Mart\'{i}nez et al. 2001).
       
 {\bf  56.} Member according to photometry indicators and radial velocity +34 $\pm$ 2.5 km/s (Riddle 1972).
       
 {\bf  57.} Magnitude variable (Johnson 1962); binary system (P\'{e}rez 1988); X-ray source (Park \& Sung 2002). 
       
 {\bf  58.} 08 Vn (Walborn 1973); X-ray emitter (Bergh\"{o}fer \& Christian 2002);
            H$_{\alpha}$ emitter (Conti 1974); strong Hydrogen absorption lines, variable radial velocity (Humphreys 1978).

 {\bf  59.} $P_{\mu} = 0.0$ (Sabogal-Mart\'{i}nez et al. 2001).
       
 {\bf  64.} X-ray emitter (Bergh\"{o}fer \& Christian 2002, Park \& Sung 2002); probable spectroscopic 
            binary (Millward \& Walker 1985).
       
 {\bf  65.} X-ray emitter (Bergh\"{o}fer \& Christian 2002, Park \& Sung 2002); variable radial velocity (Humphreys 1978);
            spectroscopic binary (Millward \& Walker 1985).
       
 {\bf  66.} Chemical peculiarity near 5200 \AA~ flux (Hensberge et al. 1998).
       
 {\bf  67.} Weak NIII 4634-4640-4642 emission (Walborn 1972); magnitude variable (Johnson 1962); PMS object (Ogura \& Ishida 1981);
            X-ray emitter (Bergh\"{o}fer \& Christian 2002).
       
 {\bf  68.} X-ray emitter (Bergh\"{o}fer \& Christian 2002, Park \& Sung 2002).
       
 {\bf  69.} Polarisation $<$ 0.2 (Hoag \& Smith 1959); X-ray emitter (Bergh\"{o}fer \& Christian 2002);
            $P_{\mu} = 0.0$ (Sabogal-Mart\'{i}nez et al. 2001).
       
 {\bf  72.} X-ray emitter (Bergh\"{o}fer \& Christian 2002).

 {\bf  75.} X-ray emitter (Park \& Sung 2002).
       
 {\bf  78.} X-ray emitter (Bergh\"{o}fer \& Christian 2002); B4Vne (P\'{e}rez 1988); foreground star (Guseva 1985).
       
 {\bf  79.} Member (Guseva 1985).
       
 {\bf  80.} V578 Mon; eclipsing variable (Kholopov et al. 1998); two line spectroscopic binary (Morgan et al. 1965); 
            X-ray variable probably due to binarity (Bergh"{o}fer \& Christian 2002); eclipsing binary with two early type B 
            stars, 2.4 days orbit (Hensberge et al. 2000).
       
 {\bf  81.} NIII multiplet 4634-4640-4642\AA~ in weak emission (Walborn 1972); X-ray emitter (Bergh\"{o}fer \& Christian 2002).
       
 {\bf  82.} Hydrogen lines very strong, may be an exception to the relation between line intensity and luminosity 
            (Morgan et al. 1965); X-ray emitter (Bergh\"{o}fer \& Christian 2002).

 {\bf  86.} $P_{\mu} = 0.20$ (Sabogal-Mart\'{i}nez et al. 2001).

 {\bf  87.} $P_{\mu} = 0.55$ (Sabogal-Mart\'{i}nez et al. 2001).
       
 {\bf  88.} H$_{\alpha}$ emission (Young 1978).
       
 {\bf  90.} Probable field and variable star (Sagar \& Joshi 1983); either a carbon or a background
            star (Mendoza \& G\"{o}mez 1980); non-member on the basis of radial velocity V$_{r}$ = +70 km/s
            (Zappala 1972); dM foreground star (Walker 1956).
       
 {\bf  93.} Probable field star (Sagar \& Joshi 1983).
       
 {\bf  95.} Broad interstellar NaI 5890-5896 (P\'{e}rez 1988); probably a member star, highly reddened (Walker 1956);
            strong interstellar circular polarization (McMillan 1977); probable field star (Sagar \& Joshi 1983).
       
 {\bf  97.} LkH$_{\alpha}$ 25; classified as Herbig Ae/Be by Herbig (1960); suspected X-ray source (Simon et al. 1985);
            abnormal extinction due to an edge-on circumstellar disk (Rydgren \& Vrba 1987); B8pe, shell, hot member 
            (Herbig \& Rao 1972); H$_{\alpha}$ emission (Herbig 1954, Young 1978); variable star (Warner et al. 1956);
            edge-on circumstellar disk (Park et al. 2002). 
       
 {\bf  98.} H$_{\alpha}$ emission (Young 1978); variable star (Warner et al. 1977); variable radial velocity, 
            spectroscopic binary (Mendoza \& G\"{o}mez 1980); PMS star (Sung et al. 1997); variable radial velocity (Walker 1956). 
       
 {\bf 100.} Variable star (Sagar \& Joshi 1983); H$_{\alpha}$ in filled emission (P\'{e}rez 1988);  LiI 6707 \AA~ present 
           (Zappala 1972).
     
 {\bf 101.} LkH$_{\alpha}$ 53; H$_{\beta}$ shows a weak emission, no H$_{\alpha}$ detected by Marcy (1980); SpT. $<$ K2 (Rydgren 
            1979); T Tauri star (Herbig \& Rao 1972); H$_{\alpha}$ emission (Herbig 1954, Young 1978); variable star (Walker 1956). 
     
 {\bf 102.} Variable star (Warner et al. 1977). 
     
 {\bf 103.} Probable field star (Sagar \& Joshi 1983).
     
 {\bf 105.} Probable field star (Sagar \& Joshi 1983).
     
 {\bf 106.} H$_{\alpha}$ emission (Young 1978); probable field star (Sagar \& Joshi 1983); no emission (P\'{e}rez 1988); 
            non-member on the basis of its radial velocity V$_{r}$ = +73 km/s (Zappala 1972). 
     
 {\bf 111.} B0 V (Boggs \& B\"{o}hm-Vitense 1989).
     
 {\bf 114.} F0-G5 (Johnson \& Borgman 1963).
       
 {\bf 118.} B1 V (Boggs \& B\"{o}hm-Vitense 1989).
       
 {\bf 119.} B1.5 V (Boggs \& B\"{o}hm-Vitense 1989); triple system (Torres 1987).
       
 {\bf 120.} A0 (Walker 1957).
       
 {\bf 121.} B0.5 V, binary (Boggs \& B\"{o}hm-Vitense 1989).
       
 {\bf 122.} LkH$_{\alpha}$112; probable Herbig Ae/Be star, intermediate mass PMS star (Boesono et al. 1987).
       
 {\bf 123.} B1 Vp (Torres 1987).
       
 {\bf 124.} Radial velocity probably variable, perhaps double line in the spectrum (Walker 1957).
       
 {\bf 127.} HeI star, strong He I absorption lines (Hiltner et al. 1965).
       
 {\bf 129.} B1 V, binary system with a secondary of the same type as of primary (Boggs \& B\"{o}hm-Vitense 1989).
       
 {\bf 132.} B1 V (Boggs \& B\"{o}hm-Vitense 1989); radial velocity variable, double line in the spectrum (Walker 1957).
       
 {\bf 133.} LSS 4621.
       
 {\bf 135.} B2.5 V (Boggs \& B\"{o}hm-Vitense 1989); double line spectroscopic binary (Hiltner et al. 1965). 
       
 {\bf 136.} B1 V (Boggs \& B\"{o}hm-Vitense 1989).
       
 {\bf 138.} B2.5 V (Boggs \& B\"{o}hm-Vitense 1989); a double system, the combined magnitude and color are given, 
           the $V$ and $(B-V)$ of the brighter component are 9.48 and 0.09 respectively (Walker 1957).
     
 {\bf 141.} B3 V (Boggs \& B\"{o}hm-Vitense 1989).
       
 {\bf 142.} Double-lined spectroscopic binary, consisting of two O6.5 V star with a period of 6.14 days (Morrison \& Conti 1978).
       
 {\bf 145.} K4 III (Buscombe 1984).
       
 {\bf 146.} B2.5 V (Buscombe 1984).

 {\bf 149.} Late type giant (Parthasarathy 1974) 

 {\bf 150.} Probable member and a close binary with a mid-B type and a cool star (van den Ancker et al. 1997) 
       
 {\bf 151.} Late type giant (Parthasarathy 1974); Variable star (Sagar \& Joshi 1978). 

 {\bf 152.} B1.5 V (Hiltner \& Morgan 1969). 
       
 {\bf 153.} V$_{sini}$ = 86 km/s (Killian-Montenbruck et al. 1994); O9 V (Walker 1961); B0.5 V (Hiltner \& Morgan 1969); 
            intrinsic polarisation (Orsatti et al. 2000).
       
 {\bf 155.} Long period spectroscopic binary (Bosch et al. 1999); Visual Binary (Duch\^{e}ne et al. 2001);
            Intrinsic polarisation (Orsatti et al. 2000); O5.5 V((f)) (Hillenbrand et al. 1993).
       
 {\bf 156.} Long period spectroscopic binary (Bosch et al. 1999); radial velocity variable (Walker 1961, Conti 
            and Frost 1977); V$_{sini}$ = 79 km/s (Penny 1996); polarimetric variable (Orsatti et al. 2000);
            O7 V((f)) (Hillenbrand et al. 1993); $P_{\mu} = 0.21$ (Belikov et al. 1999); member (van Schewick 1962).

 {\bf 157.} Blue straggler (Mermilliod 1995); large dispersion in radial velocity  (Bosch et al. 1999);
            variable radial velocity (Conti \& Frost 1977); V$_{sini}$ = 109 km/s (Penny 1996);
            no Intrinsic polarisation (Orsatti et al. 2000); visual binary (Duch\^{e}ne et al. 2001);
            O5 V((f*)) (Hillenbrand et al. 1993); $P_{\mu} = 0.17$ (Belikov et al. 1999); member (van Schewick 1962).
       
 {\bf 158.} No Intrinsic polarisation (Orsatti et al. 2000); visual binary (Duch\^{e}ne et al. 2001); 
            Be (Schild \& Romanishin 1976); B1 V (Hiltner \& Morgan 1969).
       
 {\bf 159.} Visual binary (Duch\^{e}ne et al. 2001).
       
 {\bf 160.} Be: (Walker 1961); no intrinsic polarisation (Orsatti et al. 2000); group III Herbig Ae/Be (Hillenbrand et al 1993).
       
 {\bf 161.} V$_{r}$=15 km/s (Bosch et al. 1999); V$_{r}$=6 km/s (Walker 1961); O7 Ib(f) (Walborn 1982); 
            O7 IIb(f) (Hillenbrand et al. 1993); No intrinsic polarisation (Orsatti et al. 2000).
       
 {\bf 162.} B1 V (Th\'{e} et al. 1990); V$_{r}$ = $-$49 km/s (Brown et al. 1986); V$_{r}$ = 3.8 km/s (Walker 1961); 
            intrinsic polarisation (Orsatti et al. 2000); sharp lines (Hillenbrand et al. 1993); 
            V$_{sini}$ = 38 km/s (Killian-Montenbruck et al. 1994); visual binary (Duch\^{e}ne et al. 2001).
       
 {\bf 163.} Intrinsic polarisation (Orsatti et al. 2000).
       
 {\bf 164.} O9.5 V (Hillenbrand et al. 1993); V$_{r}$ = 13 km/s (Bosch et al. 1999); radial velocity variable (Walker 1961); 
            broad lines (van Schewick 1962); fast emission rotation variable line star, 
            no intrinsic polarisation (Orsatti et al. 2000).
       
 {\bf 165.} $P_{\mu} = 0.40$ (Tucholke et al. 1986); $P_{\mu} = 0.87$ (Belikov et al. 1999); member (van Schewick 1962); 
            Intrinsic polarisation (Orsatti et al. 2000).
       
 {\bf 166.} Intrinsic polarisation (Orsatti et al. 2000); B3-4e: (de Winter et al 1997).
       
 {\bf 167.} Probable binary or multiple systems, V$_{r}$ = 19 km/s (Bosch et al. 1999);
           V$_{r}$ = 1.7 km/s, double lines (Walker 1961); no intrinsic polarisation (Orsatti et al. 2000).
       
 {\bf 168.} No intrinsic polarisation (Orsatti et al. 2000).
       
 {\bf 169.} O9.5 V (Hiltner \& Johnson 1956, Hiltner \& Morgan 1969, Hillenbrand et al. 1993); probable binary or multiple 
            systems, V$_{r}$ = 4 km/s (Bosch et al. 1999); variable radial velocity (Walker 1961); 
            V$_{sini}$ = 29 km/s (Killian-Montenbruck et al. 1994).
       
 {\bf 170.} O8.5 V (Hillenbrand et al. 1993); V$_{r}$ =18 km/s (Bosch et al. 1999); V$_{r}$ = 12.1 km/s (Walker 1961); 
            intrinsic polarisation (Orsatti et al. 2000); $P_{\mu} = 0.25$ (Belikov et al. 1999); member (van Schewick 1962). 
       
 {\bf 171.} O9.5 I (Hillenbrand et al. 1993); short period single lined spectroscopic binary (Bosch et al. 1999).
       
 {\bf 172.} B0.5 V (Th\'{e} et al. 1990); V$_{r}$ = $-$21 km/s (Brown et al. 1986).

 {\bf 173.} B1.5 V (Hiltner \& Morgan 1969).

 {\bf 174.} Member (de Winter et al. 1997).
       
 {\bf 175.} Suspected circumstellar shell, e-vector deviate significantly (Orsatti et al. 2000); 
            group III Herbig Ae/Be (Hillenbrand et al 1993).

 {\bf 176.} cB4 (Wilson 1953); $P_{\mu} = 0.43$ (Tucholke et al. 1986); $P_{\mu} = 0.93$ (Belikov et al. 1999); 
            probable member (de Winter et al. 1997).

 {\bf 177.} $P_{\mu} = 0.01$ (Tucholke et al. 1986); $P_{\mu} = 0.08$ (Belikov et al. 1999). 

 {\bf 180.} $P_{\mu} = 0.80$ (Belikov et al. 1999). 
       
 {\bf 188.} B0.5 IV (Shi \& Hu 1999); B1 III (Parthasarathy et al. 2000); 
            probable post-AGB star with cold dust $\sim$ 127 K (Gauba et al. 2003). 
       
 {\bf 190.} O8 V (Shi \& Hu 1999).
       
 {\bf 192.} B1.5 V (Shi \& Hu 1999).
       
 {\bf 194.} probable variable, anomalous position in the HR diagram (Sagar \& Joshi 1981).
       
 {\bf 195.} B0.5 Ib (Shi \& Hu 1999, Massey et al. 1995); $c_{1}$ = +0.17, B0.5 Ia (Hiltner 1956); 
            probable variable (Sagar \& Joshi 1981).
       
 {\bf 197.} B1 III (Shi \& Hu 1999); polarisation = 0.059 (Hiltner 1956); O9.5 Ia (Massey et al. 1995); 
            B0 IV (Hoag \& Applenquist 1965); B1 (Walker \& Hodge 1968); H$\alpha$ emission (Kohoutek \& Wehmeyer 1997); 
            No H$\alpha$ emission (Pigulski et al. 2000).
       
 {\bf 198.} B1 III (Shi \& Hu 1999, Massey et al. 1995); polarisation = 0.066 (Hiltner 1956); Variable star (Sagar \& Joshi 1981)
       
 {\bf 199.} K4 V (Shi \& Hu 1999); anomalous position in the HR diagram (Sagar \& Joshi 1981).
       
 {\bf 201.} B1 V (Shi \& Hu 1999).
       
 {\bf 202.} O7 V((f)) (Massey et al. 1995, Shi \& Hu 1999).
       
 {\bf 203.} B1 V (Shi \& Hu 1999); B0.5 III (Massey et al. 1995).
       
 {\bf 204.} B0.5 V (Shi \& Hu 1999, Massey et al. 1995).
       
 {\bf 205.} B1 V (Shi \& Hu 1999); B1.5 V (Massey et al. 1995); 
            EA-type eclipsing binary, $P > 0.93 d$, A-type secondary (Pigulski et al. 2000).
       
 {\bf 207.} B0 V (Shi \& Hu 1999); B1 V (Massey et al. 1995).

 {\bf 209.} Non-member on the basis of spectral type (Pigulaski et al. 2000); A3 (Egret et al. 1991); 
            anomalous position in the HR diagram (Sagar \& Joshi 1981).
       
 {\bf 210.} B2 III (Shi \& Hu 1999); B1 V (Massey et al. 1995).
     
 {\bf 211.} F8 Ib (Shi \& and Hu 1999), anomalous position in the HR diagram (Sagar \& Joshi 1981).


\begin{table*}
\caption{List of program stars with their ISOGAL, MSX and IRAS counterparts. The ISOGAL fluxes at 7 and 15 $\mu$m are
given in Jansky (Jy) along with the MSX fluxes at bands A ($\sim 8~\mu$m) and D ($\sim 15~\mu$m) and IRAS fluxes
at 12 and 25 $\mu$m are given in Jansky along with the MSX bands C ($\sim 12~\mu$m), D($\sim 21~\mu$m) so that the common
sources can be compared easily. The IRAS fluxes with letter "L" indicate an upper limit. The last column provides the spectral 
index (SpI), $s$ ($\lambda F_{\lambda}$ $\sim$ $\lambda^{s}$), in the MIR region (2.2 to 25 $\mu$m)}

\scriptsize
\begin{tabular}{rlrrrrrrc} \hline
 ID& ISOGAL-PJ../IRAS..&     F7&    F12&   F15&  F25&   F60& F100&SpI\\
   &          MSX5C-G..&      A&      C&     D&    E&      &     & ($s$)\\ 
   &             (Name)&   (Jy)&   (Jy)&  (Jy)& (Jy)&  (Jy)& (Jy)&\\ \hline

  1&        05304-0435&      -&      0.48L&       -&      4.25 &    36.45 &    32.03 & $ -0.42\pm0.74$\\
  5&        05318-0506&      -&      0.46 &       -&      0.55 &     4.79L&    46.55L& $ -0.79\pm0.01$\\
 13&        05327-0529&      -&     33.86 &       -&    366.60 &  4798.00 &    24.48 & $  1.44\pm0.25$\\
 27&        05329-0512&       &     22.77 &       -&     54.22L&   630.70L&    22.69L& $  0.52\pm0.10$\\
 30&        05330-0517&      -&     82.07 &       -&    560.70 &  1613.00L&  1600.00L& $  0.85\pm0.24$\\
 36&        05333-0543&      -&      8.52 &       -&      8.07 &   290.20L&    31.41L& $  0.14\pm0.37$\\
 41&        05341-0530&      -&      3.68 &       -&      7.64 &   122.30L&   648.20L& $  0.14\pm0.05$\\
 54&        06288+0456&      -&      1.24 &       -&      1.23 &     2.22L&    18.95L& $  0.72\pm0.53$\\
 57&        06288+0452&      -&      0.71 &       -&      0.63L&     2.28L&    11.42L& $ -0.55\pm0.13$\\
   & G206.3519-02.1984&   0.42&         - &       -&         - &        - &        - &                \\
 60&        06291+0511&      -&      0.27L&       -&      2.21 &    17.54L&    16.67L& $  0.65\pm0.37$\\
 62&        06290+0508&      -&      0.98 &       -&      1.55 &     2.16L&    19.82L& $  1.24\pm0.50$\\
 65&        06289+0504&      -&      0.30L&       -&      0.98 &     2.18L&    19.46L& $ -0.96\pm0.48$\\
 71&        06291+0456&      -&      0.43L&       -&      0.88 &     2.47L&    20.44L& $  0.36\pm0.12$\\
 87&        06311+0515&      -&      2.44 &       -&      0.71 &     2.55 &    24.10L& $ -2.55\pm0.13$\\
   & G206.2715-01.5439&   3.40&      1.74 &    1.74&         - &        - &        - &                \\
 90&        06372+0936&      -&      0.27L&       -&      0.28L&     2.53 &    15.27L& $ -2.34\pm0.48$\\
   & G203.1040+01.8224&   0.37&         - &       -&         - &        - &        - &                \\
 95& G202.9883+02.0728&   0.22&         - &       -&         - &        - &        - & $ -2.37       $\\
 97&        06379+0950&      -&      1.82 &       -&      3.77 &     1.57L&   619.80L& $  0.30\pm0.17$\\
   & G202.9947+02.1040&   1.25&         - &    1.64&      5.14 &        - &        - &                \\
 98& G202.9478+02.1479&   0.23&         - &       -&         - &        - &        - & $ -1.20       $\\
101&        06382+0939&      -&      7.06 &       -&     16.25 &   212.80 &   499.30 & $  1.54\pm0.43$\\
109& G005.9186-00.9944&   0.83&         - &       -&         - &        - &        - & $  0.77       $\\
110&        18008-2421&      -&     21.41L&       -&     32.91 &  7755.00L&  9036.00L& $ -0.09\pm0.10$\\
111&        18008-2419&      -&     17.63 &       -&     78.79 &   347.50L&  9036.00L& $  0.50\pm0.47$\\
   & G006.0569-01.1969&   1.20&      5.25 &    6.66&      9.83 &        - &        - &                \\
119& G006.1532-01.2200&   0.20&         - &    1.22&      2.89 &        - &        - & $  0.12\pm0.41$\\
121&        18012-2421&      -&      3.98 &       -&     25.68 &    26.20L&  9036.00L& $  0.80\pm0.23$\\
126&        18013-2423&      -&      6.85 &       -&     51.59 &   429.60L&  1184.00L& $ -0.10\pm0.58$\\
   & G006.0495-01.3286&   1.17&      2.87 &    4.34&     17.20 &        - &        - &                \\
129& G006.0755-01.3232&   0.43&      1.76 &    3.06&     10.77 &        - &        - & $  0.24\pm0.42$\\
132& G006.0617-01.3519&      -&         - &       -&      6.34 &        - &        - & $  0.44       $\\
134&        18015-2409&      -&      2.99 &       -&      3.46 &   513.50L&  1307.00L& $ -0.08\pm0.22$\\
146&        18006-2422&      -&    167.70 &       -&   1842.00 &  7755.00 &  9036.00 & $  2.80\pm0.16$\\
149& G006.3342-01.5543&   4.03&      2.06 &    1.55&         - &        - &        - & $ -2.78\pm0.15$\\
151& G006.4488-01.6002&   2.22&         - &    1.15&         - &        - &        - & $ -2.79\pm0.40$\\
152&  J181826.2-135005&   0.04&         - &       -&         - &        - &        - & $ -2.77       $\\
153&  J181830.0-134957&   0.02&         - &       -&         - &        - &        - & $ -3.56       $\\
154&        18156-1343&      -&     22.78L&       -&     55.62 &  1684.00 &  5317.00 & $  0.59\pm0.12$\\
155&  J181832.7-134512&   0.24&         - &    0.52&         - &        - &        - & $ -1.70\pm0.74$\\
156&  J181836.1-134736&   0.08&         - &       -&         - &        - &        - & $ -3.24       $\\
157&  J181836.4-134802&   0.24&         - &       -&         - &        - &        - & $ -2.88       $\\
160&  J181838.8-134644&   0.22&         - &       -&         - &        - &        - & $ -2.30       $\\
   & G016.9605+00.8388&   0.17&         - &       -&         - &        - &        - &                \\
161&  J181840.1-134518&   0.17&         - &       -&         - &        - &        - & $ -3.19       $\\
166&        18159-1346&      -&      5.97 &       -&     19.30 &   169.90L&  5317.00L& $  1.46\pm0.27$\\
167&  J181845.9-134631&      -&         - &    0.52&         - &        - &        - & $ -1.09       $\\
169&  J181853.2-134939&      -&         - &    0.38&         - &        - &        - & $ -0.77       $\\ 
170&  J181856.2-134830&   0.11&         - &       -&         - &        - &        - & $ -2.71       $\\ 
171&  J181858.7-135928&   0.07&         - &       -&         - &        - &        - & $ -0.40\pm1.50$\\
   & G016.8149+00.6699&      -&         - &       -&      3.38 &        - &        - &                \\
172& G016.8927+00.6807&      -&         - &    1.13&      5.06 &        - &        - & $  0.06\pm0.38$\\
173&  J181904.9-134819&   0.22&         - &    1.26&         - &        - &        - & $  0.61\pm0.34$\\
   & G016.9870+00.7339&   0.34&      1.88 &    2.67&      7.89 &        - &        - &                \\
175&  J181910.8-135649&   0.37&         - &    0.37&         - &        - &        - & $ -0.47\pm0.59$\\
   & G016.8754+00.6455&   0.41&         - &       -&      3.32 &        - &        - &                \\
176&  J181751.1-135055&   0.29&         - &    0.08&         - &        - &        - & $ -3.00\pm0.26$\\
   & G016.8083+00.9765&   0.17&         - &       -&         - &        - &        - &                \\
177&  J181806.4-133625&   0.13&         - &       -&         - &        - &        - & $ -3.14       $\\
178&  J181906.5-135745&   0.20&         - &       -&         - &        - &        - & $ -2.64       $\\
   & G016.8516+00.6530&   0.27&         - &       -&         - &        - &        - &                \\
179&  J181920.1-135420&   0.05&         - &    0.37&         - &        - &        - & $ -0.62\pm0.96$\\
188&        19399+2312&      -&      1.13 &        &      2.27 &    23.31L&    80.20L& $ -0.33\pm0.09$\\
190&  J194211.2+232604&   0.10&         - &    0.02&         - &        - &        - & $ -2.95\pm0.07$\\
191&  J194222.4+232305&   0.01&         - &       -&         - &        - &        - & $ -2.84       $\\
192&  J194227.1+232038&   0.01&         - &       -&         - &        - &        - & $ -2.67       $\\
195&  J194249.2+232754&   0.16&         - &    0.03&         - &        - &        - & $ -2.99\pm0.09$\\
196&  J194254.8+231855&      -&         - &    0.01&         - &        - &        - & $ -1.62       $\\
197&  J194306.5+231618&   0.03&         - &       -&         - &        - &        - & $ -2.93       $\\
198&  J194309.5+232621&   0.03&         - &    0.01&         - &        - &        - & $ -2.72\pm0.12$\\
199&        19410+2322&      -&      0.59 &       -&      0.93 &     2.26L&    43.78L& $ -1.65\pm0.39$\\
200&  J194310.7+231751&   0.08&         - &    0.02&         - &        - &        - & $ -1.36\pm0.63$\\
203&  J194312.2+231634&   0.02&         - &    0.05&         - &        - &        - & $ -1.64\pm0.79$\\
204&  J194313.3+231912&   0.02&         - &    0.02&         - &        - &        - & $ -1.89\pm0.24$\\
   & G059.4253-00.1490&   0.10&         - &       -&         - &        - &        - &                \\
207&  J194316.5+231915&      -&         - &    0.04&         - &        - &        - & $ -1.30       $\\
210& G059.3539-00.2348&   0.51&      1.35 &       -&         - &        - &        - & $  0.16\pm0.21$\\
\hline
\end{tabular}
\end{table*}


\begin{table*}                                                       
\caption{Adopted spectral types, color excesses and $R_{V}$ of all program stars. Asterisked ones 
are either the new or modified spectral types from the present study. The columns 11 and 12 provide equivalent widths (EW)
for the CaII triplet ($\sim$ 8500 \AA) and NaI doublet (5890/5896 \AA). The "Grp" column provides the grouping sequence 
according to their reddening behavior (see Sect. 4) while the finally adopted membership is given in the last column
in which "M", "NM" and "PM" stands for member, non-member and probable member. (see Sect. 6).}                                         

\scriptsize                                                                    
\begin{tabular}{clrrrrrrrrcccc} \hline                                                               
ID& SpT& {\it E(U-V)}& {\it E(B-V)}& {\it E(V-R)}& {\it E(V-I)}& {\it E(V-J)}& {\it E(V-H)}& {\it   
									   E(V-K)}& $R_{V}$& EW(\AA)& EW(\AA)&Grp&Memb\\                   
&&&&&&&&&&CaIIT&NaID&&\\ \hline                                                                      
  1&  B2.5 IV     &    0.33&   0.23&   0.20&    -  &   0.69&   0.71&   0.79&   3.78&     &    0.45&2&M\\
  2&  K2 IV       &    1.21&   0.57&    -  &    -  &   1.36&   1.62&   1.69&   3.26& 2.87&    1.11&1&M\\
  3&  B8 V        &    1.62&   0.80&   0.53&   1.07&   1.83&   2.11&   2.29&   3.15&     &    0.35&1&M\\
  4&  G8 V        &    0.32&   0.20&    -  &    -  &   0.41&   0.52&   0.56&    -  & 3.08&    0.93&1&M\\
  5&  F8 Ve       &    0.61&   0.33&   0.17&   0.39&   1.07&   1.55&   2.03&   4.42&$-$0.26&  0.73&3&M\\
  6&  G9 IV-Ve    &    0.05&   0.36&   0.14&   0.31&   0.95&   1.27&   1.45&   3.60& 1.80&    1.06&2&M\\
  7&  B8          &    1.47&   0.82&   0.58&   1.15&   1.90&   2.18&   2.29&   3.07&     &    0.20&1&M\\
  8&  A2 IIIe*    &    0.90&   0.52&   0.41&   0.77&   1.26&   2.01&   2.75&   3.31&     &    0.45&3&M\\
  9&  A1          &    0.82&   0.44&   0.26&   0.49&   0.85&   0.91&   1.02&   2.55&     &    0.18&3&M\\
 10&  G0 V        &    0.21&   0.21&   0.12&   0.33&   0.79&   0.97&   0.97&   5.08& 2.75&    0.61&2&M\\
 11&  A0          &    0.86&   0.58&   0.42&   0.68&   1.31&   1.54&   1.70&   3.22&     &    0.29&1&M\\
 12&  F8          &    1.46&   0.74&   0.59&   1.05&   2.11&   2.51&   2.84&   3.89& 2.65&    0.49&2&M\\
 13&  B2 V        &    0.53&   0.33&   0.28&   0.65&   1.26&   1.45&   1.68&   5.60&     &    0.43&2&M\\
 14&  B3 V        &    0.83&   0.56&   0.38&   0.88&   1.84&   2.27&   2.48&   4.48&     &    0.41&2&M\\
 15&  K0 Ve       &    0.73&   0.55&   0.30&   1.00&   1.94&   2.25&   2.42&   4.84& 3.54&    0.64&2&M\\
 16&  A2          &    0.39&   0.28&    -  &    -  &   0.71&   0.85&   0.96&   3.77&     &    0.40&2&M\\
 17&  B0 V        &    1.42&   0.53&    -  &    -  &   2.77&   3.19&   3.22&   6.68&     &    0.25&2&M\\
 18&  B0.5 V      &    0.43&   0.31&   0.34&   0.70&   2.58&   2.88&   2.80&   9.94&     &    0.38&3&M\\
 19&  A0          &    0.14&   0.22&    -  &    -  &   1.06&   1.27&   1.44&   7.20&     &    0.19&2&M\\
 20&  A1 V        &    0.46&   0.32&   0.27&   0.65&   1.27&   1.51&   1.65&   5.67&     &    0.63&2&M\\
 21&  B0.5 V      &    0.55&   0.36&   0.30&   0.66&   1.26&   1.54&   1.74&   5.32&     &    0.27&2&M\\
 22&  O7 e        &    0.52&   0.30&   0.28&   0.66&   1.35&   1.69&   1.89&   6.14&     &    0.30&2&M\\
 23&  G0 IV-IIIe  &    0.94&   0.47&   0.27&   0.65&   1.35&   1.70&   1.84&   2.92& 1.95&    0.87&2&M\\
 24&  B2 V        &    0.94&   0.56&   0.23&   0.83&   2.08&   2.64&   2.93&   5.07&     &    0.33&2&M\\
 25&  O9.5 V      &    0.36&   0.22&   0.16&    -  &   0.84&   0.96&   1.11&   5.55&     &    0.34&2&M\\
 26&  G8 V        &    0.20&   0.16&   0.16&   0.57&   1.02&   1.31&   1.34&    -  & 2.74&    1.26&2&M\\
 27&  G3 IV-Ve    &    1.91&   1.21&   0.43&   1.24&   2.07&   2.86&   3.71&   2.33&$-$1.54&  0.44&3&M\\
 28&  G3 IV-V     &    0.34&   0.31&   0.25&   0.46&   0.84&   1.07&   1.19&   4.22& 2.61&    0.64&2&M\\
 29&  G8 Ve       &    1.39&   0.41&   0.13&   0.61&   1.06&   1.39&   1.48&   3.53& 2.32&    0.88&2&M\\
 30&  B0.5 V      &    0.85&   0.54&   0.40&   0.85&   1.72&   1.99&   2.15&   4.38&     &    0.60&2&M\\
 31&  G9 IV-Ve    &    0.71&   0.37&   0.12&   0.65&   1.12&   1.47&   1.62&   4.13& 2.56&    1.22&2&M\\
 32&  K2 IVe      &    0.33&   0.25&    -  &   0.41&   1.16&   1.41&   1.76&   6.33& 1.64&    0.95&3&M\\
 33&  K2 Ve       &    0.48&   0.28&   0.12&   0.72&   1.01&   1.22&   1.51&   4.92& 1.33&    1.13&3&M\\
 34&  K1 Ve       &    0.71&   0.33&   0.22&   0.49&   1.21&   1.45&   1.53&   5.10& 1.90&    1.05&2&M\\
 35&  A3 IIIe*    &    0.66&   0.33&   0.03&   0.30&   1.58&   2.57&   3.65&   6.53&     &    0.42&3&M\\
 36&  B4 V        &    1.28&   0.81&   0.67&   1.40&   2.84&   3.44&   3.77&   4.48&     &    0.41&2&M\\
 37&  A0          &    0.15&   0.20&   0.17&   0.36&   0.71&   0.75&   0.88&   4.84&     &    0.17&2&M\\
 38&  F6 IV       &    0.46&   0.26&   0.12&   0.40&   0.80&   0.93&   0.99&   4.19& 3.12&    0.30&2&M\\
 39&  G6*         &    0.41&   0.11&   0.10&   0.40&   0.33&   0.40&   0.42&    -  & 2.70&    0.68&3&M\\
 40&  A7          &    0.03&   0.06&    -  &    -  &   0.25&   0.24&   0.29&    -  &     &    0.22&2&M\\
 41&  B5 V        &    1.41&   0.87&   0.73&   1.39&   2.40&   2.74&   2.91&   3.68&     &    0.11&2&M\\
 42&  F6          &    2.13&   1.12&   0.47&    -  &   2.65&   3.16&   3.36&   3.30& 3.03&    0.50&1&M\\
 43&  A6*         &    0.32&   0.20&   0.16&   0.26&   0.66&   0.75&   0.78&   4.29&     &    0.29&2&M\\
 44&  A0*         &    0.92&   0.70&   0.42&   0.90&   1.53&   1.75&   1.82&   2.86&     &    0.23&3&M\\
 45&  G0 III      &    0.70&   0.41&    -  &    -  &   1.10&   1.29&   1.39&   3.73& 2.50&    0.50&2&M\\
 46&  F9 V        &    0.19&   0.19&    -  &    -  &   0.47&   0.52&   0.53&    -  & 2.20&    0.41&1&M\\
 47&  G6*         &    1.29&   0.53&    -  &    -  &   1.71&   2.06&   2.23&   4.63& 2.76&    0.75&2&NM\\
 48&  F6 V        &   $-$0.04&  $-$0.02&    -  &    -  &   0.24&   0.30&   0.28&    -  & 1.65&0.37&3&NM\\
 49&  A1 IV       &    0.48&   0.21&    -  &    -  &   0.52&   0.50&   0.61&   3.20&     &    0.37&1&M\\
 50&  B9 V        &    0.98&   0.58&    -  &    -  &   1.40&   1.51&   1.65&   3.13&     &    0.37&1&M\\
 51&  F1 V        &    0.31&   0.14&    -  &    -  &   0.43&   0.40&   0.43&    -  & 2.35&    0.44&3&NM\\
 52&  B2.5 II-III &    1.62&   0.91&    -  &   1.38&   2.45&   2.80&   3.03&   3.66&     &    0.59&2&M\\
 53&  K5          &    1.20&   0.51&    -  &   0.63&   1.23&   1.45&   1.61&   3.47& 3.23&    0.96&3&NM\\
 54&  F4*         &    0.12&   0.05&    -  &   0.19&   0.35&   0.42&   0.47&    -  & 2.47&    0.38&3&NM\\
 55&  A7 V        &    0.28&   0.13&    -  &    -  &   0.36&   0.35&   0.37&    -  &     &    0.42&3&NM\\
 56&  B2 V        &    0.66&   0.40&   0.31&   0.58&   0.97&   1.11&   1.16&   3.19&     &    0.46&1&M\\
 57&  B0.5 V      &    0.78&   0.42&   0.40&   0.74&   1.13&   1.30&   1.40&   3.67&     &    0.48&3&M\\
 58&  O8 V        &    0.87&   0.46&   0.31&   0.67&   1.10&   1.27&   1.34&   3.20&     &    0.57&1&M\\
 59&  F4 V        &    0.12&   0.07&    -  &    -  &  -0.07&  -0.07&  -0.06&    -  & 2.32&    0.42&3&NM\\
 60&  F7          &    0.73&   0.34&    -  &    -  &   1.19&   1.38&   1.45&   4.69& 2.06&    0.49&2&PM\\
 61&  B8 V        &    0.53&   0.33&    -  &   0.42&   0.81&   0.90&   0.92&   3.07&     &    0.37&2&M\\
 62&  A5          &    0.53&   0.34&    -  &   0.52&   0.96&   1.03&   1.15&   3.72&     &    0.54&2&M\\
 63&  G7 III      &    0.24&   0.15&    -  &   0.23&   0.41&   0.44&   0.53&    -  & 3.00&    0.67&3&NM\\
 64&  O8.5 V      &    1.02&   0.49&   0.42&   0.75&   1.19&   1.33&   1.39&   3.12&     &    0.48&3&M\\
 65&  B0 V        &    0.71&   0.38&   0.34&   0.63&   1.10&   1.24&   1.27&   3.68&     &    0.46&2&M\\
 66&  B8 V        &    0.67&   0.43&    -  &   0.58&   1.09&   1.23&   1.34&   3.43&     &    0.32&2&M\\
 67&  O5 V        &    0.81&   0.48&   0.27&   0.58&   1.01&   1.15&   1.25&   2.86&     &    0.50&1&M\\
 68&  B4 V        &    0.74&   0.45&   0.53&   1.34&   1.15&   1.32&   1.47&   3.59&     &    0.46&2&M\\
 69&  A2 V        &    0.07&   0.03&   0.17&   0.17&   0.15&   0.10&   0.17&    -  &     &    0.24&3&NM\\
 70&  B1 V        &    0.67&   0.40&   0.26&   0.55&   1.01&   1.12&   1.21&   3.33&     &    0.58&2&M\\
\hline                                                                                              
\end{tabular}                                                                                       
\end{table*}                                                                                        
                                                                                                    
\begin{table*}                                                                                      
{{\bf Table 6.} Continued.}                                                                         
                                                                                                    
\scriptsize                                                                                         
\begin{tabular}{clrrrrrrrrcccc} \hline                                                               
ID& SpT& {\it E(U-V)}& {\it E(B-V)}& {\it E(V-R)}& {\it E(V-I)}& {\it E(V-J)}& {\it E(V-H)}& {\it   
									   E(V-K)}& $R_{V}$& EW(\AA)& EW(\AA)&Grp&Memb\\                   
&&&&&&&&&&CaIIT&NaID&&\\ \hline                                                                      
 71&  B2.5 V      &    0.77&   0.48&    -  &   0.60&   1.15&   1.29&   1.41&   3.23&     &    0.50&2&M\\
 72&  B9 V        &    0.73&   0.47&   0.88&   1.54&   1.18&   1.33&   1.49&   3.49&     &    0.35&2&M\\
 73&  B3 V        &    0.48&   0.39&   0.23&   0.48&   0.98&   1.15&   1.20&   3.38&     &    0.57&2&M\\
 74&  B2.5 V      &    0.77&   0.49&   0.36&   0.70&   1.22&   1.36&   1.48&   3.32&     &    0.44&2&M\\
 75&  O9 V        &    0.91&   0.45&   0.31&   0.66&   1.13&   1.27&   1.39&   3.40&     &    0.45&1&M\\
 76&  B2 V        &    0.85&   0.48&   0.29&   0.60&   1.14&   1.30&   1.39&   3.19&     &    0.47&2&M\\
 77&  B7 V        &    0.87&   0.58&    -  &   0.61&   1.40&   1.60&   1.79&   3.39&     &    0.45&2&M\\
 78&  B6 V        &    0.66&   0.45&   0.37&   0.70&   1.17&   1.37&   1.45&   3.54&     &    0.42&2&M\\
 79&  B8 IV       &    0.68&   0.43&    -  &   0.57&   1.14&   1.36&   1.46&   3.73&     &    0.37&2&M\\
 80&  B0.5 V      &    0.72&   0.42&   0.24&   0.55&   1.03&   1.16&   1.22&   3.20&     &    0.34&1&M\\
 81&  O4 V        &    0.93&   0.53&   0.30&   0.69&   1.22&   1.42&   1.51&   3.13&     &    0.37&1&M\\
 82&  B3 V        &    0.89&   0.50&   0.28&   0.65&   1.37&   1.67&   2.04&   3.74&     &    0.43&3&M\\
 83&  B6 V        &    0.86&   0.55&    -  &   0.74&   1.37&   1.54&   1.64&   3.28&     &    0.52&2&M\\
 84&  O9.5 V      &    1.47&   0.84&   0.49&   1.11&   2.18&   2.51&   2.67&   3.50&     &    0.43&2&M\\
 85&  F3 V        &    0.59&   0.21&    -  &    -  &   0.53&   0.54&   0.61&   3.20& 3.10&    0.47&3&NM\\
 86&  G1 V        &    0.32&   0.17&    -  &    -  &   0.48&   0.57&   0.62&    -  & 2.80&    0.43&3&NM\\
 87&  K3 Ibe      &    1.62&   0.77&    -  &    -  &   1.91&   2.66&   2.63&   3.38& 4.56&    1.44&3&PM\\
 88&  F9 V*       &    0.25&   0.06&   0.03&   0.10&   0.32&   0.36&   0.41&    -  & 2.55&    0.34&3&M\\
 89&  F9*         &    0.28&   0.18&   0.17&   0.30&   0.65&   0.82&   0.87&    -  & 2.37&    0.36&3&M\\
 90&  K1 V        &    3.93&   1.66&   0.73&   1.91&   3.26&   3.87&   4.20&   2.68& 3.58&    0.96&3&NM\\
 91&  G0 e*       &    0.02&   0.03&   0.13&   0.07&   0.27&   0.32&   0.36&    -  & 3.04&    0.20&3&PM\\
 92&  F8 Ve*      &    0.52&   0.25&   0.13&   0.35&   0.68&   0.91&   1.12&   3.71& 1.95&    0.62&3&M\\
 93&  G5*         &    1.89&   0.91&   0.57&    -  &   2.30&   2.69&   2.89&   3.49& 3.01&    0.76&3&NM\\
 94&  B3          &    1.71&   0.91&   0.67&   1.62&   3.15&   3.88&   4.36&   4.71&     &    0.24&3&PM\\
 95&  B2 V        &    1.24&   0.86&   0.55&   1.43&   2.96&   3.62&   4.04&   4.77&     &    0.59&3&PM\\
 96&  G0 IV-V     &    0.04&   0.02&  $-$0.04&   0.11&   0.27&   0.34&   0.37&    -  & 3.14&  0.61&3&M\\
 97&  B8 IIIe*    &    0.69&   0.25&   0.39&   0.93&   1.53&   2.55&   3.75&   8.35&     &    0.49&3&M\\
 98&  A2 IVe      &    0.11&   0.08&   0.07&   0.21&   0.43&   0.79&   1.47&    -  &     &    0.44&3&M\\
 99&  F8 Ve       &    0.50&   0.33&   0.11&   0.31&   0.86&   1.10&   1.17&   3.90& 1.93&    0.71&3&M\\
100&  G5 Ve       &    0.47&   0.26&  $-$0.01&   0.17&   0.51&   0.79&   1.11&   2.70& 0.06&  0.73&3&M\\
101&  G8 Ve       &    0.29&   0.21&   0.10&   0.21&   0.56&   0.87&   1.28&   3.64& 2.38&    0.45&3&M\\
102&  F9*         &    0.23&   0.10&  $-$0.05&   0.00&   0.21&   0.31&   0.32&    -  & 2.82&  0.34&3&M\\
103&  K0*         &    2.28&   1.14&   0.70&   1.48&   2.75&   3.17&   3.45&   3.33& 3.13&    0.69&3&NM\\
104&  A2*         &    0.99&   0.56&   0.15&    -  &   1.27&   1.44&   1.56&   3.06&     &    0.53&1&NM\\
105&  K0*         &    1.67&   0.86&   0.26&    -  &   1.83&   2.18&   2.34&   2.99& 2.83&    0.67&3&NM\\
106&  G8          &    1.08&   0.27&  $-$0.02&   0.07&   0.37&   0.51&   0.56&   2.28& 3.15&  0.89&3&NM\\
107&  O7          &    0.54&   0.29&    -  &    -  &   0.71&   0.84&   0.92&   3.49&     &    0.56&1&M\\
108&  B0.5        &    0.52&   0.29&   0.16&   0.37&   0.77&   0.79&   0.88&   3.34&     &    0.52&2&M\\
109&  B5 e        &    0.37&   0.29&    -  &    -  &   0.82&   1.04&   1.27&   3.86&     &    0.68&3&M\\
110&  O4 V        &    0.64&   0.35&   0.27&   0.53&   0.94&   1.10&   1.18&   3.71&     &    0.58&2&M\\
111&  B0 V        &    0.49&   0.30&    -  &   0.38&   0.74&   0.82&   0.86&   3.15&     &    0.52&2&M\\
112&  B4          &    0.61&   0.35&    -  &   0.48&   0.97&   1.13&   1.18&   3.71&     &    0.58&2&M\\
113&  B8          &    0.24&   0.36&    -  &   0.59&   1.15&   1.45&   1.60&   4.89&     &    0.29&2&M\\
114&  F4 I*       &    0.49&   0.24&   0.33&   0.73&   0.81&   0.90&   0.96&   4.40& 5.07&    0.42&3&NM\\
115&  B2 e        &    1.98&   1.05&   0.63&   1.46&   3.12&   3.99&   4.44&   4.05&     &    0.67&3&M\\
116&  B3 V        &    0.40&   0.31&   0.28&   0.53&   0.94&   1.10&   1.20&   4.26&     &    0.58&2&M\\
117&  G7          &    0.92&   0.41&    -  &   0.50&   1.00&   1.17&   1.29&   3.46&     &    0.86&3&NM\\
118&  B2 V        &    0.34&   0.28&   0.18&   0.37&   0.72&   0.80&   0.89&   3.50&     &    0.56&2&M\\
119&  B2          &    0.41&   0.33&    -  &   0.18&   0.77&   0.72&   0.83&   3.17&     &    0.63&2&M\\
120&  B8          &    0.44&   0.30&    -  &   0.58&   1.16&   1.42&   1.58&   5.79&     &    0.39&2&M\\
121&  B1.5 V      &    0.52&   0.35&   0.16&   0.44&   0.95&   1.08&   1.20&   3.77&     &    0.62&2&M\\
122&  B2 V        &    0.58&   0.42&   0.24&   0.62&   1.31&   1.63&   1.99&   4.25&     &    0.68&3&M\\
123&  B2 V        &    0.49&   0.33&   0.15&   0.42&   1.05&   1.19&   1.34&   4.47&     &    0.67&2&M\\
124&  B1 V        &    0.62&   0.33&   0.18&   0.46&   0.96&   1.13&   1.22&   4.07&     &    0.60&2&M\\
125&  B2 V        &    0.56&   0.36&    -  &   0.47&   0.96&   1.09&   1.15&   3.51&     &    0.75&2&M\\
126&  B0 IVe      &    0.78&   0.46&   0.32&   0.76&   1.45&   1.74&   2.13&   4.30&     &    0.38&3&M\\
127&  B2 V        &    0.51&   0.35&   0.16&   0.44&   0.89&   1.04&   1.12&   3.52&     &    0.91&2&M\\
128&  B2 V        &    0.59&   0.39&   0.14&   0.41&   1.02&   1.17&   1.29&   3.64&     &    0.85&2&M\\
129&  B2 IV       &    0.53&   0.32&    -  &    -  &   0.85&   0.90&   1.02&   3.51&     &    0.50&2&M\\
130&  B2.5 V      &    0.48&   0.32&    -  &   0.48&   0.93&   1.11&   1.20&   4.13&     &    0.72&2&M\\
131&  B2.5 V      &    0.22&   0.28&   0.21&   0.39&   0.81&   0.90&   1.01&   3.97&     &    0.62&2&M\\
132&  B1 V        &    0.50&   0.33&    -  &   0.49&   0.97&   1.10&   1.18&   3.93&     &    0.68&2&M\\
133&  B2.5 V      &    0.46&   0.35&   0.22&   0.45&   0.94&   1.24&   1.26&   3.66&     &    0.58&2&M\\
134&  B0.5 V      &    0.62&   0.41&    -  &    -  &   1.02&   1.14&   1.22&   3.27&     &    0.86&2&M\\
135&  B2 V        &    0.63&   0.40&   0.06&   0.35&   0.87&   1.05&   1.22&   3.36&     &    0.65&1&M\\
136&  B2 IV       &    0.38&   0.26&   0.23&   0.44&   0.67&   0.95&   1.09&   4.51&     &    0.56&3&M\\
137&  B2 V        &    0.62&   0.33&    -  &   0.51&   0.98&   1.10&   1.22&   4.07&     &    0.50&2&M\\
138&  B2 V        &       -&   0.33&    -  &    -  &   0.73&   0.85&   0.93&   3.10&     &    0.48&2&M\\
139&  B6          &    0.32&   0.31&    -  &   0.46&   0.81&   0.88&   0.95&   3.37&     &    0.40&2&M\\
140&  B3          &    0.55&   0.36&   0.26&   0.50&   1.05&   1.18&   1.27&   3.88&     &    0.55&2&M\\
\hline                                                                                              
\end{tabular}                                                                                       
\end{table*}                                                                                        
                                                                                                    
\begin{table*}                                                                                      
{{\bf Table 6.} Continued.}                                                                         
                                                                                                    
\scriptsize                                                                                         
\begin{tabular}{clrrrrrrrrcccc} \hline                                                               
ID& SpT& {\it E(U-V)}& {\it E(B-V)}& {\it E(V-R)}& {\it E(V-I)}& {\it E(V-J)}& {\it E(V-H)}& {\it   
									   E(V-K)}& $R_{V}$& EW(\AA)& EW(\AA)&Grp&Memb\\                   
&&&&&&&&&&CaIIT&NaID&&\\ \hline                                                                      
141&  B2.5 V      &    0.74&   0.47&   0.30&   0.64&   1.30&   1.57&   1.70&   3.98&     &    0.85&2&M\\
142&  O6.5 V      &    0.74&   0.41&   0.25&   0.55&   1.11&   1.28&   1.34&   3.60&     &    0.67&2&M\\
143&  K5          &    0.13&   0.13&    -  &   0.00&   0.05&   0.00&   0.01&    -  & 3.38&    1.15&3&NM\\
144&  B6          &    0.61&   0.38&    -  &   0.62&   1.27&   1.50&   1.65&   4.78&     &    0.70&2&M\\
145&  K2          &    0.44&   0.32&    -  &   0.34&   0.35&   0.34&   0.41&   1.41&     &    0.75&3&NM\\
146&  B2          &    0.67&   0.40&   0.25&   0.50&   1.31&   1.51&   1.66&   4.57&     &    0.65&2&M\\
147&  B3          &    0.76&   0.48&    -  &    -  &   1.20&   1.45&   1.58&   3.62&     &    0.82&2&M\\
148&  B8          &    0.64&   0.52&    -  &    -  &   1.45&   1.70&   1.84&   3.89&     &    0.60&2&M\\
149&  K7 I-III*   &    2.28&   1.17&    -  &    -  &   2.64&   3.33&   3.67&   3.68& 6.41&    2.63&3&PM\\
150&  B6          &    1.63&   1.00&   0.74&   1.41&   2.60&   2.94&   3.16&   3.48&     &    0.22&2&M\\
151&  K7 III-IV*  &    2.68&   1.30&    -  &    -  &   2.75&   3.42&   3.63&   3.26&     &    2.20&3&PM\\
152&  B1 V        &    1.12&   0.73&   0.44&   0.99&   1.85&   2.08&   2.23&   3.36&     &    1.17&2&M\\
153&  B0.5 V      &    1.25&   0.75&   0.48&   1.04&   1.93&   2.22&   2.33&   3.42&     &    0.76&2&M\\
154&  O8.5 V      &    2.29&   1.33&   0.88&   1.97&   3.76&   4.34&   4.70&   3.89&     &    0.45&2&M\\
155&  05.5 V      &    1.92&   1.12&   0.74&   1.68&   3.15&   3.68&   3.97&   3.90&     &    0.60&2&M\\
156&  O6.5 V      &    1.26&   0.76&   0.49&   1.08&   1.96&   2.22&   2.43&   3.52&     &    0.68&2&M\\
157&  O4 V        &    1.27&   0.77&   0.53&   1.18&   2.04&   2.40&   2.59&   3.70&     &    0.60&2&M\\
158&  B1 V        &    1.43&   0.91&   0.56&   1.28&   2.27&   2.60&   2.78&   3.36&     &    0.51&2&M\\
159&  B1.5 V      &    1.61&   0.94&   0.52&   1.17&   2.38&   2.80&   3.00&   3.51&     &    0.53&2&M\\
160&  B2          &    1.41&   1.07&   0.72&   1.61&   2.98&   3.51&   3.99&   3.80&     &    0.42&3&M\\
161&  O7 IIb      &    2.01&   1.19&   0.77&   1.65&   3.05&   3.57&   3.80&   3.51&     &    0.68&2&M\\
162&  B1 V        &    1.18&   0.70&   0.40&   0.95&   1.86&   2.15&   2.30&   3.61&     &    0.40&2&M\\
163&  B0.5 V      &    1.69&   1.01&   0.66&   1.45&   2.64&   3.05&   3.23&   3.52&     &    0.63&2&M\\
164&  O9.5 V      &    1.23&   0.79&   0.42&   1.01&   1.78&   2.00&   2.18&   3.04&     &    0.99&1&M\\
165&  B2 V        &    1.33&   0.77&   0.67&   1.28&   2.18&   2.50&   2.66&   3.80&     &    1.04&2&M\\
166&  B1.5 V      &    1.53&   0.88&   0.58&   1.30&   2.42&   2.79&   2.90&   3.63&     &    0.66&2&M\\
167&  B0 V        &    1.46&   0.92&   0.57&   1.26&   2.35&   2.64&   2.85&   3.41&     &    0.75&2&M\\
168&  B1 V        &    1.83&   1.08&   0.81&   1.66&   2.97&   3.43&   3.72&   3.79&     &    0.98&2&M\\
169&  O9.5 V      &    0.88&   0.58&   0.32&   0.71&   1.33&   1.48&   1.56&   2.96&     &    0.71&1&M\\
170&  O8.5 V      &    1.11&   0.71&   0.46&   1.00&   1.84&   2.14&   2.30&   3.56&     &    0.85&2&M\\
171&  O9.5 I      &    0.95&   0.61&   0.38&   0.85&   1.37&   1.52&   1.65&   2.98&     &    0.90&1&M\\
172&  B1 V        &    0.79&   0.55&   0.28&   0.68&   1.35&   1.51&   1.60&   3.20&     &    0.78&2&M\\
173&  B0.5 Ve     &    1.09&   0.70&   0.42&   0.94&   1.77&   2.01&   2.21&   3.47&     &    0.67&2&M\\
174&  B5 e        &    0.94&   0.59&   0.36&   0.78&   1.44&   1.61&   1.80&   3.36&     &    0.81&2&PM\\
175&  B1 e        &    0.91&   0.58&   1.07&   1.71&   2.24&   2.64&   3.03&   5.27&     &    0.61&3&M\\
176&  B2.5 I      &    2.27&   1.30&   0.82&   1.79&   3.46&   3.96&   4.30&   3.64&     &    0.83&2&PM\\
177&  G9*         &    1.36&   0.55&    -  &    -  &   1.40&   1.62&   1.76&   3.52&     &    0.84&3&NM\\
178&  G6          &    1.44&   0.56&   0.33&   0.68&   1.37&   1.65&   1.83&   3.59&     &    1.22&3&NM\\
179&  B1.5 V      &    0.99&   0.62&   0.38&   0.86&   1.52&   1.80&   1.98&   3.51&     &    0.92&2&M\\
180&  B1          &     -  &   0.64&    -  &    -  &   1.77&   1.92&   2.11&   3.63&     &    0.67&2&M\\
181&  B9 V        &     -  &   0.41&    -  &    -  &   1.06&   1.16&   1.25&   3.35&     &    0.54&3&NM\\
182&  A0 III      &     -  &   0.41&    -  &    -  &   1.19&   1.29&   1.45&   3.89&     &    0.49&3&NM\\
183&  B9.5 IV     &     -  &   0.21&    -  &    -  &   0.70&   0.75&   0.84&   4.40&     &    0.57&3&NM\\
184&  B9 IV       &     -  &   0.18&    -  &    -  &   0.90&   1.03&   1.15&    -  &     &    0.57&3&NM\\
185&  K0*         &     -  &   0.43&    -  &    -  &   0.97&   1.12&   1.23&   3.15&     &    0.74&3&NM\\
186&  A0 IV       &     -  &   0.19&    -  &    -  &   0.78&   0.88&   0.97&    -  &     &    0.63&3&M\\
187&  B1.5        &     -  &   0.59&    -  &    -  &   1.56&   1.79&   1.87&   3.49&     &    0.67&1&M\\
188&  B0.5 IV     &    1.87&   1.10&    -  &    -  &   2.34&   2.63&   2.79&   2.79&     &    0.70&1&M\\
189&  B2 V        &    1.52&   0.96&    -  &    -  &   2.23&   2.50&   2.63&   3.01&     &    0.85&1&M\\
190&  O8 V        &    1.78&   1.09&    -  &    -  &   2.43&   2.76&   2.88&   2.91&     &    1.00&1&M\\
191&  B2 V        &    1.60&   0.99&    -  &    -  &   2.20&   2.47&   2.64&   2.93&     &    0.87&1&M\\
192&  B1.5 V      &    1.41&   0.95&    -  &    -  &   1.95&   2.24&   2.36&   2.73&     &    0.87&1&M\\
193&  B1.5 V      &    1.64&   0.95&    -  &    -  &   1.96&   2.26&   2.42&   2.80&     &    0.97&1&M\\
194&  G8 III      &    1.46&   0.76&    -  &    -  &   1.46&   1.49&   1.74&   2.62& 3.39&    0.98&3&NM\\
195&  B0.5 I      &    1.93&   1.08&   0.68&   1.25&   2.32&   2.60&   2.71&   2.76&     &    0.83&3&PM\\
196&  B4 V        &    1.13&   0.77&    -  &    -  &   1.94&   2.25&   2.40&   3.43&     &    0.45&1&M\\
197&  B0 IV       &    1.22&   0.74&    -  &    -  &   1.69&   1.88&   2.00&   2.97&     &    0.90&1&M\\
198&  B1 III      &    1.86&   1.07&    -  &    -  &   2.14&   2.26&   2.51&   2.73&     &    0.82&3&M\\
199&  K8 V        &    1.38&   0.62&    -  &    -  &   1.16&   1.34&   1.52&   2.70& 2.98&    1.20&3&NM\\
200&  B2.5 V      &    1.30&   0.88&    -  &    -  &   1.87&   2.15&   2.35&   2.94&     &    0.74&1&M\\
201&  B1.5 V      &    1.11&   0.82&    -  &    -  &   1.94&   2.14&   2.31&   3.10&     &    0.91&1&M\\
202&  O7 V        &    1.46&   0.88&   0.54&   1.16&   2.08&   2.38&   2.49&   3.11&     &    0.39&1&M\\
203&  B0.5 V      &    1.72&   1.04&   0.63&   1.33&   2.44&   2.74&   2.90&   3.07&     &    0.47&1&M\\
204&  B0.5 V      &    1.63&   1.02&   0.62&   1.32&   2.31&   2.75&   2.95&   3.18&     &    0.81&1&M\\
205&  B1 V        &    1.61&   0.99&    -  &    -  &   2.43&   2.74&   2.92&   3.24&     &    0.42&1&M\\
206&  F8 V        &    0.25&   0.22&    -  &    -  &   0.63&   0.71&   0.75&   3.75& 2.84&    0.87&3&NM\\
207&  B1 V        &    1.75&   1.06&    -  &    -  &   2.65&   2.99&   3.19&   3.31&     &    0.53&1&M\\
208&  F2 IV       &    0.05&   0.11&    -  &    -  &   0.31&   0.32&   0.34&    -  & 2.75&    0.69&3&NM\\
209&  A9          &    1.83&   1.08&   0.64&   1.21&   2.10&   2.57&   2.71&   2.65&     &    0.45&3&NM\\
210&  B0.5 V      &    1.93&   1.15&    -  &    -  &   2.69&   3.04&   3.23&   3.09&     &    0.46&1&M\\
211&  F8 Ib       &    1.29&   0.66&    -  &    -  &   1.40&   1.60&   1.72&   2.87& 4.43&    0.54&3&NM\\
\hline                                                              
\end{tabular}                                                       
\end{table*}

\clearpage


\begin{table*}
\caption{List of stars with emission features and spectral peculiarities. Asterisked ones are reported in the present 
work for the first time along with some additional features reported for stars 5, 100 and 101. Stars with plus symbol are 
optical variable and stars 5, 6, 27 and 29 are NIR variable (Carpenter 2001). }

\scriptsize
\begin{tabular}{llll} \hline 
ID&    SpT&  Emission peculiarities \\  \hline
    5+ &       F8 V& CaII HK in core emission, Strong H$_{\alpha}$, totally filled H$_{\beta}$\\
    6* &    G9 IV-V& CaII HK in emission                                                      \\
    8* &     A2 III& H$_{\alpha}$ partially filled in emission                                \\
   15*+&       K0 V& CaII HK in emission on wide absorption                                   \\
   22  &       O7  & H$_{\alpha}$ partially filled in emission                                \\
   23* &  G0 III-IV& CaII HK partially filled, H$_{\alpha}$ in weak emission                  \\
   27+ &    G3 IV-V& CaII HK and Balmer lines in emission                                     \\
   29* &       G8 V& CaII HK partially filled, H$_{\alpha}$ in weak emission                  \\
   31  &    G9 IV-V& CaII HK in emission on absorption                                        \\
   32+ &      K2 IV& CaII HK in emission, H$_{\alpha}$ in strong emission                     \\
   33+ &       K2 V& CaII HK and Balmer lines in emission                                     \\
   34  &       K1 V& CaII HK in filled emission, weak H$_{\alpha}$ emission                   \\
   35+ &     A3 III& H$_{\alpha}$ in strong emission                                          \\
   87* &      K3 Ib& CaII HK in emission                                                      \\
   91* &         G0& CaII HK weak emission on wide absorption                                 \\
   92* &       F8 V& H$_{\alpha}$ partially filled in emission                                \\
   97+ &         B8& H$_{\alpha}$ in strong emission, H$_{\beta}$ partially filled in emission\\ 
   98+ &      A2 IV& H$_{\alpha}$ totally filled in emission                                  \\
   99* &       F8 V& CaII HK partially filled, H$_{\alpha}$ totally filled in emission        \\
  100+ &       G5 V& CaII HK in emission, H$_{\alpha}$ in strong emission                     \\
  101+ &       G8 V& CaII partially filled, H$_{\alpha}$ in strong emission                   \\
  109* &         B5& H$_{\beta}$ in emission, H$_{\gamma}$ partially filled in emission       \\
  115* &         B2& H$_{\beta}$ and H$_{\gamma}$ partially filled in emission                \\
  122  &       B2 V& H$_{\beta}$ in emission, H$_{\gamma}$ partially filled in emission       \\
  126  &      B0 IV& H$_{\beta}$ in emission, H$_{\gamma}$ partially filled in emission       \\
  160  &         B2& H$_{\beta}$ in emission                                                  \\
  173  &     B0.5 V& H$_{\beta}$ partially filled in emission                                 \\
  175  &        B1 & H$_{\beta}$ totally filled in emission                                   \\
  174  &        B5 & H$_{\beta}$ totally filled in emission                                   \\
\hline
\end{tabular}      
\end{table*}       


\begin{table*}
\caption{Distributions of spectral types and luminosity classes among the program stars. The luminosity class of emission objects
are discussed in the text.} 

\begin{tabular}{c|c|cccccc} \hline 
 SpT&   N& \multicolumn{5}{c}{Luninosity types}& Emission line stars\\ \cline{3-7} 
     &        &   I&  II& III&  IV&   V& (Balmer, CaII HK, both)\\  \hline
    O&      22&   1&   1&    &    &  19& 1 (1, -, -)\\ 
    B&     109&   2&    &   2&   5&  91& 9 (9, -, -)\\
    A&      20&    &    &   1&   2&  14& 3 (3, -, -)\\
    F&      20&   2&    &    &   2&  13& 3 (3, 2, 2)\\
    G&      24&    &    &   3&    &  13& 8 (5, 8, 5)\\
    K&      16&    &    &   1&   1&   9& 5 (3, 5, 3)\\ \hline
Total&    211&   5&   1&   7&  10&  159& 29\\
\hline
\end{tabular}      
\end{table*}


\begin{table*}
\caption{A comparison of the extinction laws for the program clusters with the normal interstellar extinction derived 
theoretically for the diffuse dust by van de Hulst (1949) curve No. 15 (see Johnson 1968) and derived observationally by 
Cardelli et al. (1989). The columns 4-10 contain mean $\pm$ s.e. for the corresponding cluster and group (see Sect. 4).}

\scriptsize
\begin{tabular}{lccccccccc} \hline 
              &&&&&&&&\\ 
        Source&  Groups& No.&$\frac{E(U-V)}{E(V-J)}$& $\frac{E(B-V)}{E(V-J)}$&
                          $\frac{E(V-R)}{E(V-J)}$& $\frac{E(V-I)}{E(V-J)}$&
                          $\frac{E(V-H)}{E(V-J)}$& $\frac{E(V-K)}{E(V-J)}$& R$_{V}$\\ 
        Cluster&&&&&&&&&\\ \hline
  van den Hulst&    &    &  0.75&  0.43&  &  &  1.13& 1.21&  \\ 
Cardelli et al.&    &    &  0.78&  0.44&  0.27&  0.56&  1.14& 1.24& \\ 
        &&&&&&&&&\\ 
NGC 1976&   1&  6&  $0.80\pm0.04$& $0.44\pm0.01$& $0.27\pm0.03$& $0.57\pm0.03$& $1.19\pm0.02$&    $1.27\pm0.02$& $3.18\pm0.02$\\
        &   2& 28&  $0.46\pm0.04$& $0.29\pm0.01$& $0.20\pm0.01$& $0.49\pm0.01$& $1.19\pm0.02$&    $1.31\pm0.02$& $5.11\pm0.11$\\ 
NGC 2244&   1& 10&  $0.77\pm0.05$& $0.42\pm0.01$& $0.27\pm0.01$& $0.57\pm0.01$& $1.11\pm0.02$&    $1.20\pm0.01$& $3.16\pm0.02$\\
        &   2& 19&  $0.63\pm0.01$& $0.38\pm0.01$& $0.27\pm0.02$& $0.53\pm0.01$& $1.14\pm0.01$&    $1.23\pm0.01$& $3.60\pm0.05$\\
        & 1,2& 30&  $0.68\pm0.02$& $0.39\pm0.01$& $0.27\pm0.01$& $0.54\pm0.01$& $1.13\pm0.01$&    $1.22\pm0.01$& $3.44\pm0.04$\\
NGC 6530&   2& 33&  $0.54\pm0.02$& $0.36\pm0.01$& $0.23\pm0.02$& $0.50\pm0.02$& $1.14\pm0.01$&    $1.25\pm0.01$& $3.87\pm0.05$\\
NGC 6611&   1&  3&  $0.68\pm0.01$& $0.44\pm0.00$& $0.25\pm0.01$& $0.57\pm0.03$& $1.13\pm0.01$&    $1.22\pm0.02$& $3.03\pm0.02$\\
        &   2& 23&  $0.63\pm0.00$& $0.38\pm0.00$& $0.24\pm0.00$& $0.54\pm0.01$& $1.13\pm0.01$&    $1.23\pm0.01$& $3.56\pm0.02$\\
NGC 6823&   1& 16&  $0.75\pm0.03$& $0.45\pm0.01$& $0.27\pm0.01$& $0.56\pm0.01$& $1.13\pm0.01$&    $1.22\pm0.01$& $3.00\pm0.02$\\
\hline
\end{tabular} 
\end{table*} 

\clearpage 


\begin{table*}
\caption{Near-IR flux excess and deficiency. The $\Delta$ represent differences in the spectral type calibrated 
color excess and the normal reddening based color excesses estimated using 
$E(V-J)$ e.g $\Delta (V-H) = E(V-H)_{Normal} - E(V-H)_{SpT}$.}

\scriptsize
\begin{tabular}{crrrcrrrr} \hline 
 ID&   $E(V-J)$& $\Delta(V-H)$& $\Delta(V-K)$& ID&   $E(V-J)$& $\Delta(V-H)$& $\Delta(V-K)$\\ \hline
  5&     1.07&    0.33&    0.70&  94&     3.15&    0.29&    0.45&\\
  6&     0.95&    0.19&    0.27&  95&     2.96&    0.25&    0.37&\\
  8&     1.26&    0.57&    1.19&  97&     1.53&    0.81&    1.85&\\
 12&     2.11&    0.10&    0.22&  98&     0.43&    0.30&    0.94&\\
 14&     1.84&    0.17&    0.20& 100&     0.51&    0.21&    0.48&\\
 17&     2.77&    0.03&   -0.21& 101&     0.56&    0.23&    0.59&\\
 18&     2.58&   -0.06&   -0.40& 109&     0.82&    0.11&    0.25&\\
 22&     1.35&    0.15&    0.22& 115&     3.12&    0.43&    0.57&\\
 23&     1.35&    0.16&    0.17& 119&     0.77&   -0.16&   -0.12&\\
 24&     2.08&    0.27&    0.35& 122&     1.31&    0.14&    0.37&\\
 27&     2.07&    0.50&    1.14& 126&     1.45&    0.09&    0.33&\\
 29&     1.06&    0.18&    0.17& 133&     0.94&    0.17&    0.09&\\
 31&     1.12&    0.19&    0.23& 136&     0.67&    0.19&    0.26&\\
 32&     1.16&    0.09&    0.32& 149&     2.13&    0.25&    0.33&\\
 33&     1.01&    0.07&    0.26& 151&     2.75&    0.29&    0.22&\\
 35&     1.58&    0.77&    1.69& 160&     2.98&    0.11&    0.29&\\
 36&     2.84&    0.20&    0.25& 175&     2.24&    0.09&    0.25&\\
 82&     1.37&    0.11&    0.34& 194&     1.46&   -0.17&   -0.07&\\
 87&     1.91&    0.48&    0.26& 198&     2.14&   -0.18&   -0.14&\\
 90&     3.26&    0.15&    0.16& 209&     2.10&    0.18&    0.11&\\
 92&     0.68&    0.13&    0.28&    &         &        &        &\\
\hline
\end{tabular} 
\end{table*}


\begin{table*}
\caption{List of probable PMS stars with circumstellar material. 
The letters Y and N stands for yes and no respectively in columns 3, 4 and 8 while the letters W, M and S in column 5 denote 
weak (2 to 3$\sigma$), moderate (3 to 5$\sigma$) and strong ($> 5\sigma$) excesses respectively (see Sect. 5.1).
Stars with plus symbol are reported to show variability at optical wavelengths. The MIR spectral index (SpI), $s$ 
(i.e. $\lambda F_{\lambda} \sim \lambda^{s}$), is given in column 7. The star's nature i.e. classical Be (CBe),
Herbig Ae/Be (HAB) and their groups (see Sect. 5.2), Classical T Tauri (CTT) and weak line T Tauri (WTT) is given
in the last column.}

\scriptsize
\begin{tabular}{llccccccl} \hline
ID&    SpT&  HI& CaII& Excess& EW (H$_{\alpha}$)& SpI& Memb& Remark \\ 
&&&&(NIR)& (\AA)&$(s)$&&\\ \hline
    5+ &       F8 V& Y& Y& S&  -4.0&  $ -0.79\pm0.01$& Y& CTT\\
    6  &    G9 IV-V& -& Y& W&     -&                -& Y& WTT\\
    8  &       A2 e& Y& -& S&   1.6&                -& Y& HAB I\\
   15+ &       K0 V& -& Y& -&     -&                -& Y& WTT\\
   22  &        O7 & Y& -& W&     -&                -& Y& CBe\\
   23  &  G0 III-IV& Y& Y& W&     -&                -& Y& WTT\\
   27+ &    G3 IV-V& Y& Y& S&  -7.0&  $  0.52\pm0.10$& Y& CTT\\
   29  &       G8 V& Y& Y& W&     -&                -& Y& WTT\\
   31  &    G9 IV-V& -& Y& W&     -&                -& Y& WTT\\
   32+ &      K2 IV& Y& Y& M&  -4.1&                -& Y& CTT\\
   33+ &       K2 V& Y& Y& W&  -3.0&                -& Y& CTT\\
   34  &       K1 V& Y& Y& -&     -&                -& Y& WTT\\
   35+ &       A3 e& Y& -& S&  -3.0&                -& Y& HAB I \\
   91  &         G0& -& Y& -&     -&                -& Y& WTT\\
   92  &         F8& Y& -& W&     -&                -& Y& WTT\\
   97+ &       B8 e& Y& -& S& -14.7& $   0.30\pm0.17$& Y& HAB I\\
   98+ &     A2 IVe& Y& -& S&   0.2&        $  -1.20$& Y& HAB I\\
   99  &       F8 V& Y& Y& -&     -&                -& Y& WTT\\
  100+ &       G5 V& Y& Y& S&  -4.3&                -& Y& CTT\\
  101+ &       G8 V& Y& Y& S&  -2.1&  $  1.54\pm0.43$& Y& CTT\\
  109  &         B5& Y& -& W&     -&         $  0.77$& Y& HAB I or III\\
  115  &         B2& Y& -& W&     -&                -& Y& CBe\\
  122  &       B2 V& Y& -& M&     -&                -& Y& HAB I or III\\
  126  &      B0 IV& Y& -& M&     -&  $ -0.10\pm0.58$& Y& HAB I or III\\
  160  &         B2& Y& -& M&     -&         $ -2.30$& Y& HAB III or CBe\\
  173  &       B0.5& Y& -& -&     -&  $  0.61\pm0.34$& Y& HAB I or III\\
  174  &        B5 & Y& -& -&     -&                -& Y& CBe\\
  175  &        B1 & Y& -& W&     -&  $ -0.47\pm0.59$& Y& HAB I or III\\
\hline                             
\end{tabular}      
\end{table*}       

\clearpage

\begin{table*}
\caption{List of probable non-emission MS and post-MS stars with circumstellar material. The letters Y, PY and N in 
column 5 stands for yes, probably yes and no respectively while the letters W, M and S in column 3 denote 
weak (2 to 3$\sigma$), moderate (3 to 5$\sigma$) and strong ($> 5\sigma$) excesses respectively.
Asterisked star have CaII HK in emission and the stars with plus symbol are reported to show variability at optical wavelengths. 
The mid-IR spectral index (SpI), $s$ (i.e. $\lambda F_{\lambda} \sim \lambda^{s}$), is given in column 4. The last column
tells on the type of star, "PVLS" stands for probable Vega-like star. }

\scriptsize
\begin{tabular}{llcccl} \hline
ID&    SpT& Excess& SpI& Memb& Remarks \\ 
&&(near-IR)& $(s)$&&\\ \hline

       1& B2.5 IV&-&     $ -0.42\pm0.74$&  Y& PVLS\\
      12& F8     &W&                   -&  Y& PVLS\\
      13& B2 V   &-&     $  1.44\pm0.25$&  Y& PVLS\\
      14& B3 V   &W&                   -&  Y& PVLS\\
     24+& B2 V   &W&                   -&  Y& PVLS\\
      30& B0.5 V &-&     $  0.85\pm0.24$&  Y& PVLS\\
      36& B4 V   &W&     $  0.14\pm0.37$&  Y& PVLS\\
      41& B5 V   &-&     $  0.14\pm0.05$&  Y& PVLS\\
      54& F4     &-&     $  0.72\pm0.53$&  N& PVLS\\
      57& B0.5 V &-&     $ -0.55\pm0.13$&  Y&    \\
      60& F7     &-&     $  0.65\pm0.37$& PY& PVLS\\
      62& A5     &-&     $  1.24\pm0.50$&  Y& PVLS\\
      65& B0 V   &-&     $ -0.96\pm0.48$&  Y&    \\
      71& B2.5 V &-&     $  0.36\pm0.12$&  Y& PVLS\\
      82& B3 V   &M&                   -&  Y&    \\
     87*& K3 Ib  &S&     $ -2.55\pm0.13$& PY& AGB ?\\
      94& B3 V   &W&                   -&  Y& PVLS\\
      95& B2 V   &W&                   -&  Y& VLS\\
     110& O4 V   &-&     $ -0.09\pm0.10$&  Y&    \\
     111& B0 V   &-&     $  0.50\pm0.47$&  Y& PVLS\\
     119& B2 V   &-&     $  0.12\pm0.41$&  Y& PVLS\\
     121& B1.5 V &-&     $  0.80\pm0.23$&  Y& PVLS\\
     129& B2 IV  &-&     $  0.24\pm0.42$&  Y& PVLS\\
     134& B0.5 V &-&     $ -0.08\pm0.22$&  Y& PVLS\\
     136& B2 IV  &W&                   -&  Y& PVLS\\
     146& B2     &-&     $  2.80\pm0.16$&  Y& PVLS\\
     149& K7 III &M&     $ -2.78\pm0.15$& PY& AGB ?\\
 151+& K7 III-IV &M&     $ -2.79\pm0.40$& PY& AGB ?\\
     154& O5.5 V &-&     $  0.59\pm0.12$&  Y& PVLS\\
     166& B1.5 V &-&     $  1.46\pm0.27$&  Y& PVLS\\
     172& B1 V   &-&     $  0.06\pm0.38$&  Y& PVLS\\
     179& B1.5 V &-&     $ -0.62\pm0.96$&  Y&    \\
     188& B0.5 IV&-&     $ -0.33\pm0.09$&  Y&    \\
     200& B2.5 V &-&     $ -1.36\pm0.63$&  Y&    \\
     203& B0.5 V &-&     $ -1.64\pm0.79$&  Y&    \\
     204& B0.5 V &-&     $ -1.89\pm0.24$&  Y&    \\
     210& B0.5 V &-&     $  0.16\pm0.21$&  Y& PVLS\\  
\hline
\end{tabular}      
\end{table*}       

\clearpage


\begin{figure}
\includegraphics[height=3.5in,width=3.5in]{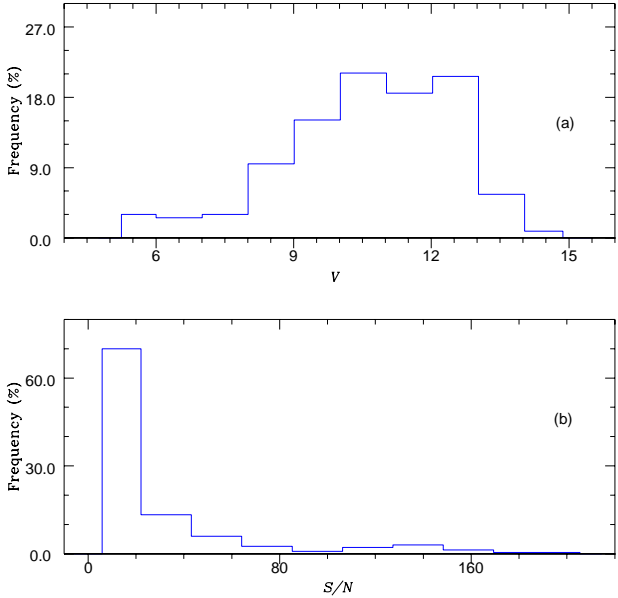}
\caption{A histogram of signal to noise ratio ($S/N$) and the visual magnitude ($V$) for the 
stars under consideration.}
\end{figure}


\begin{figure}
\includegraphics[height=3.5in,width=3.5in]{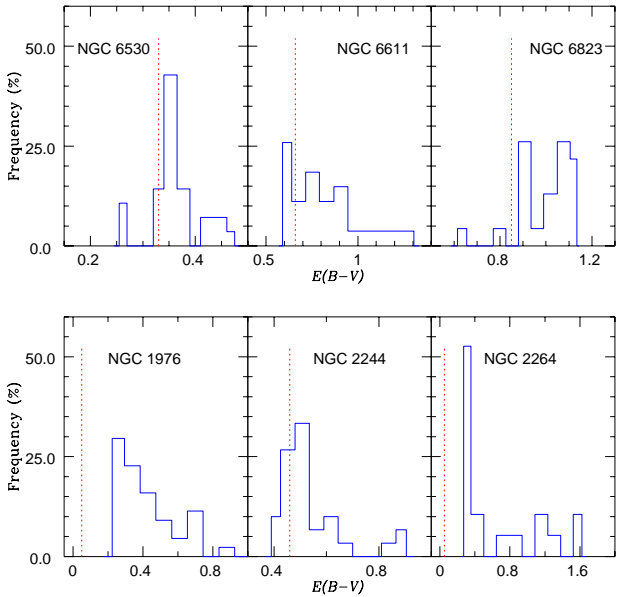}
\caption{A histogram of $E(B-V)$ for the stars under consideration. The vertical dotted lines are the mean $E(B-V)$ for the
         respective clusters.}
\end{figure}


\begin{figure}
\includegraphics[height=3.5in,width=3.5in]{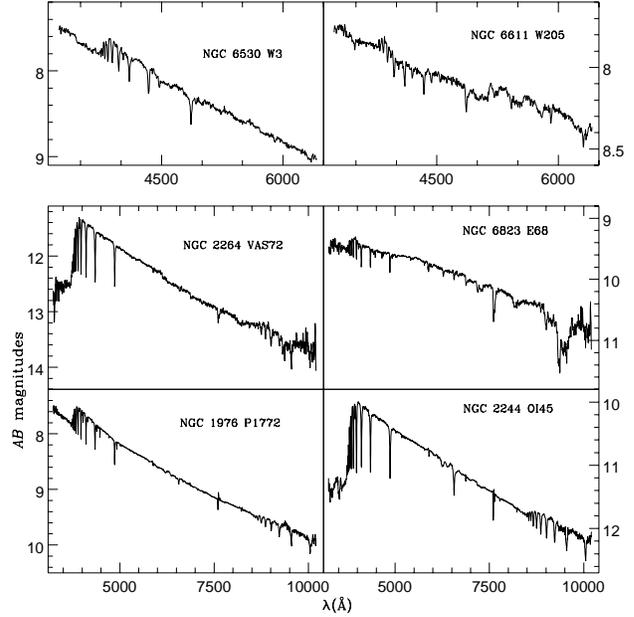}
\caption{Specimen spectra showing spectral features. The spectral types are determined by means of a cross-correlation technique as
         described in the text.}
\end{figure}


\begin{figure}
\includegraphics[height=3.5in,width=3.5in]{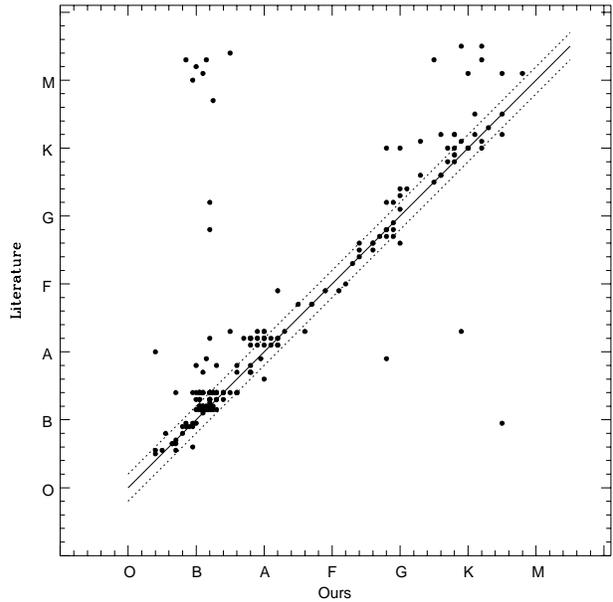}
\caption{Comparison of the spectral classification as obtained from the cross-correlation method with that available in the 
literature. The solid line represents a slope of 45 deg indicating perfect match while the dotted lines are drawn to denote an
uncertainty of two sub-spectral classes. Points showing large mismatch are discussed in the text. 
Most of the spectra were reclassified using local standards.}
\end{figure}

\clearpage

\setcounter{figure}{5}


\begin{figure}
\includegraphics[height=4.2in,width=3.5in]{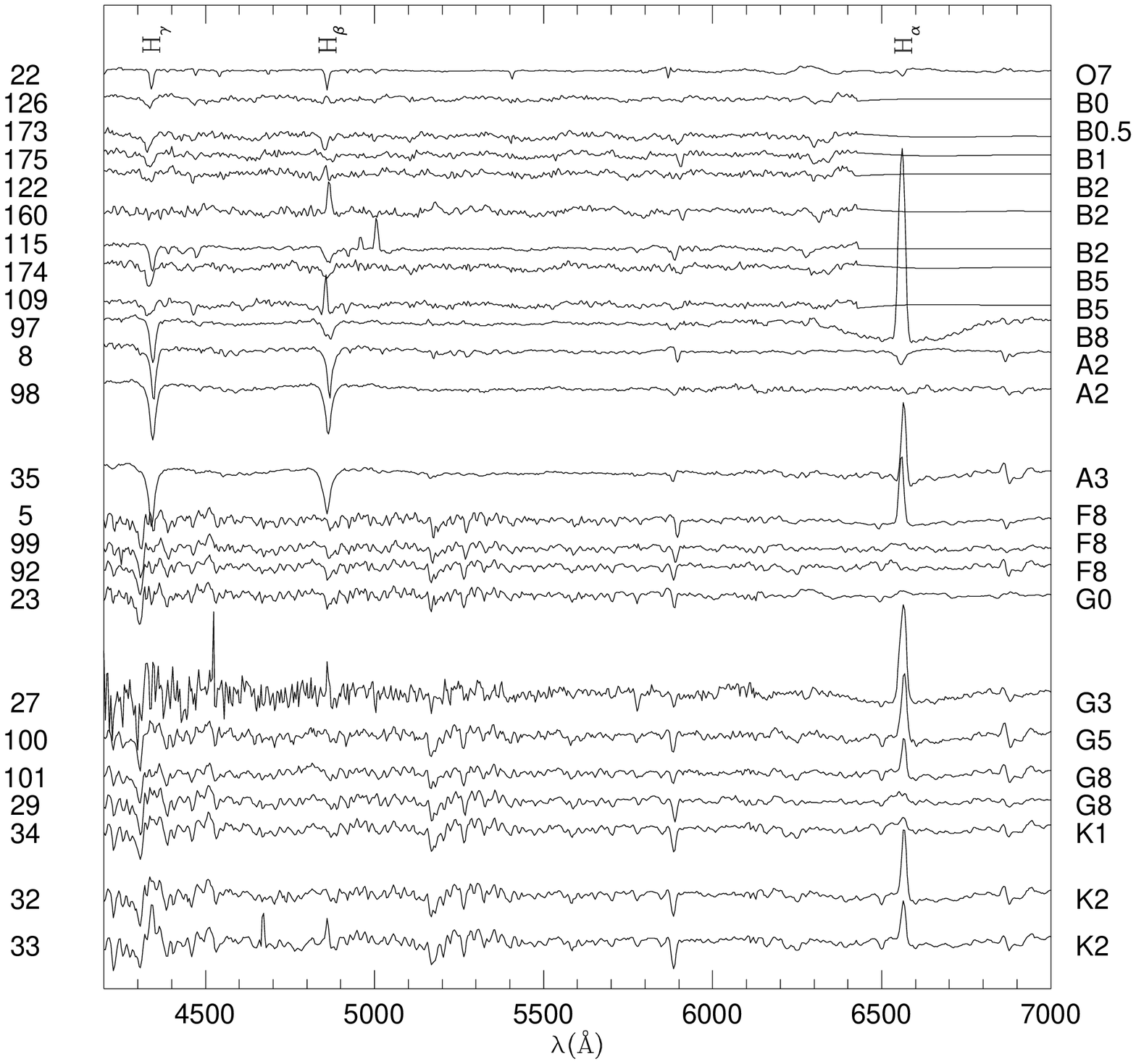}
\caption{Spectra in the $\lambda$ range 4200 to 7000 \AA~ of stars having emission features at Balmer lines.}
\end{figure}


\begin{figure}
\includegraphics[height=4.0in,width=3.5in]{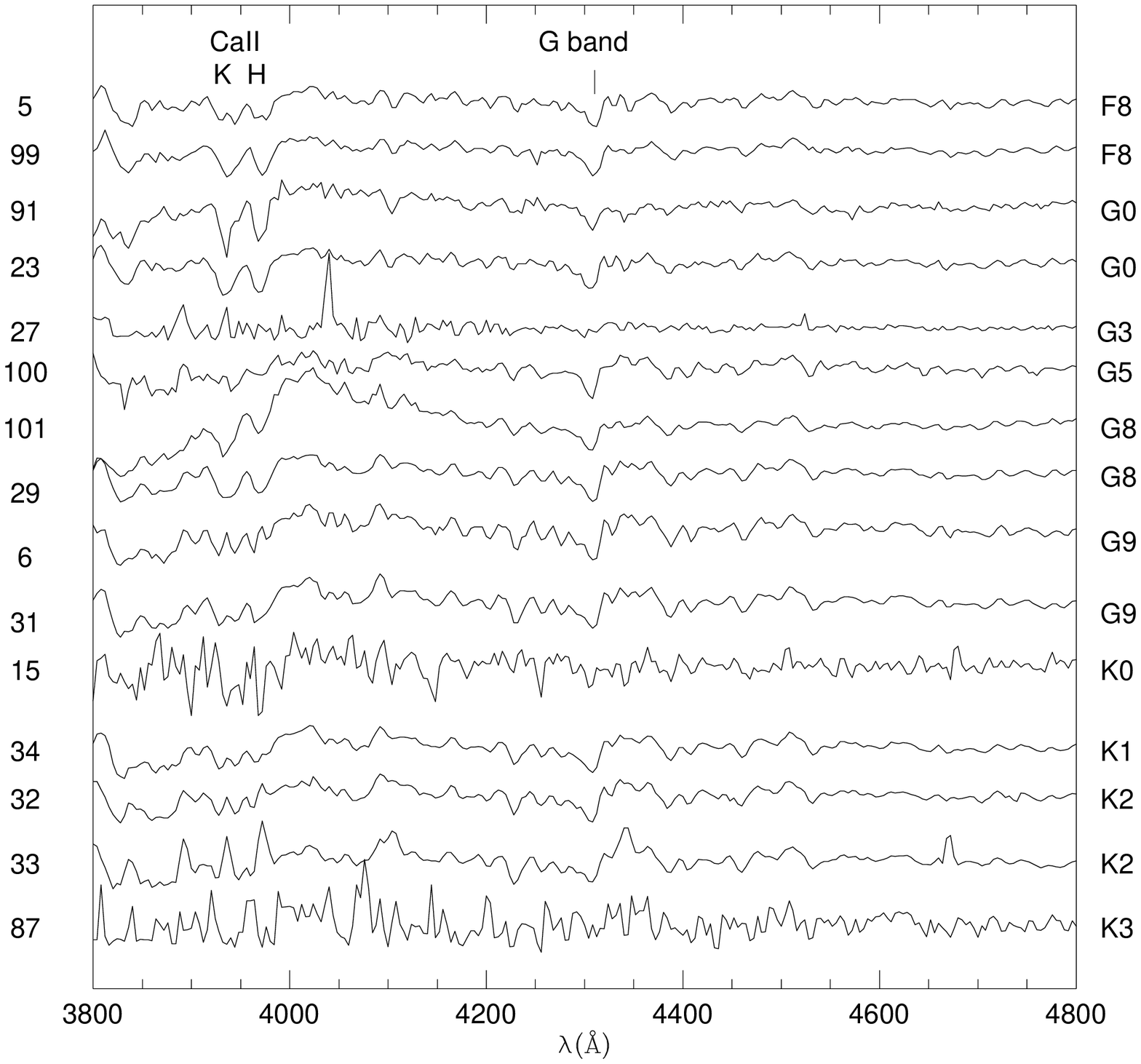}
\caption{Spectra in the $\lambda$ range 3800 to 4800 \AA~ of stars having emission features in Ca II H, K lines. The 
stars 27 and 33 have emission spectra.}
\end{figure}

\clearpage

\setcounter{figure}{8}


\begin{figure*}
\includegraphics[]{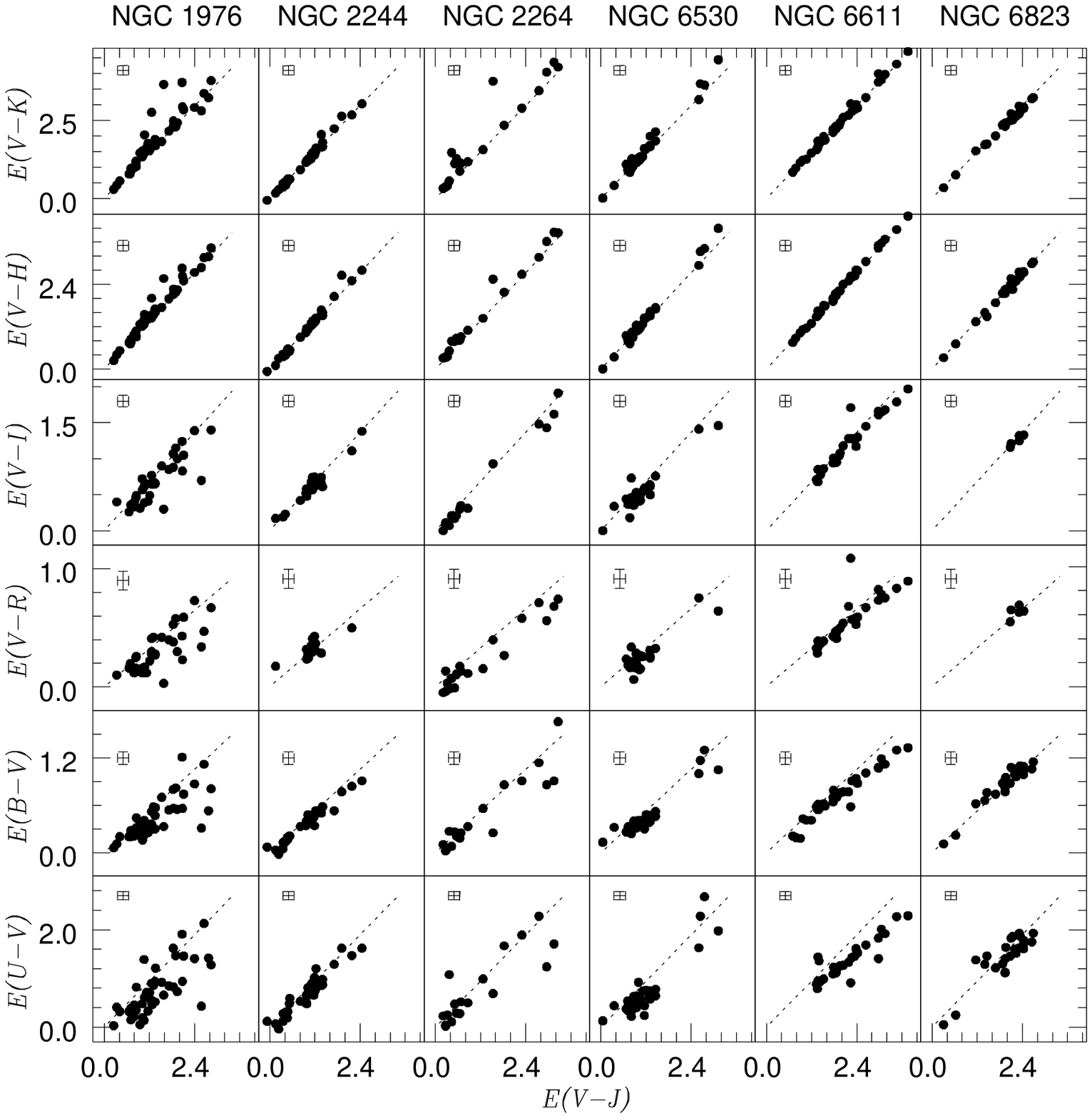}
\caption{Plots of $E(U-V)$, $E(B-V)$, $E(V-R)$, $E(V-I)$, $E(V-H)$ and $E(V-K)$ against $E(V-J)$. The dotted lines show 
         the reddening vectors characteristic of the normal interstellar extinction law. The length of the crosses represents 
         the observational errors.}
\end{figure*}


\begin{figure*}
\includegraphics[]{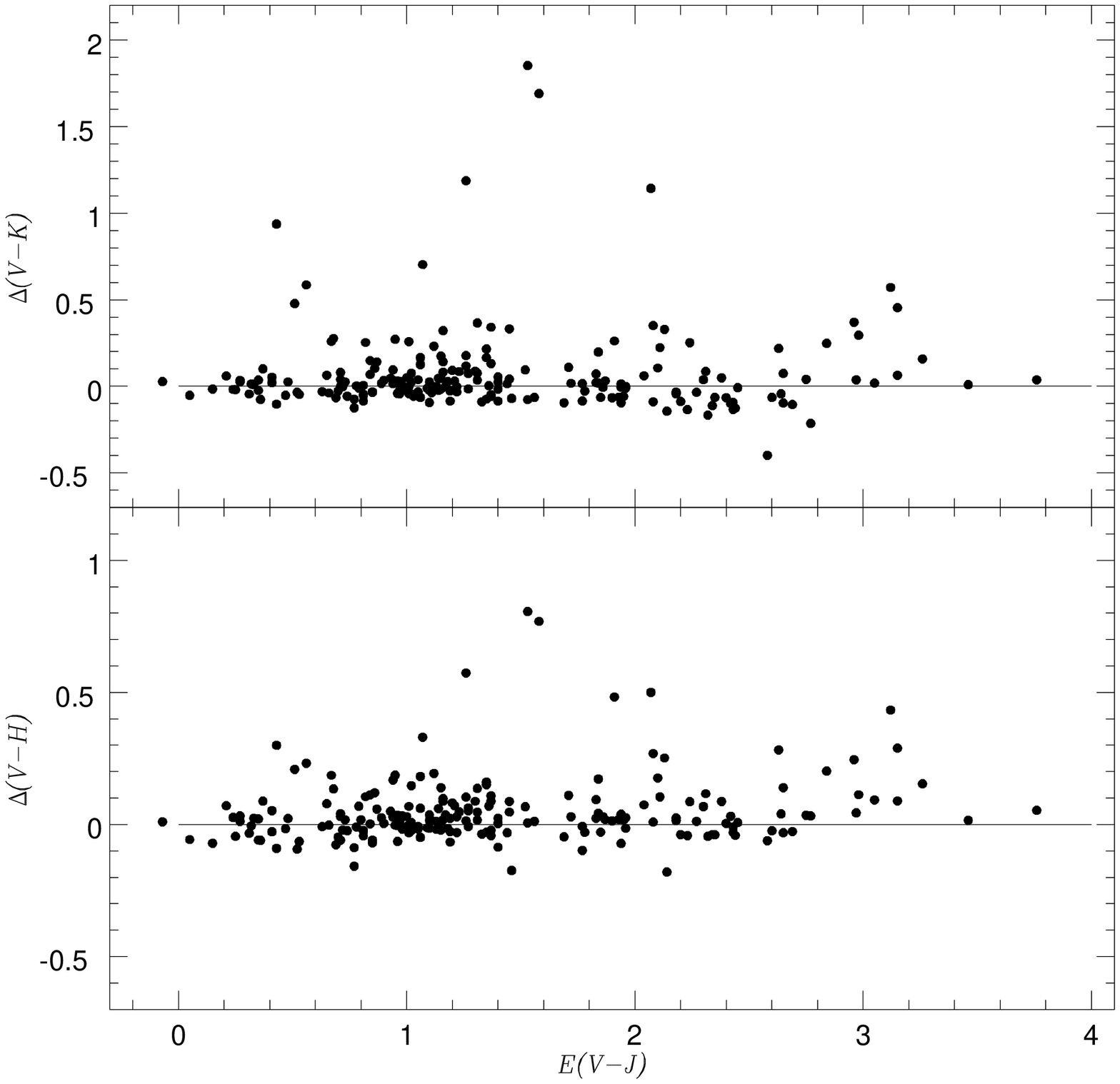}
\caption{Plots of $\Delta(V-K)$ and $\Delta(V-H)$ against the color excess $E(V-J)$. 
         Horizontal lines denote zero-level differences.}
\end{figure*}


\begin{figure*}
\includegraphics[]{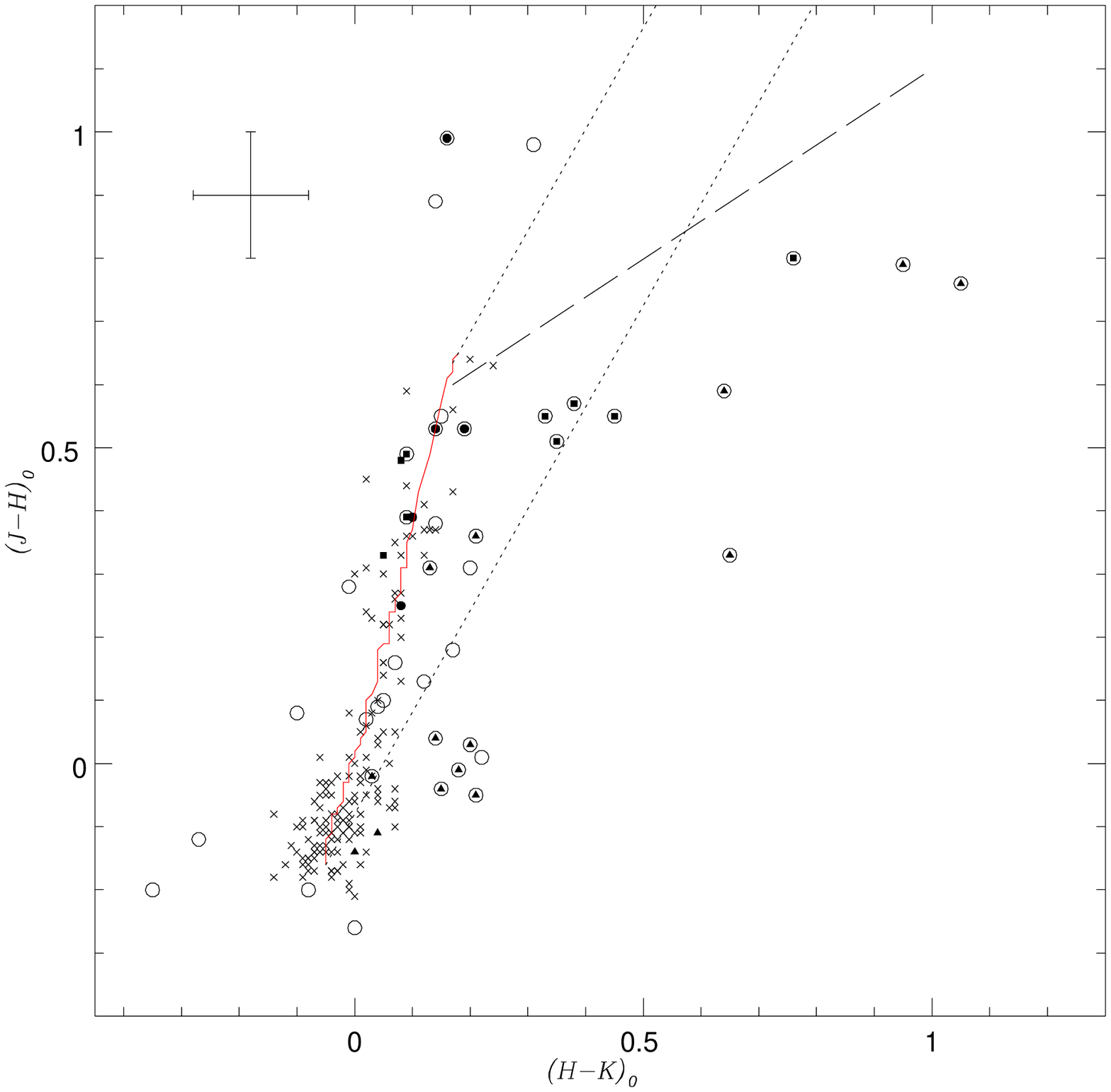}
\caption{The dereddened color-color diagram for all the program stars. Crosses denote normal stars with no NIR and spectral
peculiarities while the filled triangles, filled squares and filled circles denote stars with Balmer emissions, with
HI+CaII HK and with CaII HK emissions respectively. The circling around these symbols or an open circle denote NIR excess or 
deficiency. Solid line traces the ZAMS stars up to M0 while the dotted lines are normal reddening line and corresponds to 
$E(J-H)/E(H-K)=1.7$ (Reike \& Lebofsky 1985). The dashed line represent the locus of CTT stars (Meyer et al. 1997)}
\end{figure*}

\clearpage


\begin{figure*}
\includegraphics[height=4.0in,width=3.4in]{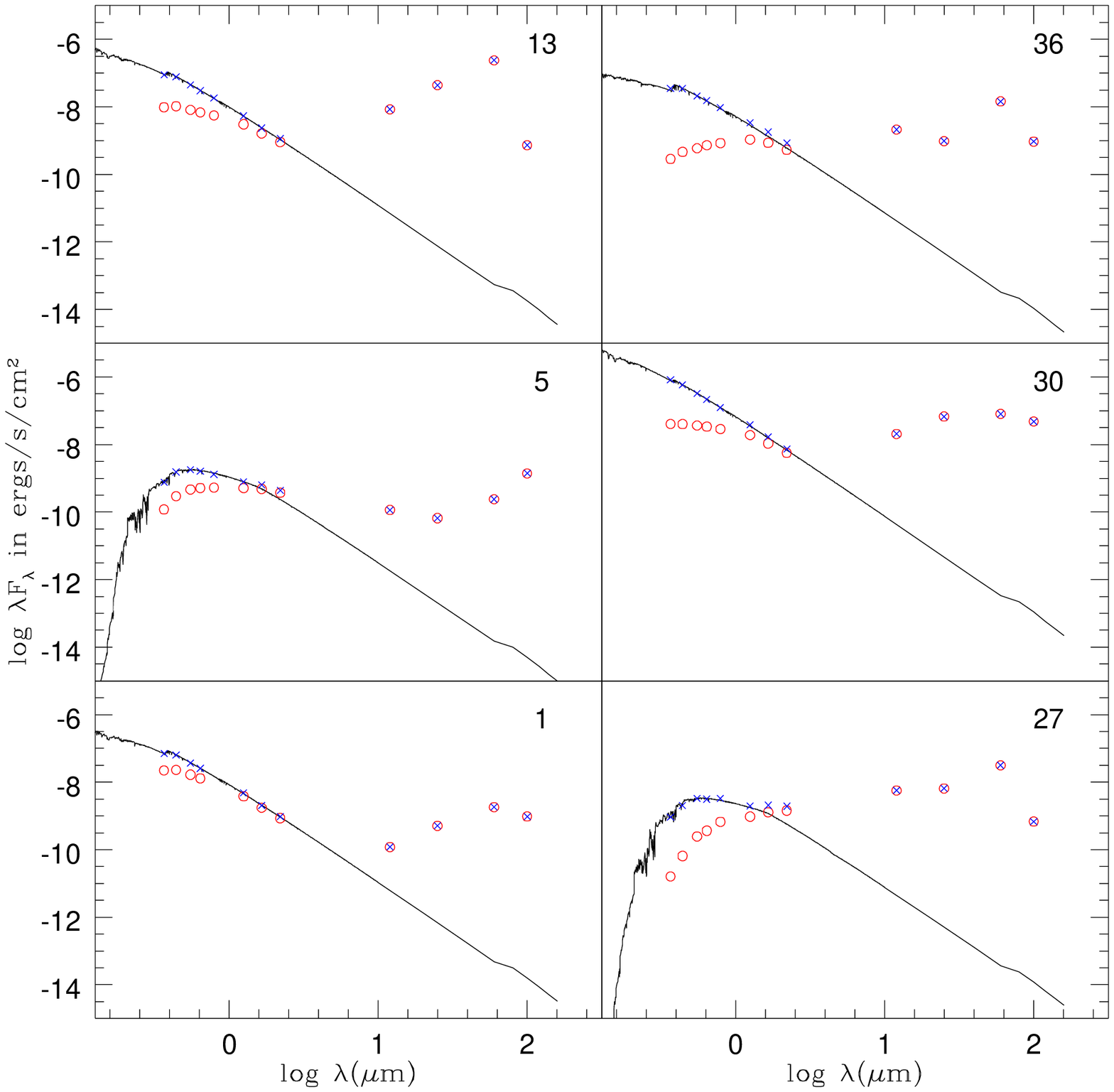}
\includegraphics[height=4.0in,width=3.4in]{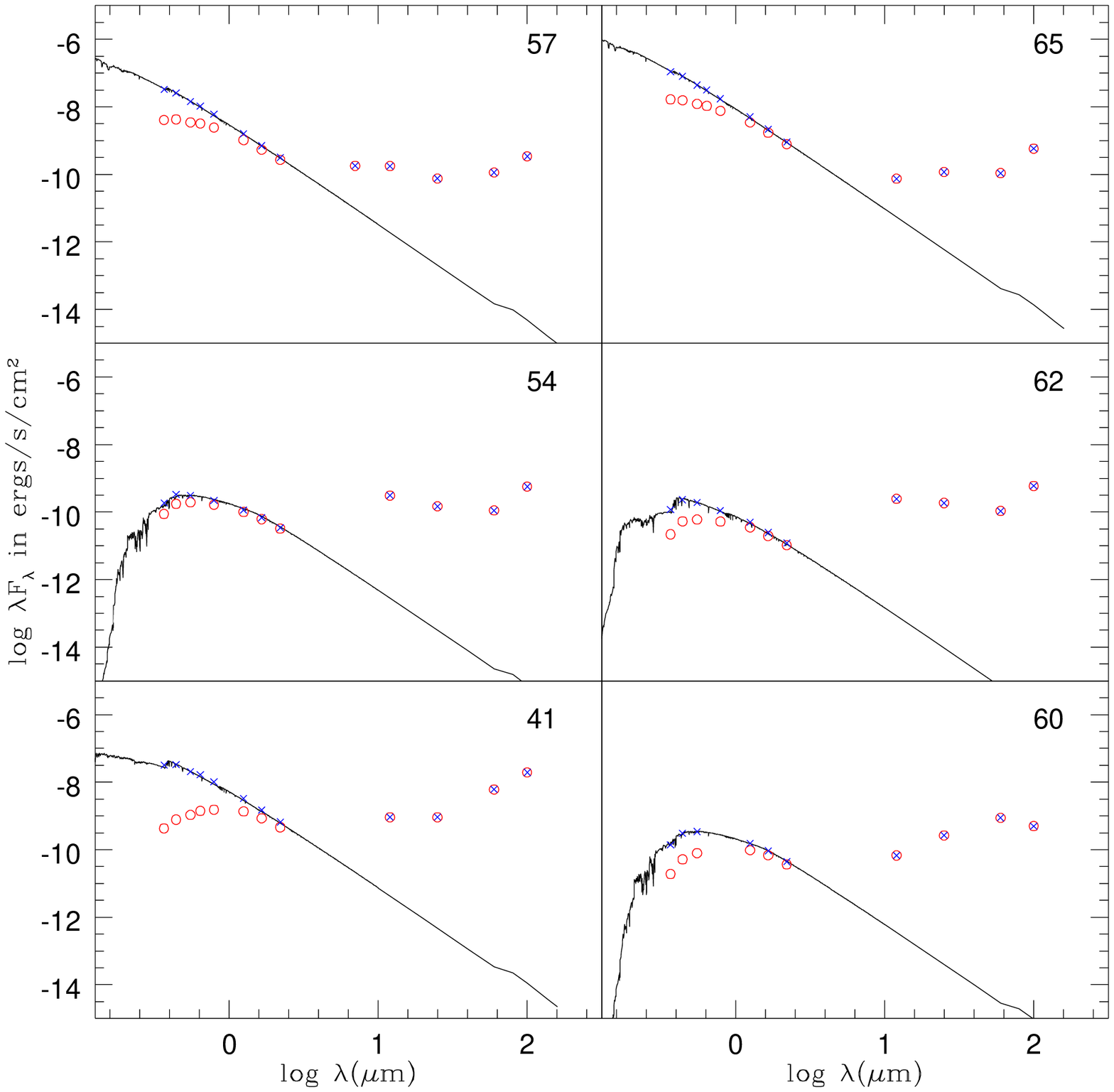}
\includegraphics[height=4.0in,width=3.4in]{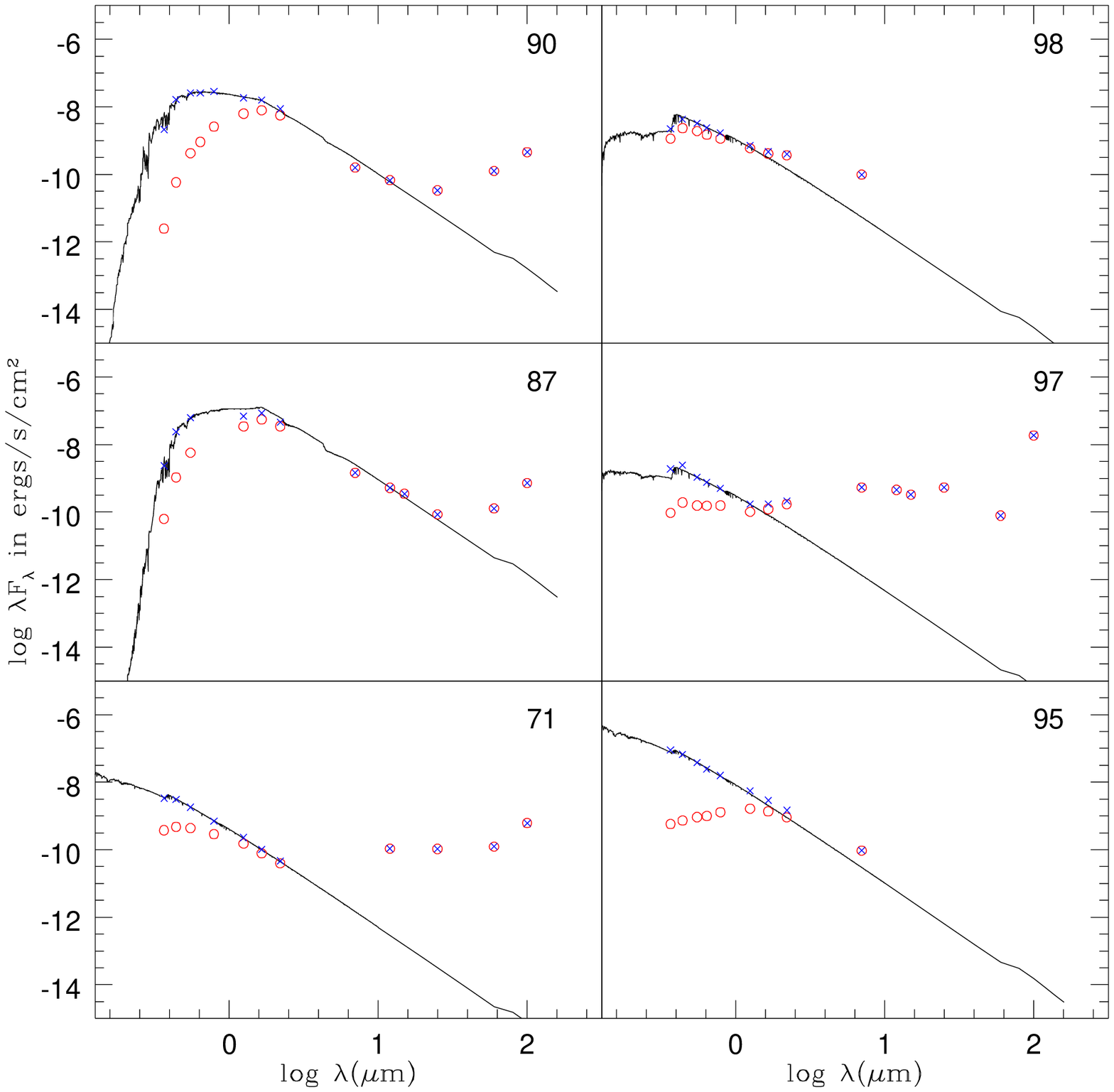}
\includegraphics[height=4.0in,width=3.4in]{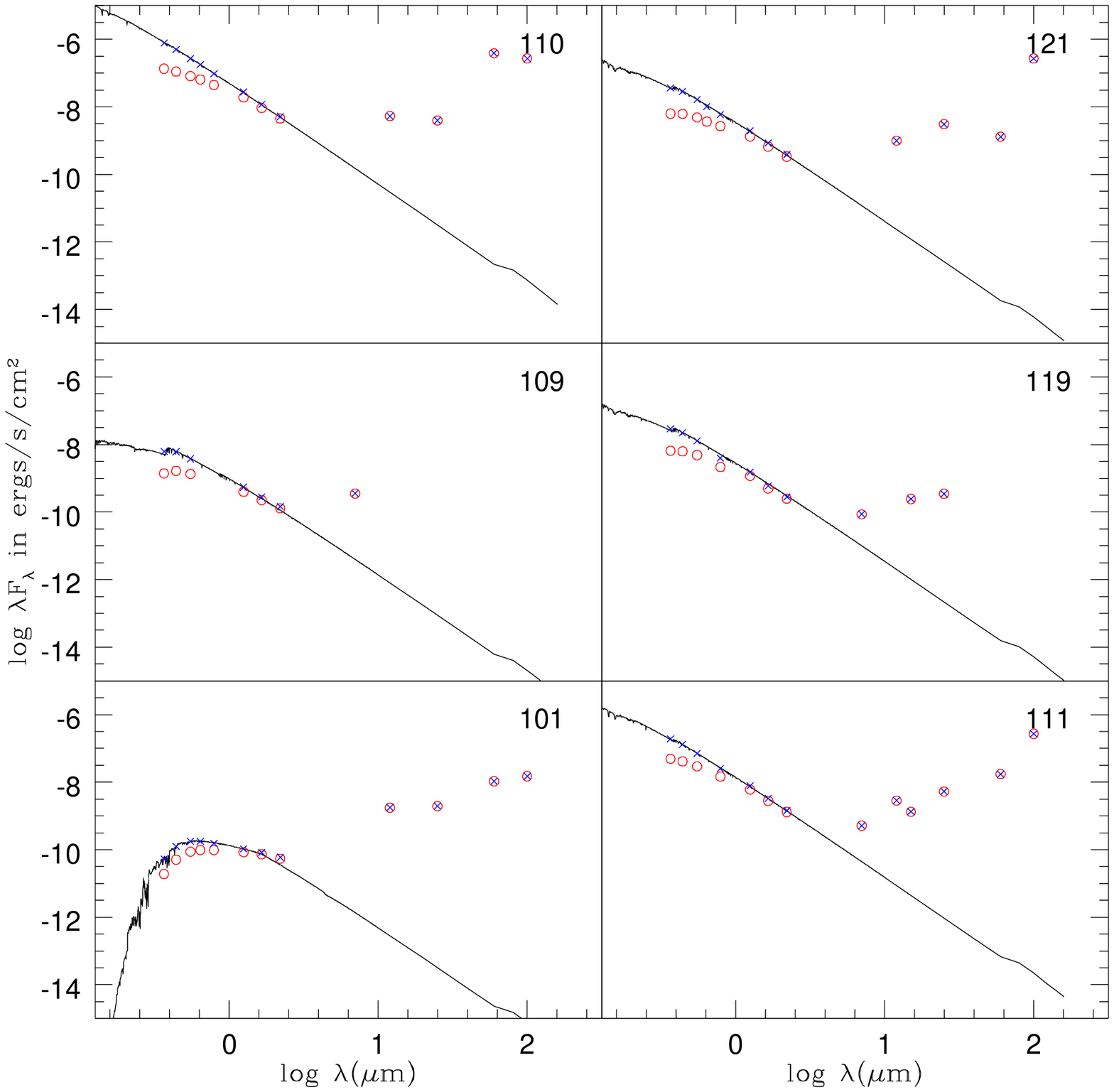}
\caption{Observed (open squares) and extinction-free (crosses) SEDs of program stars. The 
              solid lines represent the model SEDs from Kurucz (1993) as expected from the intrinsic properties of star. The model
              SEDs are adjusted to coincide at $V$ wavelength.}
\end{figure*}

\begin{figure*}
\includegraphics[height=4.0in,width=3.4in]{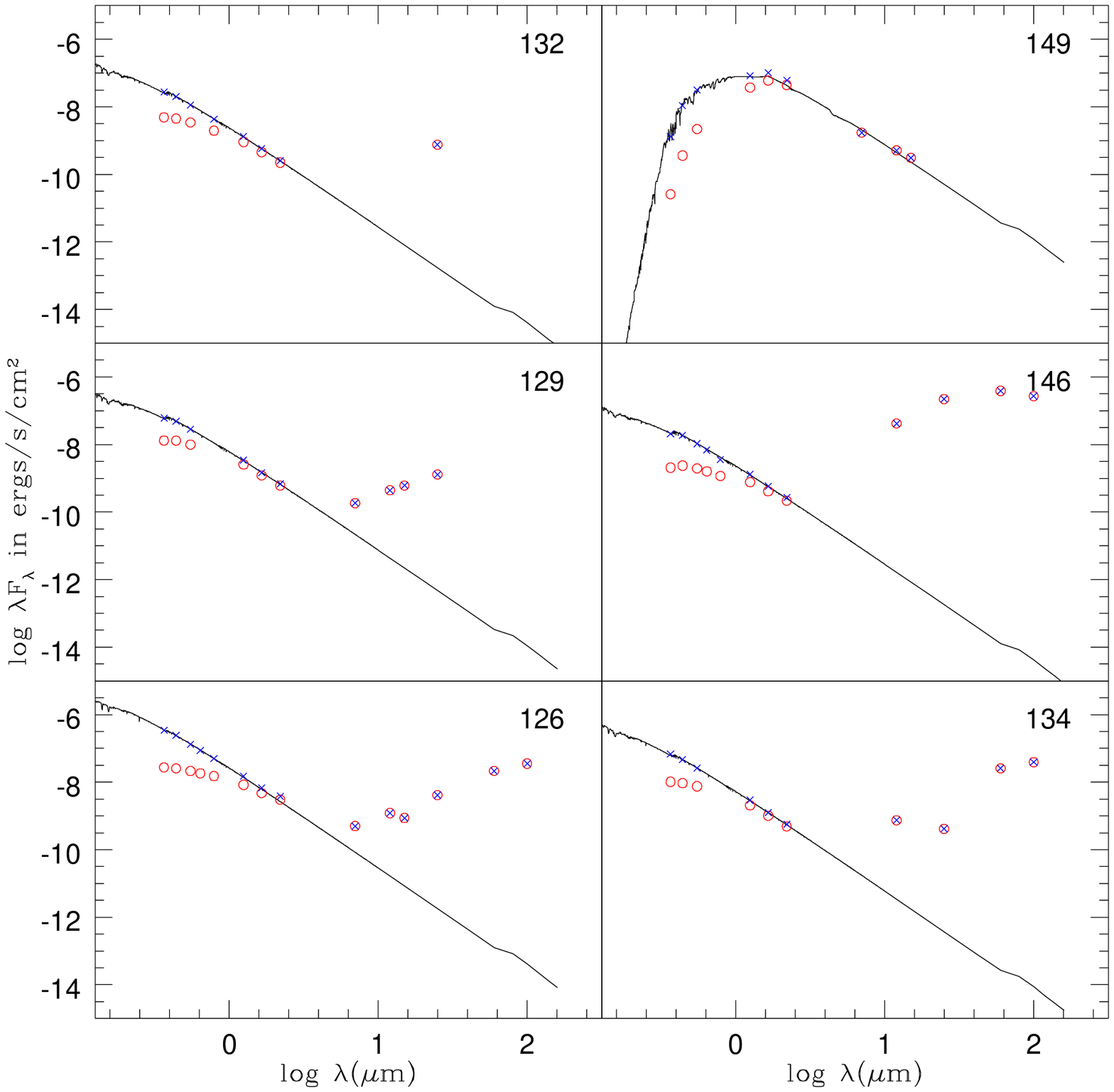}
\includegraphics[height=4.0in,width=3.4in]{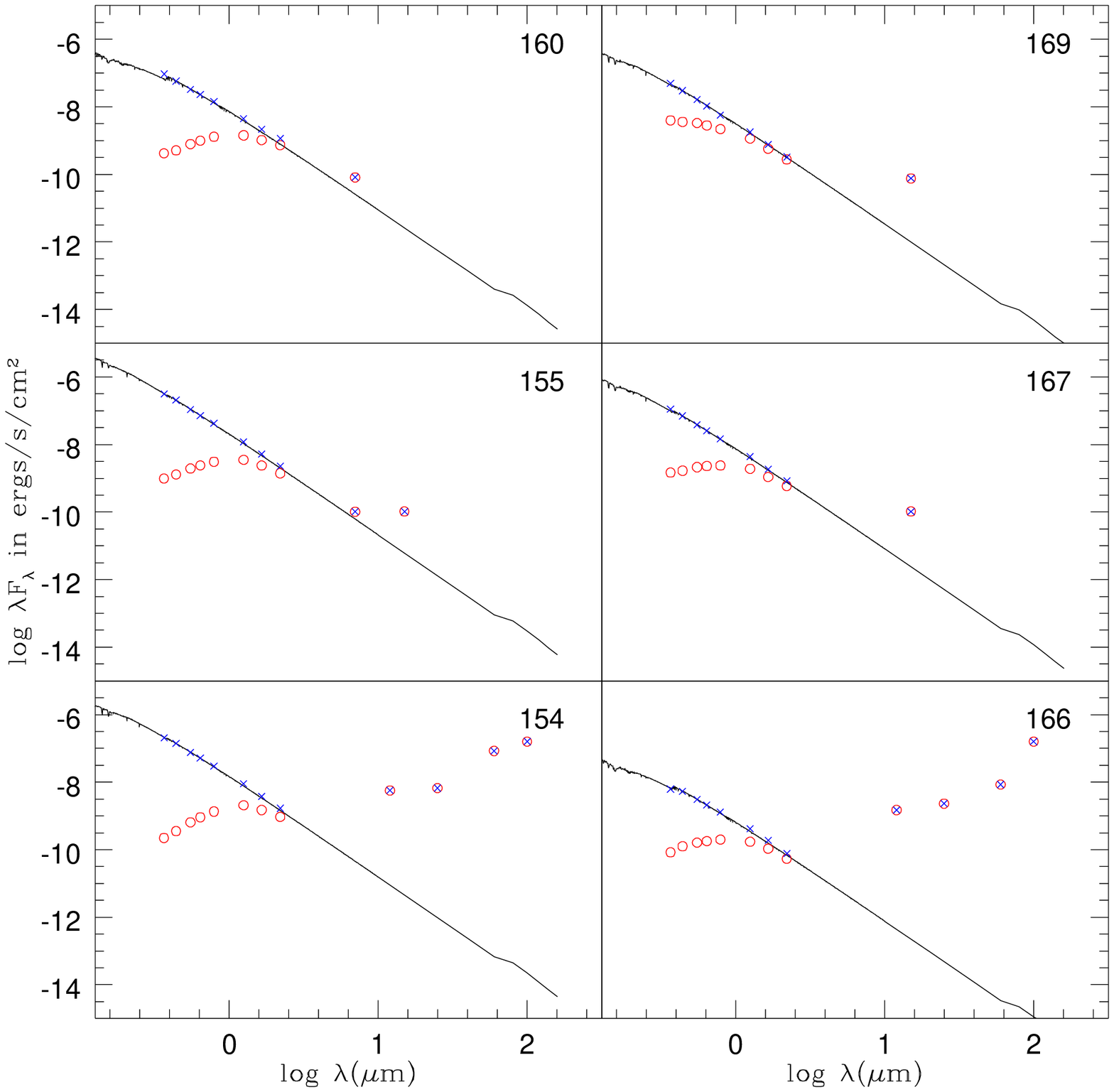}
\includegraphics[height=4.0in,width=3.4in]{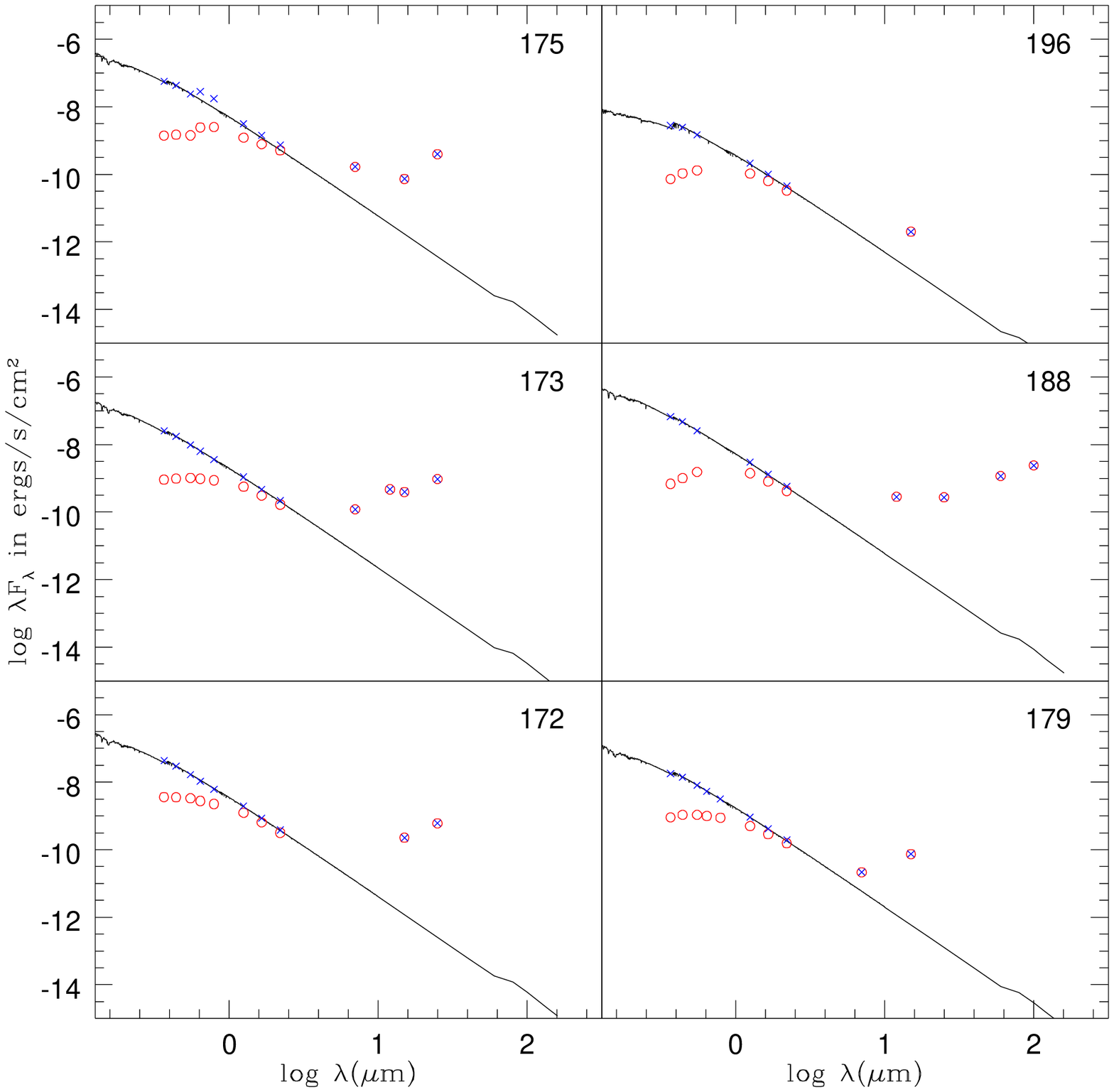}
\includegraphics[height=4.0in,width=3.4in]{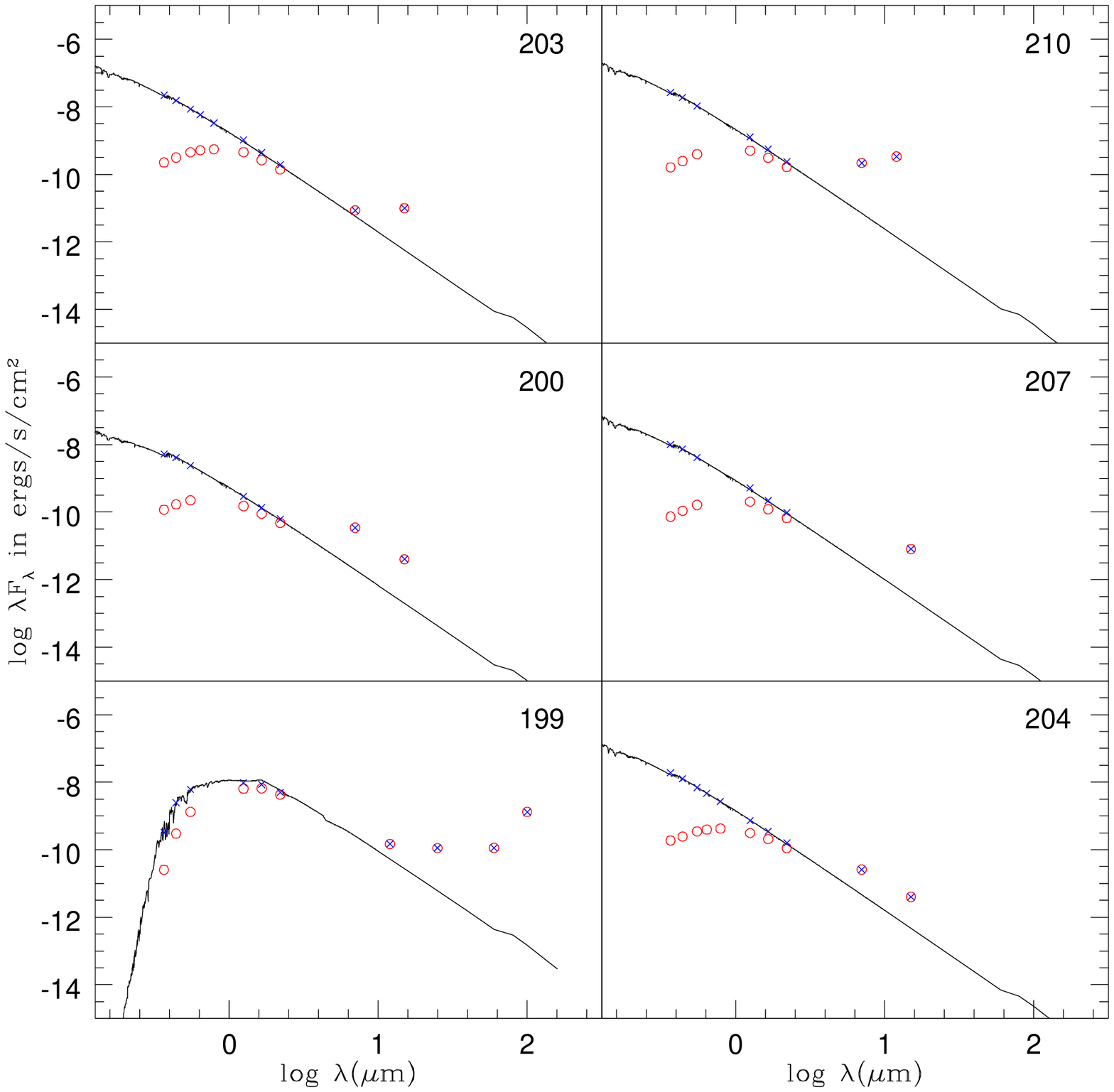}
{{\bf Figure. 12} Continued.}
\end{figure*}

\clearpage


\begin{figure*}
\includegraphics{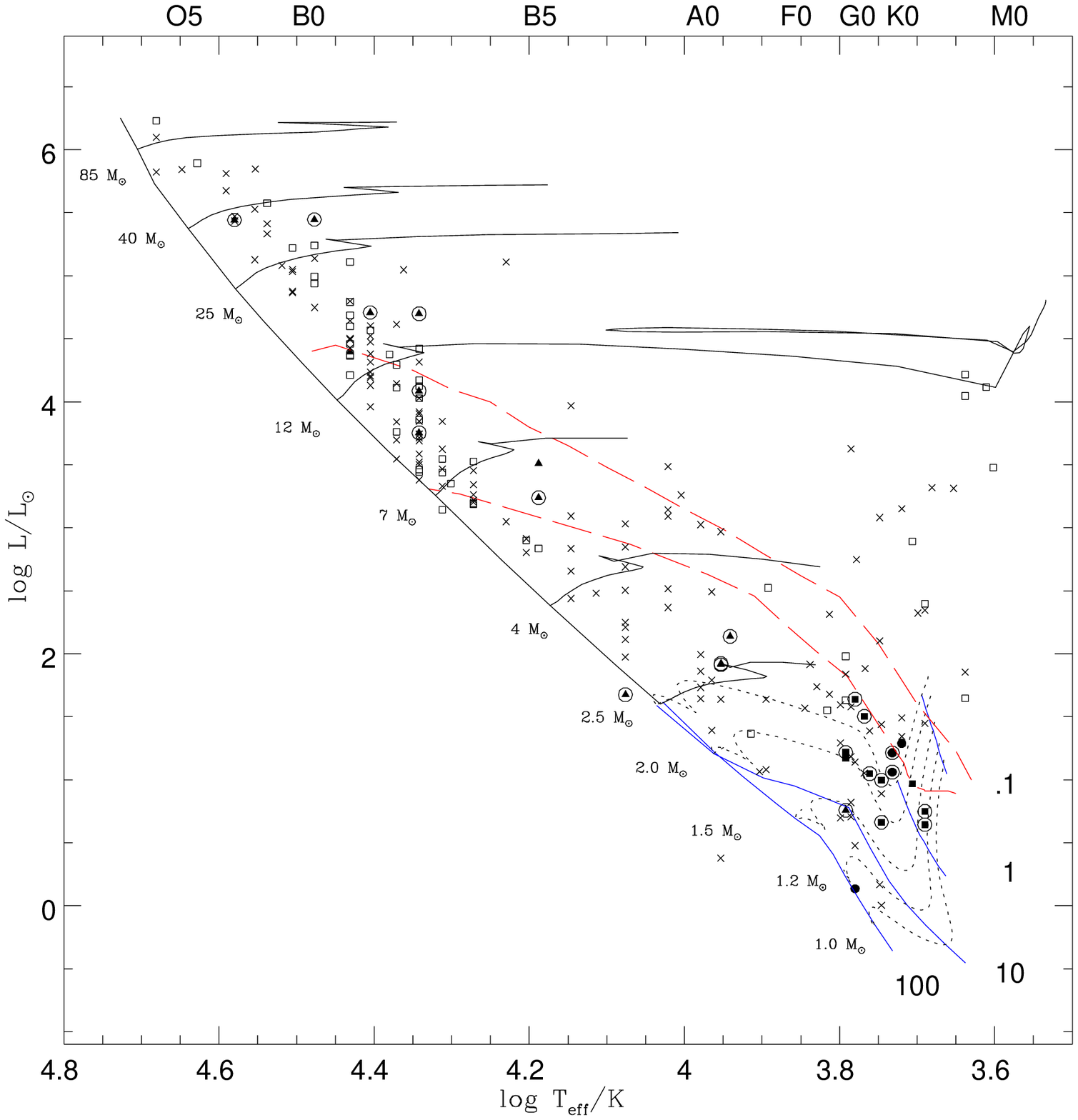}
\caption{The theoretical HR diagram of all the program stars. A group of 28 probable Herbig Ae/Be, classical Be or T Tauri 
population are shown by filled triangles, filled squares and filled circles representing stars with 
HI (Balmer), with HI+CaII HK and with CaII HK emissions respectively (see Tables 7 \& 11). The circling around these symbols 
denote NIR excess. The crosses denote normal stars while the open squares represent stars with circumstellar material as seen 
from either MIR
spectral indices or NIR excesses or from both (see Tables 10 \& 12). The dotted lines show PMS evolutionary tracks (1.0, 1.2, 
1.5, 2.0, 2.5 M$_{\odot}$) along with the isochrones (solid lines) for 0.1, 1, 10 and 100 Myr which are taken from D'Antona and 
Mazittelli (1994). The upper and lower dashed lines represent birthlines corresponding to the accretion rates 10$^{-4}$ 
and 10$^{-5}$ M$_{\odot}$/yr respectively and are taken from Palla and Stahler (1993). The ZAMS and post-MS evolutionary tracks
(85, 40, 25, 12, 7, 4 and 2.5 M$_{\odot}$) are shown with solid lines and are taken from Schaller et al. (1992).} 
\end{figure*}

\end{document}